\documentclass[%
 reprint,
nofootinbib,
 amsmath,amssymb,
 aps,
prd,
]{revtex4-2}

\usepackage{graphicx}
\usepackage{dcolumn}
\usepackage{bm}
\usepackage[normalem]{ulem}

\usepackage{color}
\usepackage{multirow}

\usepackage{hyperref}

\begin{document}

\preprint{APS/123-QED}

\title{Gravitational Radiation-Driven Chaotic Tide in a White Dwarf-Massive Black Hole Binary as a Source of Repeating X-ray Transients}

\author{Shu Yan Lau}
\email{shuyan.lau@montana.edu}
\affiliation{eXtreme Gravity Institute, Department of Physics, Montana State University,
Bozeman, MT 59717, USA}
\author{Hang Yu}
\email{hang.yu2@montana.edu}
\affiliation{eXtreme Gravity Institute, Department of Physics, Montana State University,
Bozeman, MT 59717, USA}




\date{\today}

\begin{abstract}
Eccentric white dwarf-massive black hole binaries can potentially source some extreme X-ray transients, including tidal disruption events and the recently observed quasi-periodic X-ray eruptions at the galactic nuclei. Meanwhile, they are one of the target gravitational wave sources with extreme mass ratios for future millihertz gravitational wave missions.
In this work, we focus on the tidal evolution and orbital dynamics of such binaries under the influence of the tidal backreaction and the dissipative effect from gravitational wave emission. We find that the latter can cause the dynamical tide to evolve chaotically after more than one orbital harmonics encounter the mode resonance through orbital decay. Different from the tidal-driven chaos, which only occurs for pericenter distances within around two times the tidal radius ($R_t$), this gravitational wave-driven chaos can happen at much larger pericenter distances. The growth scales similarly to a diffusive process, in which the tidal energy grows linearly with time on average over a long duration. 
If the tidal energy eventually approaches the stellar binding energy, achievable when the pericenter distance is less than $4R_t$, it can cause mass ejection as the wave breaks due to nonlinear effects. We show that this can potentially lead to the repeating partial tidal disruptions observed at galactic centers. That means these disruptions can occur at a much larger pericenter distance than previous analytical estimates. Furthermore, if the system can evolve to a close pericenter distance of about 2.7$R_t$, the white dwarf can further lose mass via tidal stripping at each pericenter passage. This provides a mechanism for producing the quasi-periodic eruptions. However, it requires at least a certain amount of angular momentum to escape the binary during the mass transfer process to match the observations, which we do not have a justification based on first-principle calculations.
\end{abstract}

\maketitle


\section{\label{sec:intro} Introduction}

Stellar mass objects orbiting a massive black hole (MBH) at the galactic nucleus are interesting sources of some rare electromagnetic (EM) transients.
If the orbit is tight enough, the binary can also be a potential gravitational wave (GW) source for the Laser Interferometry Space Antenna (LISA) \cite{AmaroSeoane_2017}, offering opportunities for multi-messenger astronomy study of such extreme environments. Since the MBHs have masses more than 5 orders of magnitude greater than the orbiting object, they are called extreme-mass-ratio-inspirals (EMRIs) if orbital period $\lesssim 10^{4}s$ \cite{AmaroSeoane_2018}.

EMRIs with significantly eccentric orbits can be formed via tidal disruption of a binary by the MBH or dynamical interactions with other stars. The former scenario is known as the Hills mechanism \cite{Hills_1988, Yu_2003, Miller_2005, Bromley_2012}, where the tidal force from the MBH unbinds a binary star system flying by at close distance, ejecting one of the stars and capturing the other one. The latter one corresponds to relaxations, in which the scattering processes cause exchange of angular momentum of the stars around the MBH, bringing them to close pericenter distances \cite{Hopman_2005, AmaroSeoane_2007, Gair_2017, AmaroSeoane_2018}.
Both channels can form EMRIs with high eccentricities, but the rates are still uncertain \cite{Babak_2017}. Besides, EMRIs can also be formed via interactions with the accretion disk of an active galactic nucleus. The resulting orbits are expected to have low eccentricities \cite{Pan_2021, Pan_2021_a, Li_2025}.

Out of the various transient signals observed at galactic centers, the recently discovered quasi-periodic eruptions (QPEs) have attracted much attention in the community on observation and theoretical modelling of X-ray phenomena.
QPEs are roughly periodic, repeating, bright X-ray bursts with a recurrence time of $\mathcal{O}(10)$ hours, whose origin is currently uncertain. To date, there are 9 confirmed QPE sources \cite{ Miniutti_2019, Giustini_2020, Arcodia_2021, Arcodia_2022, Arcodia_2024, Evans_2023, Guolo_2024, Nicholl_2024, Chakraborty_2025, Kara_2025} and 2 more candidates \cite{Chakraborty_2021, Quintin_2023, Bykov_2024}. For the high-confidence events, the recurrence time of the X-ray bursts ranges from $\sim 5$-$90$ hours. Numerous theoretical models have been proposed for the underlying mechanism of QPEs, including star/compact object-disk interactions \cite{Suková_2021, Xian_2021, Franchini_2023, Linial_2023, Zhou_2024, Zhou_2024_2, Miniutti_2025}, Roche lobe overflow of an orbiting star or white dwarf (WD) \cite{King_2020, Krolik_2022, Wang_2022}, two interacting circular EMRI systems \cite{Metzger_2022}, disk instability model \cite{Pan_2022, Pan_2023}, self-lensing of a binary MBH \cite{Ingram_2021}, etc. Further observations and long-term monitoring \cite{Miniutti_2023, Miniutti_2023_a, Pasham_2024} are needed to reveal the detailed properties of QPEs and narrow down their physical origin.

Tidal disruption events (TDEs) and partial TDEs (pTDEs) are another type of X-ray transients with different spectral and temporal properties from those of QPEs \cite{Suková_2024}.
While a full TDE results from a complete destruction of a star by the tidal field, a pTDE can happen multiple times until the star is fully disrupted or unbinds from the system. A pTDE is also expected to have a different mass fallback rate \cite{Coughlin_2019, Gezari_2021} than the power law dependence on time with an index $-5/3$ for full TDEs \cite{Rees_1988, Phinney_1989}.
There are observations of TDEs or pTDEs in at least four of the QPE sources \cite{Nicholl_2020, Chakraborty_2021, Miniutti_2023, Quintin_2023, Nicholl_2024, Jiang_2025}. In GSN 069, the pTDE repeats after a time gap of $\sim 9$ years \cite{Miniutti_2023}. While the origin of the QPEs is currently uncertain, there is increasing evidence showing that they are produced from EMRIs \cite{Miniutti_2023, Nicholl_2024} and have TDE associations. 
If the EMRI consists of a compact object, it can also be a strong GW emitter in LISA's frequency band \cite{AmaroSeoane_2007, Babak_2017}.  One example of such a multimessenger source is a WD-MBH binary \cite{Zalamea_2010}.


Wang \textit{et al.} \cite{Wang_2022} proposed a unified model for both QPEs and TDEs, consisting of a highly eccentric WD-MBH binary formed by the Hills mechanism. 
In this model, the dynamical tidal interaction of a fly-by encounter \cite{Press_1977} causes tidal heating in the WD envelope, triggering a tidal nova that inflates the WD envelope \cite{Fuller_2012}. This can be captured by the MBH, sourcing a TDE. As the pericenter distance becomes close enough, the WD loses mass to the MBH via Roche lobe overflow. The accreted matter then causes QPEs.

The event rate leading to this unified scenario is still under debate.
Metzger \textit{et al.} \cite{Metzger_2022} previously approximated the rate from the Hills channel to be $\sim10^{-10}\text{yr gal}^{-1}$, which is too low to be a QPE source.
However, this rate can be underestimated by orders of magnitude as pointed out by \cite{Wang_2022}, which we briefly summarize below:
The estimation of the rate of the Hills mechanism in \cite{Metzger_2022} does not consider the cusp galaxies, which have a much higher event rate \cite{Yu_2003}.
Moreover, they estimate that the lifetime of the close double WD binaries required to form the EMRI via Hills mechanism is very short only based on the timescale of GW decay, which ignores the possible mass transfer between the two WDs that can counteract the GW orbital decay effect and extend the lifetime by 1-2 orders of magnitude. On top of that, the population of close double WD systems is also underestimated by $\sim 3$ orders of magnitude \cite{Lamberts_2019} due to the assumption that they are homogeneously distributed in their age.

 Several aspects of the mass transfer process and the dynamical tidal interactions have been overlooked in the model by Wang \textit{et al.} \cite{Wang_2022}. First, the treatment of tidal stripping from Roche lobe overflow follows Eggleton's fit \cite{Eggleton_1983}, which is derived from synchronous, circular binary systems and is expected to deviate significantly from the highly eccentric, asynchronous cases \cite{Sepinsky_2007}. The nonzero tidal amplitude excited by previous passages is also not accounted for in the calculation of tidal stripping, which only considers the Roche lobe filling by the initially spherical WD. At such high eccentricities and close pericenter distances, the tidal amplitude developed in the previous orbits can be so large that this approximation fails.
 Second, the effect of mass transfer on the orbital evolution is significantly underestimated due to a mistake in their Eqs.~(7) and (8) (see Appendix~\ref{app:mapping_MT}). Moreover, the mass transfer model is limited to cases where the total mass and angular momentum are conserved within the binary.
 Third, the calculation of the dynamical tide that induces the tidal nova only includes the tidal amplitude from a single pericenter passage. In this paper, we mainly focus on the last two points.
 The first point corresponds to the deficiency of extending the analytical Roche's model to highly eccentric, asynchronous systems, which are known to deviate from the simulation results  \cite{Guillochon_2011, Liu_2012}. To provide an analysis of the mass transfer that accounts for these discrepancies, we use a parametrized, analytical model with reasonable prescriptions based on the simulation results.

If the tidal amplitude evolution over multiple orbits is taken into account, it can deviate from the single passage result by orders of magnitude for highly eccentric binaries as the evolution becomes chaotic for small pericenter separations \cite{Mardling_1995, Ivanov_2004, Ivanov_2007, Vick_2018, Wu_2018, Yu_2021, Yu_2022}.
The chaotic behavior comes from the fact that the previously excited tide, in the form of internal mode oscillations, can either be further excited or de-excited at the future pericenter passages, depending on the mode phase. When the tidal backreaction on the orbit is strong, or when the mode frequency is affected by anharmonicity due to large mode amplitude \cite{Yu_2021}, the mode phase is approximately a random quantity \cite{Ivanov_2004}.
This causes the tidal energy to evolve stochastically, with a linear scaling in the number of pericenter passages. The tidal amplitude is expected to grow by orders of magnitude from the initial tidal kick. When it approaches a significant fraction of the stellar radius, the nonlinear effect can cause the oscillations to break at the surface, leading to a mass ejection and dissipation of the tidal energy \cite{Wu_2018, Vick_2019, MacLeod_2022}. 

The purpose of this paper is to incorporate the effect of the dynamical tide and a more flexible mass transfer model in the highly eccentric WD-MBH binary.
We show that the chaotic evolution of dynamical tide can lead to both pTDEs and QPEs, when coupled to a non-conservative mass transfer model, and can potentially explain the observed features in both. That includes the repeated pTDEs observed in GSN 069 \cite{Miniutti_2023}, separated by $\mathcal{O}(10)\,{\rm years}$, the variations in the amplitude and recurrence time of QPEs, and the reappearance of the QPEs after a quenched phase \cite{Miniutti_2023_a, Chakraborty_2024}. However, for the QPE part, the mass transfer model requires specific fine-tuning of parameters to match the quantitative properties of the observed signals. That takes further justification from more fundamental calculations.

In the following, we first give a brief summary of the major results of this paper in Sec.~\ref{sec:summary}. In Sec.~\ref{sec:formulation}, we present an efficient iterative formalism to calculate the dynamical tide and orbital evolution, with or without mass transfer. We then present the numerical results of the evolution in Sec.~\ref{sec:chaos}, where we demonstrate the chaotic growth of the tide driven by GW backreaction on the orbit. In Sec.~\ref{sec:observations}, we discuss the relations of the WD-MBH system with pTDEs and QPEs, as well as being a GW source. Finally, we conclude the results and provide some future prospects of extending this work in Sec.~\ref{sec:conclusion}

We denote the masses of the WD and the MBH as $M_1$ and $M_2$, respectively. The WD radius is denoted by $R_1$. The total mass of the binary is denoted by $M$. We also define the characteristic tidal radius by $R_t = (M_2/M_1)^{1/3}R_1$.

\section{\label{sec:summary} Executive summary}

Let us briefly summarize the major results. For the first time, we show that the dynamical tide of a highly eccentric WD-MBH binary can evolve chaotically due to GW backreaction for $r_p/R_t \gtrsim 2.2$. The chaotic behavior arises from a series of resonances of the WD internal oscillation with the orbital harmonics as the orbit decays, 
and can occur for any $r_p/R_t$ as long as the orbit is eccentric enough, damping of the nonradial fundamental mode ($f$-mode) can be ignored, and there is sufficient time for the system to decay through at least two resonances.
The energy of the tide grows in a stochastic manner with a roughly linear scaling with time, similar to the case driven by tidal backreaction \cite{Mardling_1995, Ivanov_2004, Ivanov_2007}. 
Consequently, a large portion of the orbital energy is converted to tidal energy if the system evolves long enough before being circularized. The accumulated amount of tidal energy can significantly affect the orbital motion for $r_p/R_t\lesssim 4.2$, assuming the $f$-mode damping is negligible except for catastrophic wave breaking when the mode energy approaches the stellar binding energy. 
If only the tidal backreaction is incorporated, the chaotic evolution occurs within a restricted parameter range ($r_p/R_t \lesssim 2.2$, see Sec.~\ref{sec:chaos} below). We further demonstrate that the GW-driven chaos can lead to several scenarios closely related to astrophysical observables at galactic nuclei.

\begin{figure}[!tp]
    \includegraphics[width=1.\linewidth]{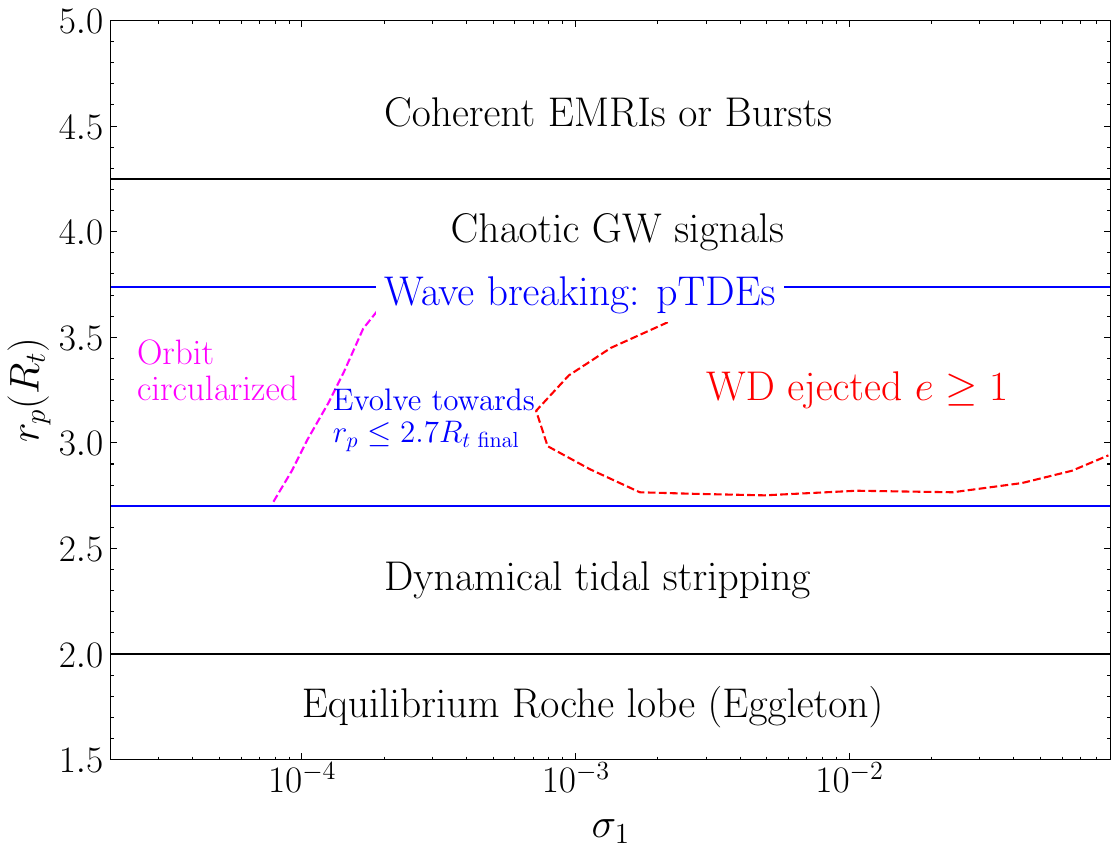}
    \caption{The various behavior of the highly eccentric WD-MBH binary in the parameter space formed by the initial $r_p$ and the fractional mass loss of the WD at wave breaking. From top to bottom, the regions correspond to where the system remains as a coherent GW source for LISA, a source for pTDEs, and a potential QPE source due to tidal stripping. There is also a region where the WD becomes unbound from the binary before reaching the tidal stripping stage due to mass transfer.}
    \label{fig:par_space}
\end{figure}

In Fig.~\ref{fig:par_space}, we provide a parameter space plot summarizing the several possible observables. The parameter $\sigma_1$, defined in Eq.~\eqref{eq:sigma_1} of Sec.~\ref{ssec:mass_transfer_formulation}, quantifies the fractional mass loss of the WD as the tidal energy reaches the wave-breaking limit.

\begin{itemize}
    \item $r_p/R_t \gtrsim 4.2$: The orbit circularizes enough before the tidal energy reaches the wave-breaking limit. Moreover, the shift in the orbital period due to the chaotically growing tidal amplitude is not large enough to be resolved in the GW signal for a 4-year mission. These systems behave as coherent EMRIs, as if there is no chaotic influence from the tides. However, the signals may be burst-like instead of continuous if the residual eccentricity is high, and are unlikely to be detected by LISA unless they occur within the nearby galaxies \cite{Rubbo_2006, Hopman_2007, Berry_2012, AmaroSeoane_2018}.

    \item $3.7 \lesssim r_p/R_t \lesssim 4.2$: The mode energy reaches a point where it can cause a phase shift larger than the resolution in the GW signal. Even with a continuous signal, it is chaotic and cannot be detected through coherent processing methods.

    \item $2.7 \lesssim r_p/R_t \lesssim 3.7$: the maximum tidal energy can get close to the stellar binding energy. The nonlinearity becomes significant and leads to wave breaking. The WD ejects a portion of its mass in this process, which leads to an increase in both the semi-major axis and eccentricity. The tidal energy is dissipated or released through mass ejection during wave breaking. The mass loss also causes an increase in $R_t$ as ($R_t\propto M_1^{-1/3} R_1\propto M_1^{-2/3}$), further enhancing the buildup of tidal energy. 
    We show that this process can lead to pTDEs at a much larger $r_p/R_t$ than the critical value from previous results for a single encounter \cite{Maguire_2020, Coughlin_2022}.

    Within this region, there are several possible fates of the system depending on the parameter $\sigma_1$: (1) Generally, for large $\sigma_1\gtrsim 10^{-3}$, the eccentricity of the binary grows larger than unity and the orbit becomes parabolic/ hyperbolic, which ejects the WD from the MBH. This is relevant to the previous findings about stars being kicked into a positive-energy orbit due to tidal disruptions \cite{Manukian_2013, Coughlin_2025} (while the speed is too low to be classified as hypervelocity stars). (2) The system reaches $r_p/R_t \leq 2.7$ before the binary unbinds, thus initiating dynamical tidal stripping \cite{Yu_prep}. (3) For small $\sigma_1$, the GW dissipation effect dominates the orbital evolution, so it circularizes before the WD or the binary gets fully disrupted. This ends the chaotic growth of the tide. Note that the above three outcomes are sensitive to the angular momentum change per mass loss during a non-conservative mass transfer. We assume this quantity equals the angular momentum per mass of the WD \cite{Huang_1963, MacLeod_2018}. See  Sec.~\ref{ssec:pTDE} and Appendix~\ref{app:angular_momentum_loss} for details.

    \item $2 \leq r_p/R_t \leq 2.7$: The large amplitude tidal bulge, built up through multiple pericenter passages, can lead to dynamical tidal stripping of the envelope. The critical $r_p/R_t$ is unknown for the WD-MBH system. Here, the value $2.7$ is assumed based on the simulation results of a Jupiter-like planet-star system \cite{Guillochon_2011}. This process can potentially lead to QPEs (see Sec.~\ref{ssec:QPE} below). However, for a valid mass transfer model that can survive long enough and have the correct recurrence time to match the observations, it requires a portion of angular momentum loss from the binary under a non-conservative mass transfer and fine adjustments on $r_p/R_t$. These are big unknowns in our current model and pose a caveat to interpreting our results.

    \item $r_p/R_t \leq 2$: Based on Roche's model, the Roche lobe overflow for a synchronous, circular binary with $M_1 \ll M_2$ occurs as the orbital separation reaches $2R_t$ (derived from Eggleton's fitting formula \cite{Eggleton_1983}). In this range, the mass shredding happens from the first pericenter passage.
    It is also worth noting that \cite{Coughlin_2022} estimated the distance for a partial tidal disruption to occur for a star by a MBH to be $\sim 1.67 R_t$ by equating the maximum internal gravity of the WD with the external tidal field from the MBH, though this value is much lower than our prediction with the dynamical tide built up over multiple orbits.
\end{itemize}

\section{\label{sec:formulation} Formulation}

In this section, we summarize the method used to calculate the dynamical tide coupled to the orbital evolution. 
For large $r_p/R_t$, the orbital motion is mainly affected by the backreaction from GW emission, which enters at 2.5 Post-Newtonian (PN) order. At close $r_p/R_t$, the tidal backreaction dominates. In the scenarios where the WD loses mass, e.g., from wave breaking, the mass transfer process also affects the orbital motion through a (partial) transfer of angular momentum between the WD and MBH.


\subsection{\label{ssec:dyn_tide_formulation} Dynamical tide in highly eccentric binaries}
We first carry out a phase space expansion of the fluid Lagrangian displacement vector $\boldsymbol{\xi}(t, \mathbf{x})$ and its time derivative \cite{Schenk_2001}:
\begin{align}
    \begin{bmatrix}
        \boldsymbol{\xi} \\
        \dot{\boldsymbol{\xi}}
    \end{bmatrix}
    = \sum_{\substack{a \\ {\omega_a} > 0}} q_a(t)
    \begin{bmatrix}
        \boldsymbol{\xi}_a(\mathbf{x}) \\
        - i \omega_a \boldsymbol{\xi}_a(\mathbf{x})
    \end{bmatrix} + (c.c.), \label{eq:phase_space_expansion}
\end{align}
where $\omega_a$ and $\boldsymbol{\xi}_a$ are, respectively, the eigenfrequency and eigenfunction of mode $a$, characterized by the set of indices $(n_a, \ell_a, m_a)$, representing the radial, polar, and azimuthal order, $(c.c.)$ is the complex conjugate\footnote{From the reality requirement on $\boldsymbol{\xi}$, a $\omega_a<0$ mode is equivalent to the complex conjugate of the $\omega_a>0$ mode with $-m_a$.
}. For conciseness, we drop the mode index $a$ in the subscripts of these individual indices in the following. The WD is assumed to be non-rotating. We also ignore nonlinear fluid effects, hence the anharmonicity-triggered chaos \cite{Yu_2021}. The eigenfunctions are normalized such that the mode energy (containing both the positive-frequency mode and its complex conjugate) is given by \cite{Weinberg_2012}
\begin{align}
    E_a = 2\omega_a^2 |q_a|^2 \int d^3x \rho |\boldsymbol{\xi}_a|^2 = E_*|q_a|^2,
\end{align}
where $E_* = G M_1^2/R_1$ is the stellar binding energy of the WD. The mode amplitude follows the equation of motion \cite{Schenk_2001}:
\begin{align}
    \dot q_a(t) = - i \omega_a q_a(t) + i \omega_a U_a(t), \label{eq:q}
\end{align}
where $\Phi$ is the azimuthal angle of the BH in the coordinate system positioned at the WD center, and the tidal force component is given by
\begin{align}
U_a(t) = \frac{M_2}{M_1} W_{\ell m} Q_a \left(\frac{R_1}{r}\right)^{\ell+1} e^{- i m \Phi}, \end{align}
where $r$ is the separation between the WD and MBH, 
and the coefficient $W_{\ell m}$ and tidal overlap integral $Q_{a}$ are defined by
\begin{align}
W_{\ell m} =&  \frac{4\pi}{2\ell+1}Y_{\ell m}\left(\pi/2,0\right),\\
Q_{a} =& \frac{1}{M_1 R_1^{\,\ell}}\int d^3 x \rho \boldsymbol{\xi}^*_a \cdot \boldsymbol{\nabla}\left[r^{\ell} Y_{\ell m}(\theta,\phi) \right].
\end{align}
For highly eccentric orbits, the tidal deformation is dominated by the ($\ell, m$) = (2,2) $f$-mode. This allows us to focus on one single mode. Moreover, the modulus of the complex mode amplitude and orbital elements can be well approximated with piecewise step functions except near pericenter. Hence, an efficient method has been developed to iteratively calculate the quantities orbit-by-orbit to accurately capture long-duration evolutions \cite{Ivanov_2004, Vick_2018}.

The $k$-th mode amplitude can be approximated with the iterative map:
\begin{align}
    \bar q_{a,k} = \left(\bar q_{a,k-1} + \Delta q_{a, k-1}\right) e^{- i \omega_a P_k}, \label{eq:map}
\end{align}
where
\begin{align}
    \bar q_{a,k} =&  q_{a, k} e^{- i \omega_a P_k/2}, \\
    \Delta q_{a, k} =& \int_{t_{k}}^{t_{k+1}} i \omega_a U_a(t^\prime) e^{i\omega_a t^\prime} dt^\prime,
\end{align}
$t_{k-1}$ and $t_{k}$ represent the time of the $k-1$-th and the $k$-th apocenter passages, respectively.
The tidal kick amplitude of the ($\ell, m$) = (2,2) $f$-mode can be well approximated by the asymptotic expansion in the parabolic limit \cite{Lai_1997}:
\begin{align}
    \Delta q_{a, k} =& i \frac{4\sqrt{\pi}}{3}W_{22} Q_a \left(\frac{M_2}{M_1}\right)\left(\frac{R_1}{r_p}\right)^3 \nonumber\\
    &\times z^{5/2} e^{-2z/3}\left(1 - \frac{\sqrt{\pi}}{4\sqrt{z}} + \mathcal{O}(z^{-1})\right), \label{eq:tidal_kick_approx}
\end{align}
where $z = \omega_a \sqrt{2r_p^3/(GM)}$. Notice that $\Delta q_{a, k}$ only depends on $r_p/R_t$ for $M_1 \ll M_2$ and $\omega_a \propto \sqrt{M_1/R_1^3}$.

The orbital period and eccentricities are also updated after each pericenter crossing based on the change in orbital energy and angular momentum. In this subsection, we focus on the tidal backreaction and GW backreaction.
\begin{align}
    a_k =& a_{k-1} + \Delta a_{\text{T}, k-1} + \Delta a_{\text{GW}, k-1}, \label{eq:a_backreaction}\\
    e_k =& e_{k-1} + \Delta e_{\text{T}, k-1} + \Delta e_{\text{GW}, k-1}, \label{eq:e_backreaction}
\end{align}
where $a_k = [GM P_k^2/(2\pi)^2]^{1/3}$ is the semi-major axis of the $k$th orbit, the subscripts T and GW represent tidal and GW backreaction, respectively. The tidal backreaction follows energy and angular momentum balancing, where we have neglected the change of tidal angular momentum due to the high eccentricity \cite{Vick_2018, Yu_2022}:
\begin{align}
    \frac{\Delta a_{\text{T}, k}}{a_{ k}} =& -2 \left(\frac{a_k}{R_1}\right)\left(\frac{M_1}{M_2}\right)\left(|q_{a, k+1}|^2-|q_{a, k}|^2\right),\\
    \Delta e_{\text{T}, k} =& \frac{1-e_k^2}{2e_k} \frac{\Delta a_{\text{T}, k}}{a_k},
\end{align}
and the GW backreaction at 2.5 PN order is given by \cite{Peters_1963}
\begin{align}
    \frac{\Delta a_{\text{GW}, k}}{a_k} =& -\frac{128\pi}{5}\eta \left(\frac{GM}{c^2 a_k} \right)^{5/2} \frac{1+\frac{73}{24}e_k^2 + \frac{37}{96}e_k^4}{(1-e_k^2)^{7/2}}, \label{eq:Peters_a}\\
    \Delta e_{\text{GW}, k} =& -\frac{608 \pi}{15}\eta \left(\frac{G M}{c^2 a_k} \right)^{5/2} \frac{e_k+\frac{121}{304}e_k^3}{(1-e_k^2)^{5/2}} \label{eq:Peters_e},
\end{align}
where the asymmetric mass ratio $\eta$ is defined as $M_1 M_2/M^2$. 

\begin{table*}
\begin{ruledtabular}
\begin{tabular}{ccccccc}
$M_1 (M_\odot)$ & $M_2 (M_\odot)$ & $R_1 (10^3 km)$ & $\omega_a (\sqrt{GM_1/R_1^3})$ & $Q_a$ & $P_0$ (hours) & $|E_*/E_{\text{orb}, 0}|$\\\hline
0.5 & $10^5$ & $10.6$ & $1.455$ & $0.243$ & $9$ & $0.066$ \\
\end{tabular}
\end{ruledtabular}
\caption{\label{tab:sources} The physical parameters of the WD-MBH model. The WD $f$-mode parameters, $\omega_a$ and $Q_a$ follows \cite{Wang_2022} (also see \cite{Lee_1986}).
}
\end{table*}

We have also ignored the damping of $f$-mode. The dominant dissipative effect comes from GW emission of the $\ell = 2$ $f$-mode, which has a time scale of $\mathcal{O}(10^3)$ years \cite{GarcaBerro_2006}. This is significantly longer than the growth timescale of the mode energy from the chaotic evolution discussed in Sec.~\ref{sec:chaos}, and therefore can be ignored.

The conservative PN corrections, including the orbital precession and Doppler shift on the mode frequency, are also neglected. Although these effects can be significant at the pericenter and have a less steep dependence on the orbital separation
than both the tidal backreaction and GW backreaction, they do not qualitatively affect the chaotic evolution.

\subsection{\label{ssec:mass_transfer_formulation} Non-conservative Mass Transfer in Highly Eccentric Binaries}
The orbital evolution due to non-conservative mass transfer in eccentric binaries is derived by Sepinsky \textit{et al.} \cite{Sepinsky_2009} based on angular momentum balancing.
In the highly eccentric limit, we can assume the mass transfer rate is well approximated by a Dirac delta function centered at the pericenter passage, which leads us to (see also Appendix~\ref{app:mapping_MT}):
\begin{align}
    \Delta e_{\text{MT},k} =& -2 \sigma_1 (1+e_k)\bigg[\left(\sigma_2 \frac{M_{1, k}}{M_{2, k}} - 1\right) \nonumber\\
    &+ (1-\sigma_2)\left(\sigma_3 + \frac{1}{2}\right) \frac{M_{1, k}}{M_k}\bigg], \label{eq:de_MT}\\
    \frac{\Delta a_{\text{MT},k}}{a_k} =& \frac{\Delta e_{\text{MT},k}}{1-e_{k+1}}, \label{eq:da_MT}
\end{align}
where $\sigma_1$, $\sigma_2$, and $\sigma_3$ are parameters governing the fractional mass ejection by the WD, the fraction of ejected mass accreted by the MBH, and the fractional change in angular momentum per mass, respectively, defined by
\begin{align}
    \sigma_1 =&  - \frac{\Delta M_{1, k}}{M_{1, k}}, \label{eq:sigma_1}\\ 
    \sigma_2 =& - \frac{\Delta M_{2, k}}{\Delta M_{1, k}}, \\
    \sigma_3 =& \; \frac{\Delta J_{\text{orb},k}/\Delta M_k}{J_{\text{orb},k}/M_k}. \label{eq:sigma_3}
\end{align}
Here, $\Delta M_i$ is the change in mass of object $i = 1, 2$ over one orbit, while $J_{\text{orb},k}$ and $\Delta J_{\text{orb},k}$ are the orbital angular momentum and its per-orbit change. We have the WD as the donor and the MBH as the accretor, meaning $\Delta M_1 < 0$ and $\Delta M_2 >0$.

Notice in Eq.~\eqref{eq:da_MT} that we employ an implicit iteration instead of a forward iteration like in Eqs.~\eqref{eq:a_backreaction}-\eqref{eq:Peters_e}. That is, the right hand side of Eq.~\eqref{eq:da_MT} depends on $e_{k+1}$ instead of $e_{k}$. This ensures that the mass transfer does not change the pericenter distance. As explained in Appendix~\ref{app:mapping_MT}, the forward iteration breaks down quickly as the system approaches the parabolic limit due to the increasing error in $\Delta P_{\text{MT}, k}$.

Equations~\eqref{eq:da_MT} and \eqref{eq:de_MT} are added to Eqs.~\eqref{eq:a_backreaction} and \eqref{eq:e_backreaction} during wave breaking only when the mode energy reaches a threshold, which we set at $E_{a, k} \geq 10\% E_*$ (see Sec.~\ref{ssec:pTDE}, and \cite{Wu_2018, Vick_2019}). The mode energy of the next step, $E_{a, k+1}$, is then reduced to 0.1\% $E_*$.

We treat $\sigma_1$ as a free parameter, and set
\begin{align}
    \sigma_2 = 
    \begin{cases} 
        \text{exp}\left[\alpha \left(\frac{r_p}{R_t} - 2\right)\right] & \mbox{for } r_p > 2 R_t, \\ 
        1 & \mbox{otherwise}.
    \end{cases}
\end{align}
This form of $\sigma_2$ is inspired by the simulation result of the accreted mass by a star from the tidal stripping or disruption of an orbiting Jovian planet \cite{Guillochon_2011, Liu_2012}. The parameter $\alpha$ is expected to be negative and of order unity based on their results. We further choose $\alpha = -1.43$, such that $\sigma_2 = \exp[-1]$ at $r_p/R_t = 2.7$. When setting the upper limit for $\sigma_2$, we use the equilibrium Roche lobe radius (for synchronous, circular systems), which is approximately $0.5 r_p q^{1/3}$ for $q \ll 1$ \cite{Eggleton_1983, Zalamea_2010, Wang_2022}. This gives $r_p/R_t = 2$ as a critical separation where we assume mass lost by the WD is fully accreted by the BH. 
We also assume the majority of the mass that escapes the binary is ejected at the WD's orbit due to wave breaking (i.e., the ``intermediate mode" of mass ejection in \cite{Huang_1963}). This implies the angular momentum loss per (total) mass is the same as that of the WD, which gives $\sigma_3 = M_2/M_1$ \cite{Huang_1963, MacLeod_2018, Yu_2024} (see Appendix~\ref{app:angular_momentum_loss}). We find that the evolution outcomes in Fig.~\ref{fig:par_space} are not very sensitive to $\sigma_2$, but can depend on the choice of $\sigma_3$. Further discussions on $\sigma_3$ are provided in Sec.~\ref{ssec:pTDE}.

For dynamical tidal stripping, we also employ Eqs.~\eqref{eq:de_MT} and \eqref{eq:da_MT} to describe its effect on the orbital dynamics, with modifications to $\sigma_1$ and $\sigma_3$. The parameter choice is phenomenological and will require further justifications from hydrodynamics results, which we leave as future work. This is discussed in detail in Sec.~\ref{ssec:QPE}.

Other than $a_k$ and $e_k$, the masses of the WD and MBH, as well as the radius of the WD, are also updated. This also affects the mode frequency, which is updated through the change in the dynamical frequency, $\sqrt{GM_1/R_1^3}$. The dimensionless overlap integral can also change within a factor of unity. However, we keep it fixed for simplicity.

\section{\label{sec:chaos} GW-driven chaotic growth}
In this section, we demonstrate that the GW backreaction can induce chaotic growth of the dynamical tide in a similar way to the tidal backreaction or mode anharmonicity. This will lead to the linear growth of the average tidal energy with the number of orbits, and the eventual mass transfer due to wave breaking if the tidal energy reaches unity of the stellar binding energy. 
In the following, we focus on a WD-MBH model with the parameters listed in Table~\ref{tab:sources}.
\begin{figure*}[!htp]
    \centering
    \includegraphics[width=0.48\linewidth]{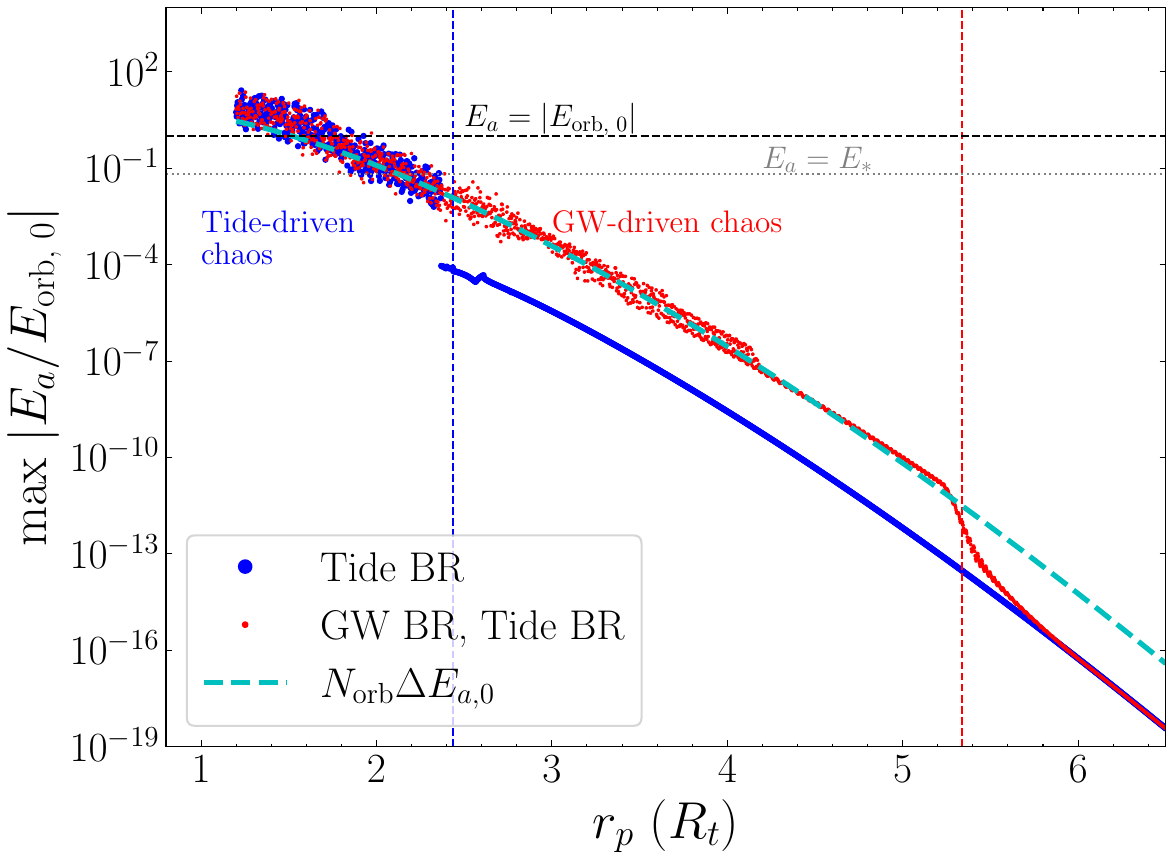}
    \includegraphics[width=0.48\linewidth]{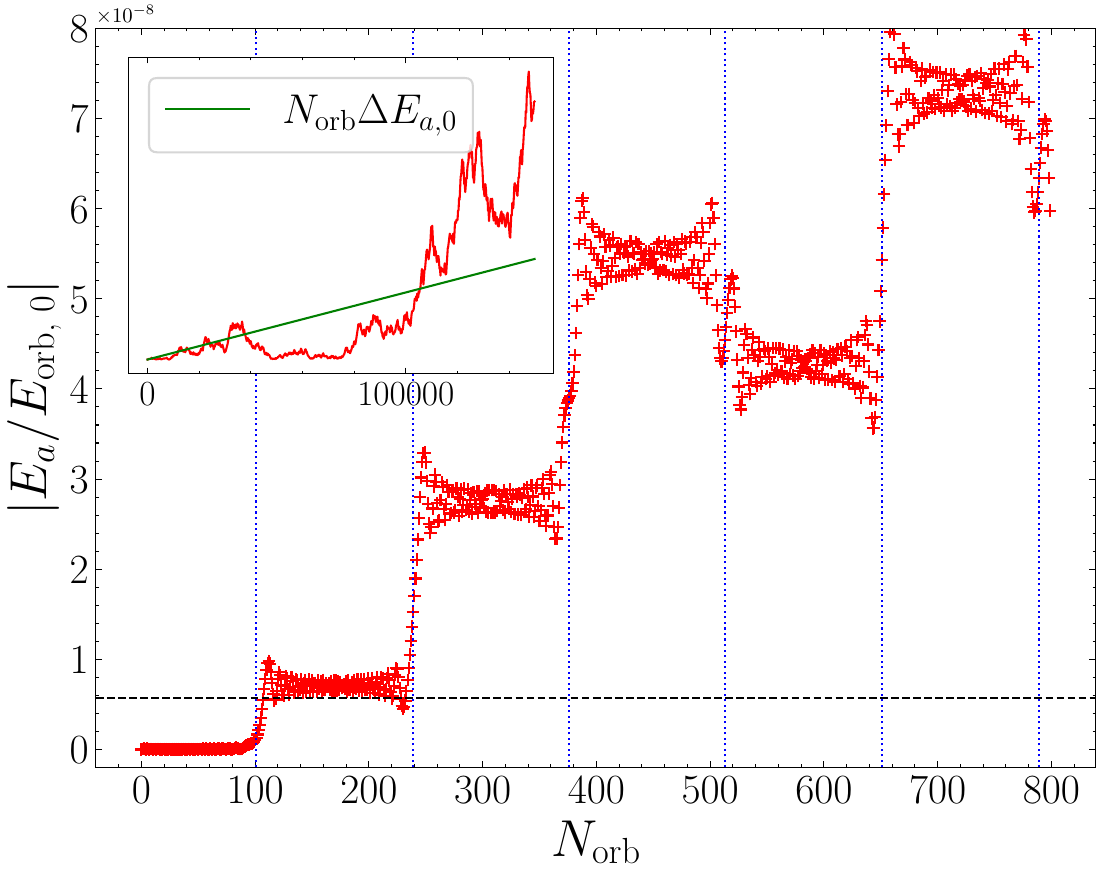}
    \caption{Left panel: The maximum mode energy (in units of the initial orbital energy $E_\text{orb, 0}$) over 200 orbits with initial $P_\text{orb} = 9$~hr and different $r_p$, obtained with the iterative method in Sec.~\ref{sec:formulation}. The blue dots are the case that only includes tidal backreaction, and the red dots include both tidal and GW backreaction. A blue dashed line indicates the initial $r_p$ where the change in mode phase at the first pericenter crossing equals unity. A red dashed line indicates where the GW orbital decay effect causes one resonance at the 200th orbit. Keep in mind that its position depends on the number of orbits. A cyan dashed line shows that the averaged mode energy roughly follows $N_\text{orb} \Delta E_{a, 0}$. The initial orbital energy and the stellar binding energy are shown with the horizontal lines.
    Right panel: The mode energy at different numbers of orbits, $N_\text{orb}$, with initial $r_p$ at $4.4 R_t$. The effects of tidal back reaction and GW backreaction are included. The vertical blue dotted lines correspond to $\omega_a P_k$ being integer multiples of $2\pi$. The inset shows the large $N_\text{orb}$ evolution of the mode energy, and the green line illustrates the linear growth of its averaged value. The horizontal black dashed line gives the SPA of the first resonance energy \eqref{eq:dq_res}.}
    \label{fig:Emode_chaos}
\end{figure*}

In the left panel of Fig.~\ref{fig:Emode_chaos}, we show the maximum mode energy over 200 orbits for the WD-MBH system at different pericenter distances but fixed initial orbital period $P_0=9\,{\rm hr}$. We present one case with tidal backreaction only, and another with both the tidal and GW backreaction.

For the case with only the tidal backreaction, it shows a distinctive chaotic region and a periodic region, roughly separated by a critical $r_p/R_t = 2.2$ where the tidal backreaction in a single orbit causes a change in mode phase larger than unity \cite{Ivanov_2004}:
\begin{align}
    \omega_a |P_1 - P_0| \gtrsim 1.
\end{align}
This causes the mode phase at each pericenter passage to behave as a random quantity, and the resulting evolution resembles that of a random-walk process \cite{Mardling_1995, Ivanov_2004, Ivanov_2007, Vick_2018}. The averaged mode energy has a linear scaling with the number of orbits, $N_\text{orb}$, as indicated by the dashed cyan line in the left panel of Fig.~\ref{fig:Emode_chaos}. Meanwhile, the maximum mode energy stays within the same order as the initial mode energy $\Delta E_{a, 0}$ when $r_p/R_t$ is in the periodic region.

When we include the GW backreaction, as illustrated by the red dots, the chaotic region extends to large $r_p$. Clearly, this chaotic behavior in the extended range of $r_p$ is driven by the GW-induced orbital decay. We further show that the randomness comes from the multiple $f$-mode resonances encountered by various orbital harmonics of the eccentric orbit. In the right panel of Fig.~\ref{fig:Emode_chaos}, we illustrate one example of the mode energy evolution against the number of orbits, $N_\text{orb}$, for $r_p/R_t = 4.4$. At this $r_p$, the tidal backreaction is negligible and the GW backreaction causes a continuous decay in $P_k$. While the mode energy evolves periodically within segments separated by the dashed blue lines that correspond to $\omega_a P_k$ being integer multiples of $2\pi$, i.e., when an orbital harmonic hits the $f$-mode resonance frequency. At those resonances, the mode amplitude changes drastically within several orbits. This causes a substantial change in the mode phase, $\omega_a P$, leading to chaotic behavior in the subsequent resonances in a similar manner to how tidal backreaction causes chaos.

The amount of change at the first resonance can be well estimated by the stationary phase approximation (SPA)
\begin{align}
    |\Delta q_\text{a, res}| \approx&  \left|\int_{-\infty}^{\infty} i \omega_a U_a(t^\prime) e^{i\omega_a t^\prime} dt^\prime\right|_{\omega_a P_k = 2 \pi \nu} \nonumber\\
    \approx& \omega_a P_k C_a \sqrt{\frac{a_k}{\nu |\Delta a_{\text{GW}, k}|}}X_{\ell m \nu}, \label{eq:dq_res}
\end{align}
where $\omega_a P_k = 2 \pi \nu$, with $\nu$ being an integer, $X_{\ell m \nu}$ is the modified Hansen coefficient
\begin{align}
    X_{\ell m \nu} =& \frac{1}{P_k} \int_{-P_k/2}^{P_k/2} \left(\frac{r_p}{r}\right)^{\ell+1} e^{-im \Phi + i 2 \nu \pi t/P_k} dt,\\
    C_a =& \frac{M_2}{M_1} W_{\ell m} Q_a \left(\frac{R_1}{r_p}\right)^{\ell+1}.
\end{align}
The first SPA resonance energy is shown with the black dashed line in the right panel of Fig.~\ref{fig:Emode_chaos}.

We also see that
\begin{align}
    |\Delta q_{a, 0}| \sim \omega_a P_0 C_a X_{\ell m \nu},
\end{align}
which gives us 
\begin{align}
    N_\text{res} |\Delta q_\text{a, res}|^2 \sim N_\text{orb} |\Delta q_{a, 0}|^2, \label{eq:gw_chaos_scaling}
\end{align}
where $N_\text{res}$ is the number of resonances the system goes through over a large $N_\text{orb}$. The scaling in Eq.~\eqref{eq:gw_chaos_scaling} requires the relation for the GW backreaction dominated case, 
\begin{align}
    N_\text{res} \approx \left(\frac{2\pi}{\omega_a \Delta P_{0, \text{GW}}}\right)^{-1} N_\text{orb},
\end{align}
with the change in the period over the first orbit caused by GW decay given by $\Delta P_{0, \text{GW}} \approx 1.5 P_0 \Delta a_{0, \text{GW}}/ a_0$.

Notice, in the left panel, that as $r_p/R_t$ increases above $4.2$, the maximum mode energy is no longer scattered, but still stays at around $N_\text{orb} \Delta E_{a,0}$ for $r_p/R_t \lesssim 5.5$. 
In this range, the system only goes through one $f$-mode resonance throughout the 200 orbits.
Since the chaotic behavior requires changes due to the resonance to back-react onto the orbit which affects the subsequent resonances, it does not emerge when there is only a single resonance.
For $r_p/R_t \gtrsim 5.5$ (indicated by the red dashed line), i.e., the range where the system does not go through one resonance, the maximum energy becomes the same order of magnitude as $\Delta E_{a, 0}$. If we further increase the number of orbits, the chaotic region can extend to larger $r_p$.

When $r_p$ is a few factors larger than $R_t$, the orbital evolution is dominated by the GW-backreaction. The number of orbits the system is then limited by the GW circularization timescale.
We can estimate the condition for the GW-driven chaos to happen by requiring that there are at least two resonances before the orbit circularizes enough such that the random-walk behavior can no longer exist, while the orbital decay is not too fast so that it contains a considerable number of orbits in the vicinity of resonance:
\begin{align}
    1 \ll \frac{4\pi}{\omega_a \Delta P_{0, \text{GW}}} \leq N_\text{max}, \label{eq:condition_chaos}
\end{align}
with
\begin{align}
    N_\text{max} = \frac{e_0 - e_\text{f}}{\Delta e_\text{GW}} \label{eq:nmax},
\end{align}
where the final eccentricity $e_f$ is set as 0.9. This leads to
\begin{align}
    \omega_a P_0 \geq \frac{38\pi}{9(e_0-e_f)}\frac{e_0 (1-e_0^2)\left(1+\frac{121}{304} e_0^2\right)}{1+\frac{73}{24}e_0^2 + \frac{37}{96}e_0^4} \approx 41.9 (1-e_0^2).
\end{align}
At high eccentricities, this condition sets a lower bound on $\omega_a P_0$, which is always satisfied by the WD-MBH system ($\omega_a P_0 \sim 10^5$). The other bound in Eq.~\eqref{eq:condition_chaos} is also satisfied as $\omega_a \Delta P_{0, \text{GW}} \sim \mathcal{O}(0.01)$.

Therefore, we have shown that the averaged mode energy will grow linearly with the $N_\text{orb}$ if we include the GW backreaction in the highly eccentric WD-MBH system, even for those with $r_p/R_t$ significantly larger than $2$. The chaotic behavior and the iterative method results have been checked against a numerical integration, which shows good agreement (see Appendix~\ref{app:numerical_check}).

\section{\label{sec:observations} Relations with Observations}

\subsection{\label{ssec:pTDE} Partial Tidal Disruptions from Wave Breaking}
As the mode energy comes close to the binding energy of the WD due to chaotic growth, the $f$-mode can break at the surface, resulting in energy dissipation and mass loss \cite{Wu_2018, MacLeod_2022}. The portion accreted by the MBH can source the transient signal observed as a pTDE. In this subsection, we consider the evolution of the system, including the effect of non-conservative mass transfer, where the total mass and angular momentum of the system are not conserved.

We assume the $f$-mode wave breaking occurs once it reaches 10\% of $E_*$, as the WD passes the pericenter. Meanwhile, the mode energy is instantaneously reduced to 0.1\% of $E_*$, while the mode phase remains the same. 
Note that the choice of the mode phase immediately after wave breaking does not affect the results qualitatively.
A similar prescription is employed in \cite{Wu_2018, Vick_2019} when studying the role of chaotic tides in hot Jupiter migrations. As noted in \cite{Vick_2019}, the orbital evolution is not sensitive to the choice of the wave-breaking limits, but can depend on the residual energy in the mode after breaking. Nonetheless, the overall growth in a random-walk manner is not affected.

The linear growth of mode energy in chaotic evolution is limited by several factors. They determine the fate of the system shown in Fig.~\ref{fig:par_space}. Other than the nonlinear dissipation from wave breaking that eventually suppresses the growth, the orbital circularization by GW backreaction also sets a limit to the maximum number of orbits, $N_\text{orb} \leq N_\text{max}$ (see Eq.~\eqref{eq:nmax}). This condition is applied for the parameter range in which the GW backreaction dominates the orbital evolution ($r_p/R_t \gtrsim 3.7$).
When the system is inside the parameter space where wave breaking can occur, the eccentricity is affected by both the GW backreaction and mass transfer. For $\sigma_1 \gtrsim 10^{-4}$, the mass transfer effect dominates, causing the orbit to migrate towards the parabolic limit (see Fig.~\ref{fig:par_space}). The system is then evolved until $e \geq 1$ or $e \leq 0.9$.

\begin{figure}[!tp]
    \includegraphics[width=1.\linewidth]{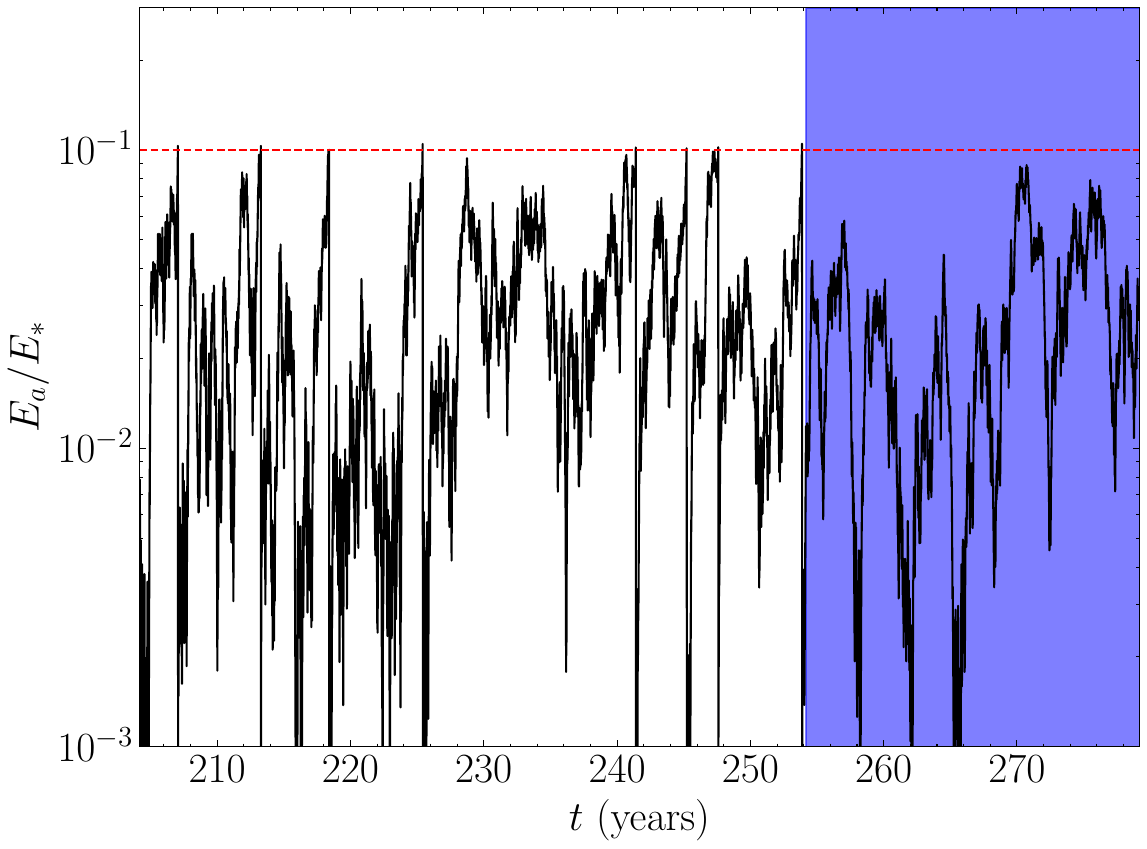}
    \caption{The mode energy evolution normalized by the WD stellar binding energy. The initial $r_p/R_t$ is $2.9$, and the fractional mass ejected by the WD per orbit, $\sigma_1$, is chosen as $5\times 10^{-4}$. The purple region corresponds to $r_p/R_t \leq 2.7$, where we expect the mass loss via dynamical tidal stripping further alters the orbital dynamics. The horizontal dashed line is used to indicate the threshold for wave breaking, i.e., $E_a = 10\% E_*$.}
    \label{fig:Emode_MT}
\end{figure}

In Fig.~\ref{fig:Emode_MT}, we show the mode energy evolution in a case with initial $r_p/R_t = 2.9$. The effect of mass transfer due to wave breaking is incorporated through including Eqs.~\eqref{eq:de_MT} and \eqref{eq:da_MT} in the orbital element evolution (Eqs.~\eqref{eq:a_backreaction} and \eqref{eq:e_backreaction}), with $\sigma_1 = 5\times 10^{-4}$. 
The angular momentum loss from the binary is parametrized by $\sigma_3$, which is chosen to be $M_{2,k}/M_{1,k}$ as explained in Sec.~\ref{ssec:mass_transfer_formulation} and Appendix~\ref{app:angular_momentum_loss}.
The mode energy grows chaotically, with a roughly linear scaling with time, until it reaches the wave-breaking limit. With the above-specified mass transfer parameters, we find that the time between wave-breaking is of $\mathcal{O}(10)$ years, which matches the recurrence time of the pTDEs detected in GSN 069 \cite{Miniutti_2023} (around 9~years).

We can also estimate the recurrence time using the linear scaling of the mode energy with $N_\text{orb}$:
\begin{align}
    T_\text{pTDE} \sim \frac{0.1}{|\Delta q_{a, 0}|^2} P_0.
\end{align}
For $r_p$ ranging from $2.7$-$2.9 R_t$, we have $T_\text{pTDE} \sim \mathcal{O}(1)$-$\mathcal{O}(10)~\text{years}$.

To estimate the brightness of the signal produced by the mass ejected from each wave breaking, we use \cite{Rees_1988, Phinney_1989}
\begin{align}
    L \approx \epsilon \frac{\Delta m_\text{acc} c^2}{3 P_\text{min}} \label{eq:luminosity}
\end{align}
where $\epsilon\approx 0.1$ is the efficiency of converting gravitational energy to electromagnetic energy, $\Delta m_\text{acc} = \sigma_1 \sigma_2 M_1$ is the mass that falls towards the MBH, and $P_\text{min}$ is the period of the most bound ejected material given in \cite{Ulmer_1999}:
\begin{align}
    P_\text{min} = \frac{2\pi r_p^3}{\sqrt{G M_2 (2 R_1)^3}}.
\end{align}
We find that for $r_p = 2.9 R_t$, $\sigma_1 = 5\times 10^{-4}$, it has a peak luminosity of $\sim 10^{43}~\text{erg s}^{-1}$, which matches the observed pTDE in GSN 069 \cite{Miniutti_2019}.


As the WD loses mass, its radius increases. We employ the fit by \cite{Zalamea_2010, Wang_2022} to estimate the evolution of the radius:
\begin{align}
    R_1 = 9.04\times 10^8 \text{cm} \left(\frac{M_1}{M_\text{Ch}}\right)^{-1/3} \left(1 - \frac{M_1}{M_\text{Ch}}\right)^{0.447},
\end{align}
where $M_\text{Ch} = 1.44 M_\odot$. This causes $R_t$ and $\sigma_2$ to increase. As $r_p/R_t$ reaches a threshold value, which we set as $2.7$, dynamical tidal stripping occurs (the purple region of Fig.~\ref{fig:Emode_MT}). This has the periodicity of $\mathcal{O}(10)$ hours and therefore can potentially cause QPEs, which we further discuss in Sec.~\ref{ssec:QPE}. For now, we do not include the effect of tidal stripping on the orbital dynamics within this region.

\begin{figure}[!tp]
    \centering
    \includegraphics[width=1.0\linewidth]{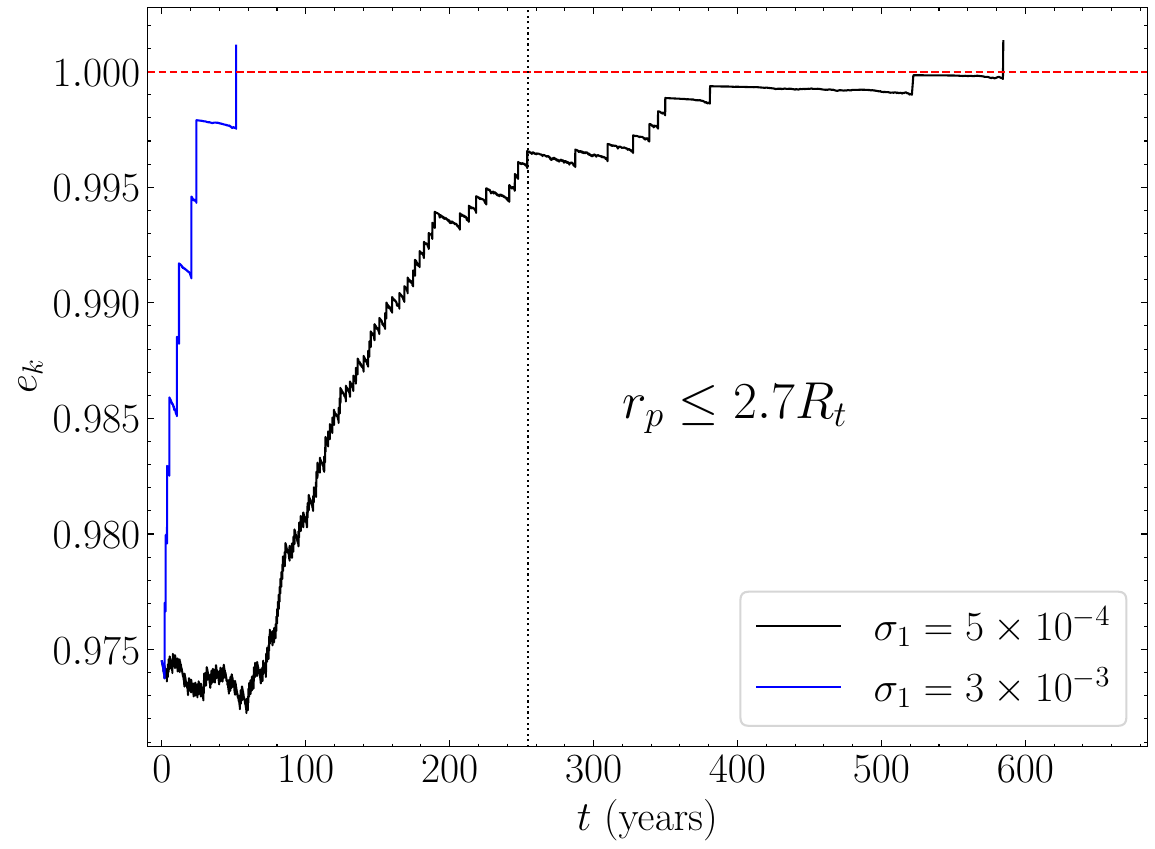}
    \caption{The eccentricity evolution under the effects of tide, GW backreaction, and wave breaking. The black curve corresponds to the same system in Fig.~\ref{fig:Emode_MT}. A vertical black dotted line shows where $r_p/R_t = 2.7$ for the case with $\sigma_1 = 5\times 10^{-4}$. For $r_p/R_t$ smaller than that, we expect dynamical tidal stripping. The blue curve has the same initial pericenter distance and period, but a different $\sigma_1$. The red dashed line indicates the $e = 1$ limit where the orbit unbinds.}
    \label{fig:e_vs_t_MT}
\end{figure}

Another possible fate of the system is the unbinding of the binary as the eccentricity increases above $1$ before $r_p/R_t$ reaches $2.7$, the prescribed threshold for dynamical tide stripping. This occurs in a region with large $\sigma_1$ as shown in Fig.~\ref{fig:par_space}. An example is shown in Fig.~\ref{fig:e_vs_t_MT}, where the eccentricity increases over multiple wave breaking for both cases. For $\sigma_1 = 3\times 10^{-3}$, the binary system unbinds before it can reach the distance 
for dynamical tidal stripping. For $\sigma_1 = 5\times 10^{-4}$, the system reaches the critical $r_p/R_t$ for dynamical tidal stripping. 
The effects on the orbital dynamics require detailed modelling of the stripping process, which is not within the scope of this paper. In Sec.~\ref{ssec:QPE}, we utilize our mass transfer model (Eqs.~\eqref{eq:de_MT} and \eqref{eq:da_MT}), with some crude assumptions on the mass transfer parameters Eqs.~\eqref{eq:sigma_1}-\eqref{eq:sigma_3}, to roughly predict the orbital motion during this process.

Lastly, we remark that the magnitude and sign of $\Delta e_\text{MT}$ and $\Delta a_\text{MT}$ depend on the fractional loss of angular momentum per mass, $\sigma_3$. In Sec.~\ref{ssec:mass_transfer_formulation} and Appendix~\ref{app:angular_momentum_loss}, we explain the choice of $\sigma_3 = M_{2, k}/M_{1, k}$ for wave breaking due to the location where we expect the mass to escape from the binary in the idealized scenario. If some of the mass is lost at other locations, we have a different $\sigma_3$, and the outcome of the resulting orbital evolution can change drastically. For instance, a larger $\sigma_3$ may give a smaller or even negative $\Delta e_\text{MT}$, and the binary will not be unbound as illustrated in Fig.~\ref{fig:par_space}, even for large $\sigma_1$. In this case, the WD can be fully disrupted after encountering multiple wave breakings.

\subsection{\label{ssec:QPE} Quasi-periodic Eruptions from Tidal Stripping}

\begin{figure*}[!tp]
    \includegraphics[width=0.49\linewidth]{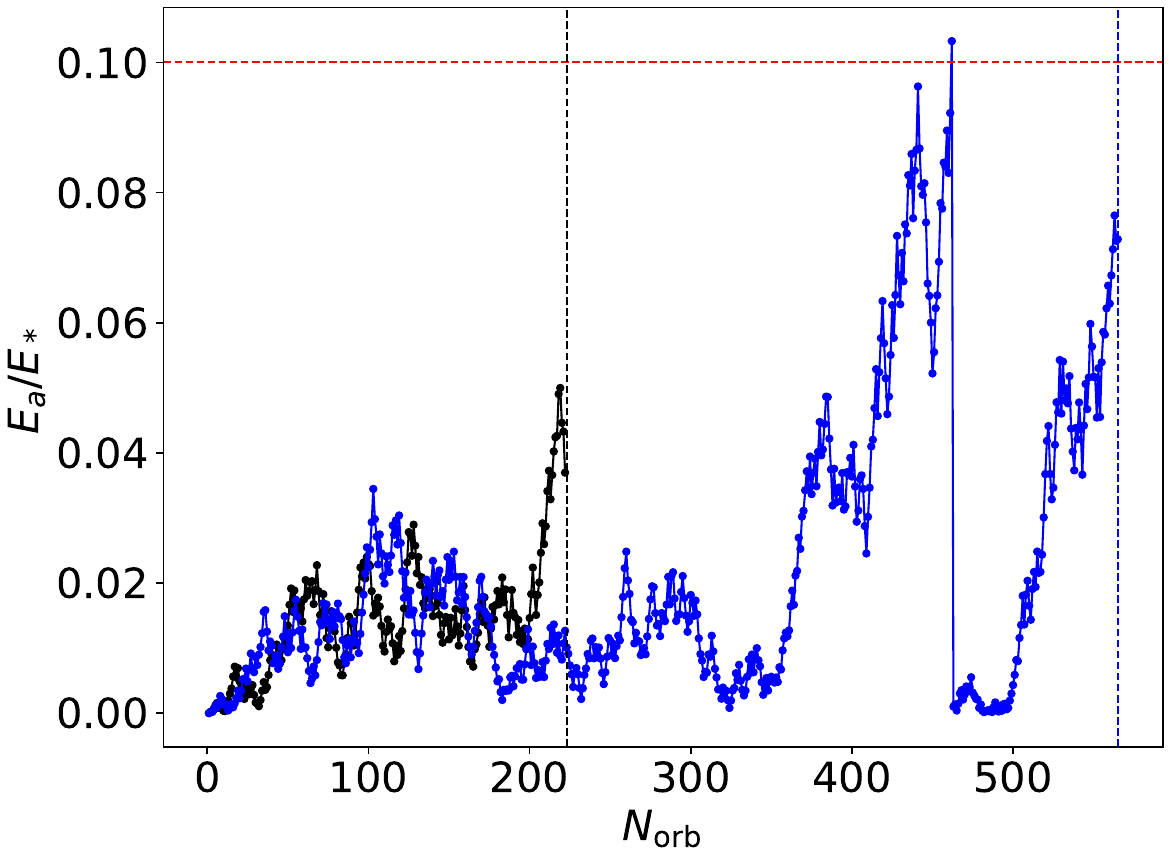}
    \includegraphics[width=0.49\linewidth]{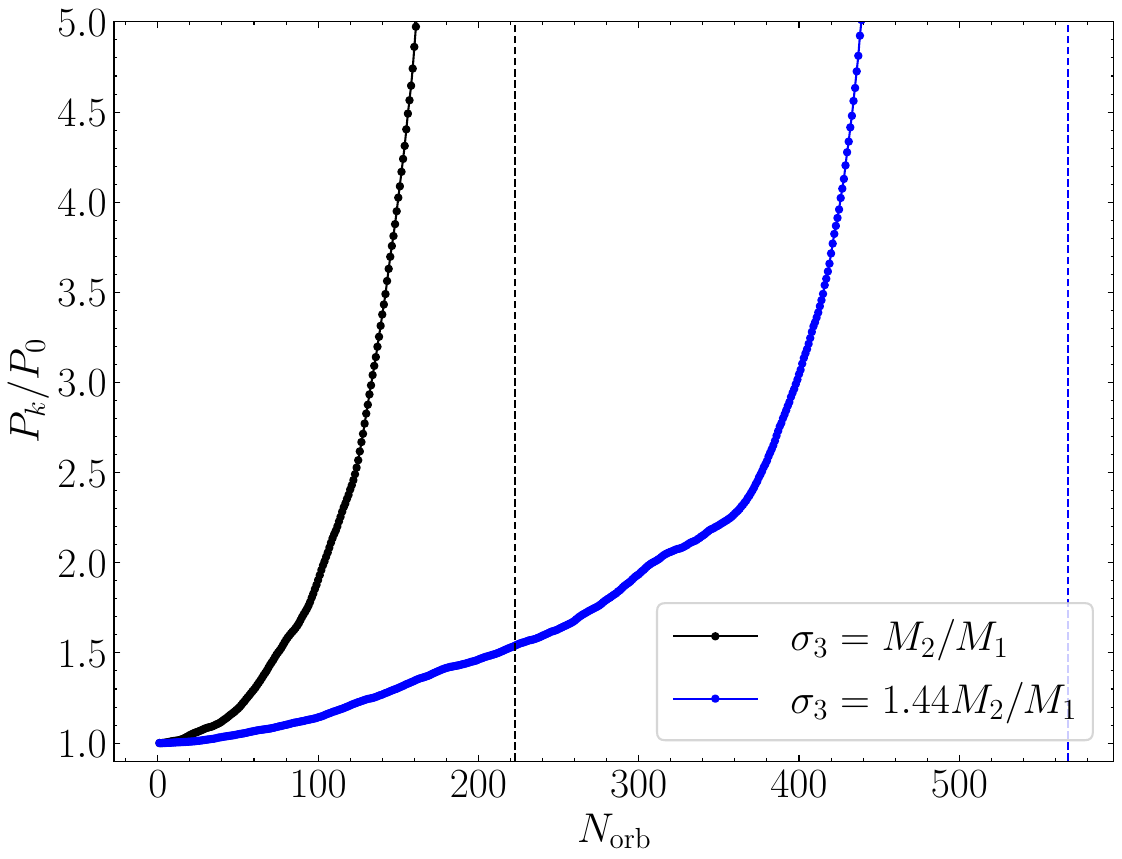}
    \caption{The mode energy (left panel) and orbital period (right panel) at different numbers of orbits during dynamical tidal stripping. The effects of tidal and GW backreaction are included. We choose the initial values $r_p/R_t = 2.7$, $q_a = 0$, and $\sigma_1=2\times10^{-4}$. We end the evolution when the binary unbinds, which is indicated by the vertical dashed lines. When mode energy is above 10\%$E_*$, we assume wave breaking to happen. We use the parameters $\sigma_1 = 5\times 10^{-4}$ and $\sigma_3 = M_2/M_1$ for the mass transfer during wave breaking.
    }
    \label{fig:dyn_tidal_strip}
\end{figure*}

In the previous unified model of a WD-MBH EMRI explaining QPE and TDE by \cite{Wang_2022}, the evolution of the dynamical tide amplitude over multiple orbits is not included, and the tidal stripping calculation is based on the approximate formula by Eggleton \cite{Eggleton_1983} for synchronous, circular binary systems. 
Significant deviation is expected for eccentric binaries outside synchronicity \cite{Sepinsky_2007}.
Moreover, as we have seen in Sec.~\ref{sec:chaos}, the long-term evolution of the tidal amplitude is crucial due to the effect of chaos. In this subsection, we discuss the qualitative effects of asynchronicity, eccentricity, and large tidal amplitude on tidal stripping and how they can explain the observed features in QPEs. 
We also employ the mass transfer model described in Sec.~\ref{ssec:mass_transfer_formulation}, with modifications on the parameters $\sigma_1$ and $\sigma_3$, to predict the evolution of the system. Our results show that the model can explain certain features of the QPEs.
However, it requires a specific parameter range for the angular momentum loss ($\sigma_3$), and an $r_p/R_t$ close to 2.7, or the binary cannot survive long enough to match observations.

For synchronous, circular binaries, the Roche lobe volume-equivalent radius can be well approximated by Eggleton's fitting formula \cite{Eggleton_1983}, which simply relates it to the orbital separation and the mass ratio. This formula gives us a Roche lobe filling distance at $2R_t$ for $M_1 \ll M_2$ (see Sec.~\ref{ssec:pTDE}). For asynchronous, highly eccentric systems, it is common to extrapolate this result by replacing the orbital separation by the pericenter distance $r_p$, as the Roche lobe radius during the closest passage.
The effect of the deviations from synchronicity and circularity has been investigated in \cite{Sepinsky_2007} through a Monte Carlo integration, which shows that the actual Roche lobe radius at pericenter can be larger by $20\%$ for $M_1 \ll M_2$, meaning that it requires a smaller pericenter distance for tidal stripping to occur. This comes from the fact that an asynchronous WD is more tightly bound without the centrifugal force. However, this analysis has not included the initial tidal amplitude from previous pericenter passages. Moreover, mass loss can occur at the Roche limit, a critical separation corresponding to hydrodynamic instability when subjected to the tidal field\footnote{The Roche lobe is determined from the extremum of the (gravitational and centrifugal) potential in the rotating frame, assuming the star is in synchronous spin with the orbit. The Roche limit accounts for the hydrodynamical stability of the star under the external tidal field. Hence, the latter is a more generic treatment for mass loss.}.
This separation does not necessarily coincide with that for Roche lobe overflow, and is shown to be larger for uniform density stars \cite{Chandrasekhar_1963}. Although a systematic study with realistic stellar models is not available at the moment, the results for polytropic models indicate that increasing the compressibility would reduce this critical separation \cite{Diener_1995}.

As we have shown in Sec.~\ref{sec:chaos}, the tidal amplitude can be orders of magnitude greater than the initial tidal kick due to chaotic growth.
Numerical simulations have shown that close fly-by objects in highly eccentric binaries can lose mass as they deviate significantly from a spherical shape due to tidal interactions, which is not captured in the calculations with Eggleton's formula. In the simulation by \cite{Guillochon_2011} on a Jupiter-like planet around the host star, the mass loss happens at $r_p/R_t \approx 2.7$.
This extends the Jupiter exclusion zone, where the mass transfer causes the planet to either fully disrupt or get ejected from the binary, way beyond the prediction with Eggleton's formula. Inspired by this result, we take $r_p/R_t = 2.7$ as the distance where dynamical tidal stripping occurs. In our separate work \cite{Yu_prep} (in preparation), we study the conditions for tidal stripping considering the onset of instability due to dynamical deformations of the star in a binary.
Our preliminary results agree with the above simulation that the critical $r_p/R_t$ for dynamical tidal stripping is well above $2$.


Due to the chaotic nature of the dynamical tide, the orbital period and the amount of mass loss can vary when the system is within the parameter space for tidal stripping. This provides a natural explanation for the observed variations of the recurrence time of the QPE flares \cite{Miniutti_2019, Giustini_2020, Chakraborty_2024}. Furthermore, since the tidal energy can be reduced significantly during the chaotic evolution due to its random nature, it also allows the tidal stripping to pause for a prolonged period, which is in agreement with the quenched phase of GSN 069 \cite{Miniutti_2023_a, Chakraborty_2024}, where the QPE reappears after 2 years of absence.

Without the first principle results of the dynamical stripping process, we model it using a variation of the mass transfer formalism introduced in Sec.~\ref{sec:formulation}. As we expect the tidal stripping repeats almost every $\sim\mathcal{O}(10)$~hours except during the quenched phase, a positive $\Delta e_\text{MT}$ larger than $|\Delta e_\text{GW}|$ can easily cause a runaway disruption of the binary, unlike the mass transfer in wave breaking which happens every $\mathcal{O}(10)$ years. For the system to stay long enough to explain the QPE, we need a different amount of angular momentum loss compared to the wave-breaking process ($\sigma_3 = M_2/M_1$). Also, the amount of mass loss by the WD should depend on the mode amplitude. In the following, we make several changes to the parametrized model for $\Delta e_\text{MT}$ and $\Delta a_\text{MT}$ in Sec.~\ref{ssec:mass_transfer_formulation} and apply it to the dynamical tidal stripping process:
(1) The mass transfer occurs at every pericenter passage.
(2) The fractional mass loss by the WD, $\sigma_1$, is proportional to the mode amplitude, i.e.
\begin{align}
    \sigma_1 = \sigma_{1,0} |q_{a, k}|/\sqrt{0.1},
\end{align}
where $\sigma_{1,0}$ is fixed at $2\times 10^{-4}$, such that it roughly gives a luminosity of $\sim \mathcal{O}(10^{42})~\text{erg s}^{-1}$ (Eq.~\eqref{eq:luminosity}).
(3) The tidal stripping has a negligible impact on the mode amplitude, unlike the wave breaking process in which the mode energy reduces to 0.1\%$E_*$. During wave breaking, the $f$-mode cannot maintain its wave-like pattern and there is non-linear dissipation of the mode energy.
(4) We consider values of $\sigma_3$ larger than $M_2/M_1$ for $\sigma_2 \approx \exp[-1]$ such that $\Delta e_\text{MT}$ is less positive.
One example corresponds to the mass escaping the binary through the L2 Lagrangian point, where $\sigma_3 \approx 1.44 M_2/M_1$ \cite{Pribulla_1998, MacLeod_2018}. Note that an even larger $\sigma_3$ can make $\Delta e_{\text{MT}}$ reach zero or even change sign.

As an illustration of the concept, we show, in Fig.~\ref{fig:dyn_tidal_strip}, the orbital period against the number of orbits under the effect of dynamical tidal stripping with the parameter choice mentioned above. The initial values chosen are $r_p/R_t = 2.7$ and $q_a = 0$. 
This gives roughly periodic bursts with a $0.1~\%$ variation in recurrence time due to chaos. 
The actual variation can be much larger, as the amount of mass stripped off and successfully accreted by the MBH should depend on the mode phase encoded in $q_a$.
For the case with $\sigma_3 = 1.44 M_2/M_1$, the orbital period increases by about 10\% over 100 orbits. This is close to the increase in recurrence time of GSN 069 over 50 days \cite{Miniutti_2019}. There are also regions where the mode energy decreases substantially, which can quench the tidal stripping process. This qualitatively captures the observed features in several QPEs. 

\begin{figure*}[!tp]
    \includegraphics[width=0.49\linewidth]{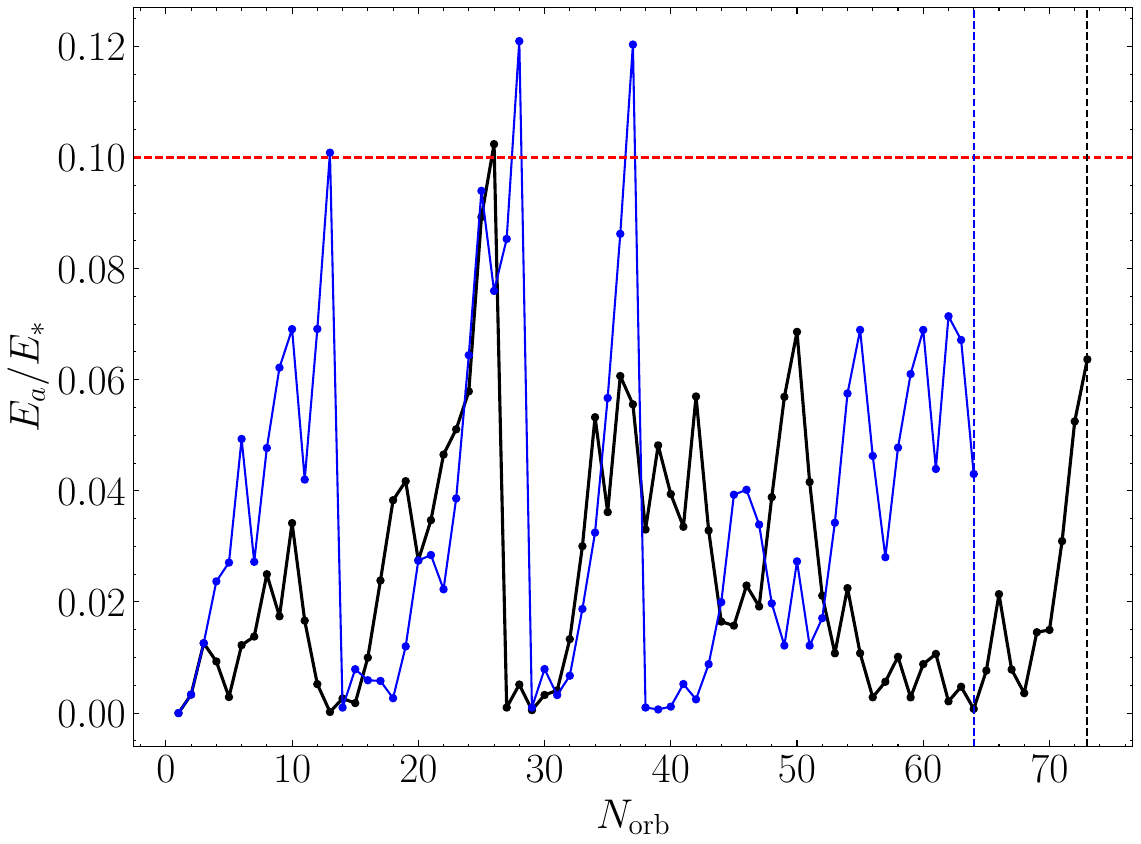}
    \includegraphics[width=0.49\linewidth]{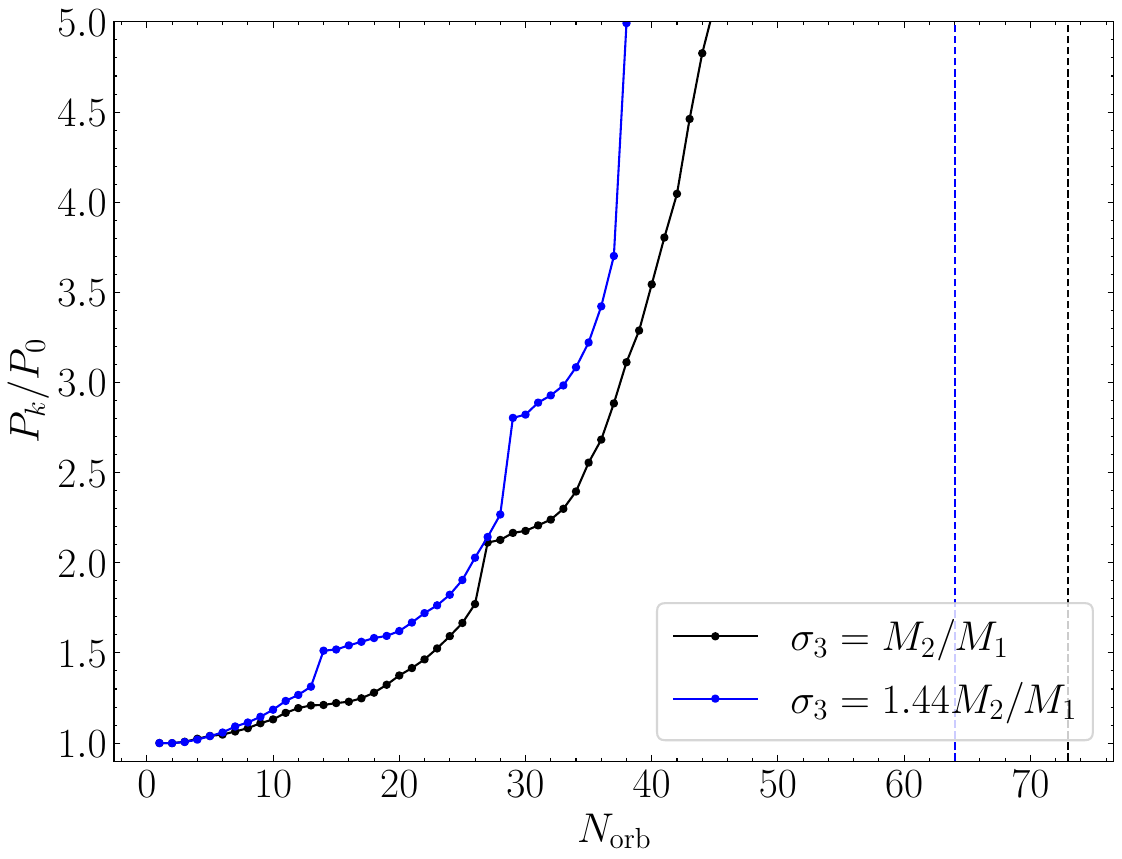}
    \caption{The mode energy and orbital period of the same system as Fig.~\ref{fig:dyn_tidal_strip}, but with initial $r_p/R_t = 2.2$.}
    \label{fig:dyn_tidal_strip_close}
\end{figure*}
For comparison, we illustrate the mode energy evolution and the orbital period for initial $r_p/R_t = 2.2$ in Fig.~\ref{fig:dyn_tidal_strip_close}. In this case, the increase in $\sigma_3$ does not reduce $\Delta e_\text{MT}$ (and $\Delta a_\text{MT}$) significantly since $1-\sigma_2$ is small (see Eq.~\eqref{eq:de_MT}). The case with $\sigma_3 = 1.44 M_2/M_1$ even unbinds slightly earlier than that with $\sigma_3 = M_2/M_1$ due to the random fluctuations. In both cases, the orbital period grows rapidly with the number of orbits, and the system unbinds within 2 years. For $\sigma_3 = 1.44 M_2/M_1$, the successive wave breakings are separated by $\sim 10$ days, which is much shorter than the observed pTDEs. However, the time gap is expected to extend if we include the damping of the mode amplitude due to tidal stripping, especially if the damping coefficient has non-linear scaling with the amplitude. Here, we ignore this effect due to its difficulty in modeling without knowledge of the hydrodynamics of the WD during mass transfer, though we expect it to be important at large amplitudes.

Based on this varied mass transfer model for dynamical tidal stripping, we reproduce some of the observed features of the QPE.
However, we emphasize again that the actual dynamics during the tidal stripping requires a detailed calculation accounting for the hydrodynamics of the deformed star under the tidal field from the MBH in an asynchronous, eccentric orbit. 
Also, the strong dependence on $\sigma_3$ for $r_p/R_t$ close to 2.7 and the limited lifetime of the binary for small $r_p/R_t$ pose extra challenges to this model.

\subsection{\label{ssec:EMRI} LISA EMRI sources}

LISA can detect GW signals from EMRIs in the mHz range with a high signal-to-noise ratio. 
However, the chaotic growth of the mode energy will introduce random variations in the period for a WD-MBH EMRI system with sufficiently high eccentricity. Even for loud sources within $\mathcal{O}$(10)~Mpc \cite{Wang_2022}, they might not be detectable as a coherent signal when there is a strong effect of chaos. In this subsection, we estimate the range of $r_p$ where the WD-MBH binary remains a coherent GW source throughout a 4-year observation time. 

\begin{figure}[!tp]
    \centering
    \includegraphics[width=1.0\linewidth]{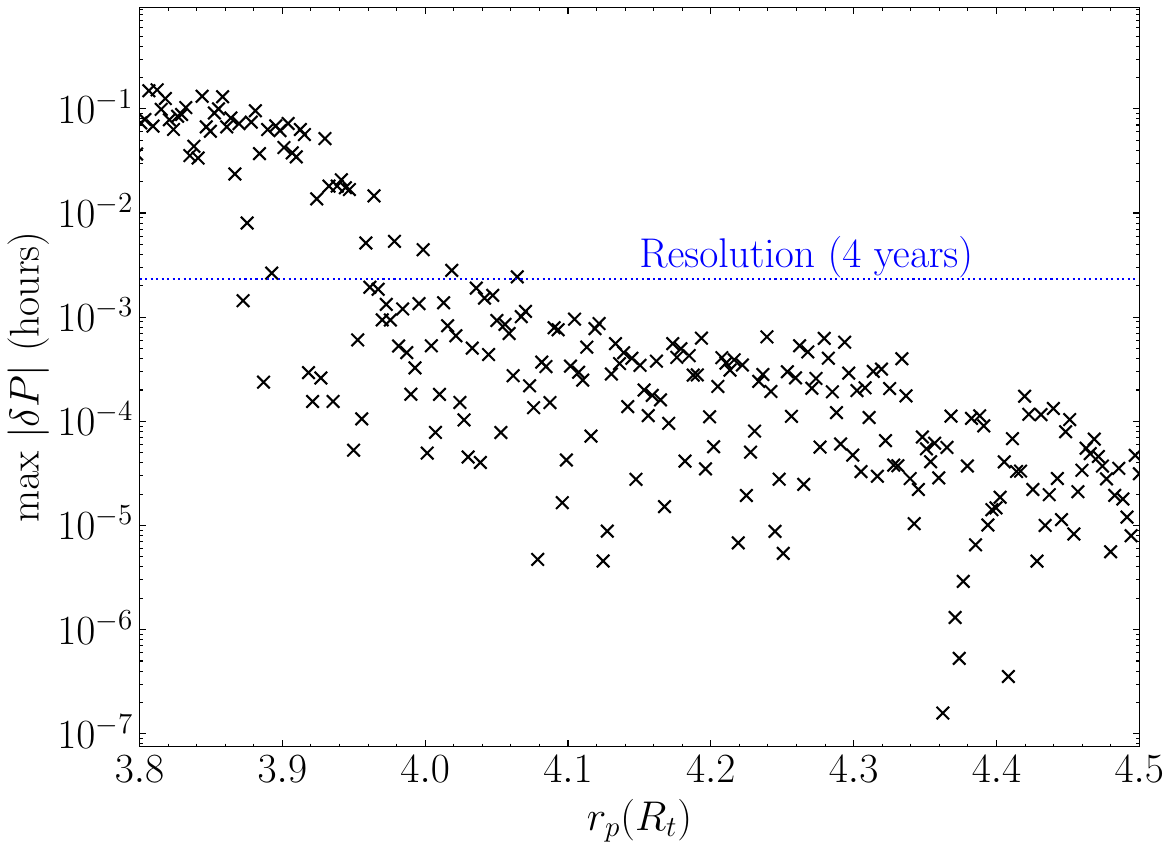}
    \caption{The maximum shift in the orbital period due to tidal backreaction with respect to that caused by GW orbital decay. 
    The blue dotted line indicates the resolution in the period for a 4-year observation. The maximum number of orbits for computing the maximum period shift is determined using Eq.~\eqref{eq:nmax}, with $e_\text{f} = 0.9$.}
    \label{fig:period_shift}
\end{figure}

In Fig.~\ref{fig:period_shift}, we show the maximum shift in period due to the dynamical tide from that caused by GW orbital decay. This quantity is defined using Kepler's third law
\begin{align}
    \delta P_k = 2\pi \left(\sqrt{\frac{a_k^3}{GM}} - \sqrt{\frac{a_{k \text{,GW}}{}^3}{GM}}\right), \label{eq:period_shift}
\end{align}
where $a_k$ is the semi-major axis of the $k$th orbit including effects of tide and GW backreaction, while $a_{k \text{,GW}}$ includes the GW backreaction only. Note that the deviation accumulates with increasing $k$.
With Eq.~\eqref{eq:period_shift}, we define the maximum period shift for $N_\text{max}$ orbits (Eq.~\eqref{eq:nmax}) as $\text{max}|\delta P|$.

The variation in period is resolvable only if it causes a frequency shift larger than the size of a frequency bin, i.e., the inverse of the observation time. In Fig.~\ref{fig:period_shift}, we indicate the resolution of the period shift, which is given by $P_0^2/(4~\text{years})$. For initial $r_p/R_t \gtrsim 4.2$, the shift is unresolvable and the system behaves effectively as a coherent EMRI source or repeating GW bursts with regular time intervals depending on the eccentricity.

Note that for $r_p/R_t \lesssim 3.7$, we also expect wave breaking as the mode energy reaches 10\% $E_*$ within $N_\text{max}$. Depending on the mass ejected by the WD, it can introduce significant variations in the orbital period and further disrupt the coherence of the GW signal.

\section{\label{sec:conclusion} Conclusion}

In this paper, we investigate the tidal evolution and orbital dynamics of a highly eccentric WD-MBH binary system and its EM and GW observables. Using the iterative map method, we evolve the system by coupling the $(\ell,m)=(2,2)$ $f$-mode with the orbital equation of motion with the leading 2.5 PN effect from GW radiation reaction. We show that the GW-induced orbital decay can cause the mode to evolve chaotically like a ``random walk" process. Unlike the tidally-driven chaos, which occurs only for $r_p/R_t \lesssim 2.2$, this GW-driven chaos appears in a much wider range of $r_p/R_t$, depending on the number of orbits $N_\text{orb}$.

Next, we demonstrate that the WD-MBH binary system can lead to pTDEs and QPEs observed at galactic centers. As the tidal energy approaches the stellar binding energy, the $f$-mode oscillations break at the stellar surface. This leads to a mass ejection from the WD. We incorporate this effect in the iterative map, where the mass loss by the WD is parametrized by $\sigma_1=-\Delta M_{1,k}/M_{1,k}>0$. We show that there exists a region in the ($r_p$, $\sigma_1$) parameter space where the wave breaking recurrence time and peak luminosity from mass accretion match those of the pTDEs observed in GSN 069. Due to the chaotic growth, the pTDE caused by wave breaking can occur at a $r_p/R_t$ much greater than the critical value previously estimated without accounting for the dynamical tide accumulated over multiple orbits. This means the event rate of pTDEs may be higher, but we leave a quantitative investigation to future studies.

As the WD loses mass via wave breaking, $R_t$ also increases. This can cause systems with $2.7 \lesssim r_p/R_t \lesssim 3.7$ to evolve towards $r_p/R_t \lesssim 2.7$. For these systems or those initially within this range, the WD can lose mass via dynamical tidal stripping. 
We again apply the parametrized model for the orbital dynamics under mass transfer to study the evolution of the system, with some modifications from the wave-breaking scenario, as described in Sec.~\ref{ssec:QPE}.
We assume the mass loss happens at each pericenter passage, which is of $\mathcal{O}(10)$ hours, matching that of the QPEs. Moreover, the stochastic nature of the dynamical tide can explain the recurrence time variations and quenched phase of the QPEs. Nevertheless, it requires a $r_p/R_t$ close to 2.7, and a certain amount of angular momentum loss from the system during mass transfer, which is yet to be justified through more detailed calculations of the underlying dynamics. For $r_p/R_t \approx 2.7$, a small angular momentum loss will result in the prompt unbinding of the binary, making it unable to match the observations. For $r_p/R_t \approx 2$, the mass transfer effect can quickly unbind the system in less than a year. The mode energy also grows rapidly and causes the system to reach wave breaking every several days. As a sidenote, we have ignored the dissipation of mode energy due to dynamical tidal stripping, which may suppress the growth of the mode and extend the time between wave breaking. But the amount of dissipation is also unknown without the detailed calculations involving the hydrodynamics of the WD tidal deformations.

The WD-MBH EMRI system can also be a GW emitter in the LISA frequency band. However, the signal only stays coherent without significant random variation in frequency from the chaotic tide for sufficiently large $r_p/R_t$.
We find that for $r_p/R_t \gtrsim 4.2$, the maximum mode energy from chaotic growth is well below the orbital energy. The shift in the orbital period is less than that marginally resolvable by a 4-year observation by LISA.

There are several more caveats regarding this study:
(1) We have not accounted for the conservative GR corrections with a lower PN order than the 2.5 PN order GW backreaction, which lead to the orbital precession, redshift, and frame-dragging corrections to the mode frequency. Given that $G M_2/(r_p c^2) \sim \mathcal{O}(0.1)$, these effects cause significant differences in the relations between orbital elements from the Keplerian solutions, but they do not have a direct influence on the chaotic nature of the mode evolution and its linear scaling with time. Besides, we also do not include the WD rotation and nonlinear hydrodynamics of the tide, which can directly affect the mode resonance and chaotic growth.
(2) We propose that the dynamical tidal stripping at $r_p/R_t \leq 2.7$ can lead to the observed QPEs due to the matching timescales and the variations in recurrence time due to chaotic evolution. However, some features of the QPEs still pose challenges to our model. One of such is the alternating brightness and recurrence time of every consecutive burst in some QPEs (mainly GSN 069 \cite{Miniutti_2019}, given the limited number of bursts detected in other sources). Even though the segment-wise periodic mode evolution between resonances (see Sec.~\ref{sec:chaos} and Sec.~\ref{ssec:QPE}) can explain the alternating behaviors if $\omega_a P_k/(2\pi)$ is close to a half-integer, it still requires further justifications from the tidal stripping dynamics, mass transfer process, and the mass fallback rate-luminosity relation.
(3) Even though our model can reproduce some features of the QPEs quantitatively, the prescribed forms of the mass transfer parameters are not based on fundamental hydrodynamic calculations. It also requires the fine-tuning of some undetermined parameters to match the observed data.


As an extension to this work, we plan to perform a first-principles calculation of the mass loss process due to dynamical tide. To establish a robust linkage of the highly eccentric WD-MBH binary with the pTDEs and QPEs, a detailed model of the mass loss process accounting for the internal hydrodynamics of the WD is required. This will be (partially) addressed in our separate work as we develop an analytical model by considering the hydrodynamical instabilities of a uniform density star under an external tidal field \cite{Yu_prep} (in preparation).


\vspace{2mm}
\begin{acknowledgments}

This work is supported by Montana NASA EPSCoR Research Infrastructure Development under award No. 80NSSC22M0042 and NSF grant No. PHY-2308415.  

\end{acknowledgments}

\appendix

\section{\label{app:mapping_MT} Orbital evolution during mass transfer}
In this appendix, we provide some details on obtaining the iteration formulas Eqs.~\eqref{eq:de_MT} and \eqref{eq:da_MT} from the orbital equation of motion. We also justify the use of an implicit iteration scheme instead of an explicit forward iteration scheme, where we write the difference between the quantities of the $(k+1)$th orbit and that of the $k$th orbit as a function of the $k$th orbit quantities only.

The secular evolution of the orbital elements of an eccentric binary undergoing non-conservative mass transfer at pericenter passage is given by \cite{Sepinsky_2009}
\begin{align}
    \left\langle \dot{e} \right\rangle =& 2 \bigg\langle \frac{\dot{M}_1}{M_1} (1+e)\bigg[\left(\sigma_2 \frac{M_1}{M_2} - 1\right) \nonumber\\
    &+ (1-\sigma_2)\left(\sigma_3 + \frac{1}{2}\right) \frac{M_1}{M}\bigg] \bigg\rangle, \label{eq:Sepinsky_de}\\
    \left\langle \frac{\dot{a}}{a} \right\rangle =& \left\langle\frac{ \dot{e}}{1-e}\right\rangle, \label{eq:Sepinsky_da}
\end{align}
where the overhead dot represents a time derivative, $a$ is the semi-major axis, related to the period via Kepler's third law $P = 2\pi a^{3/2}/\sqrt{GM}$.The angled brackets mean averaging over one complete orbit. The mass transfer rate is modeled as a Dirac-delta function.
\begin{align}
    \dot{M}_1  =& \frac{\Delta M_1}{P} \delta(t-t_p) \nonumber\\
    =& \frac{\Delta M_1}{2\pi} \sqrt{\frac{(1-e)^{3}}{1+e}} \delta(\phi),
\end{align}
\footnote{The calculation of $\Delta a_\text{MT}$ and $\Delta e_\text{MT}$ in \cite{Wang_2022} is missing the conversion factor when changing from $\delta(t-t_p)$ to $\delta(\phi)$. The effect on the orbit is lowered by $(1-e)^{3/2}/\sqrt{1+e}$.}where $t_p$ is the time of pericenter passage, $\phi$ is the true anomaly.
Note that Eqs.~\eqref{eq:Sepinsky_de} and \eqref{eq:Sepinsky_da} lead to 
\begin{align}
    \left\langle \dot{r}_p \right\rangle = 0. \label{eq:Sepinsky_drp}
\end{align}
In our system with $M_1/M_2 \ll 1$ and $\sigma_3 = M_2/M_1$, Eq.~\eqref{eq:de_MT} can be approximated by
\begin{align}
    \Delta e_{\text{MT},k} \approx 2\sigma_1 \sigma_2(1+e_k). \label{eq:de_MT_approx}
\end{align}
Therefore, we can approximate the per-orbit change in $e$ using the forward iteration (Eq.~\eqref{eq:de_MT}) as far as the product of the parameters is small.

The same does not hold for Eq.~\eqref{eq:Sepinsky_da} due to the $(1-e)^{-1}$ dependence, which diverges at the parabolic limit. Therefore, it is more appropriate to replace Eq.~\eqref{eq:Sepinsky_da} with Eq.~\eqref{eq:Sepinsky_drp}, resulting in the implicit iteration formula Eq.~\eqref{eq:da_MT}.

Next, we demonstrate that an accurate treatment of Eq.~\eqref{eq:Sepinsky_da} gives the exact same formula as Eq.~\eqref{eq:da_MT}.
Integrating Eq.~\eqref{eq:Sepinsky_da} by treating $a$ and $e$ as piecewise step functions between pericenter passages, we obtain
\begin{align}
    \Delta a_k = \frac{(a_{k+1}+a_k)/2}{1 - (e_{k+1}+e_k)/2} \Delta e_k, \label{eq:Sepinsky_da_integrate}
\end{align}
where we have used the properties of the Heaviside step function, $\Theta$, with the half-maximum convention and the Dirac delta function
\begin{align}
    \int_{-\infty}^\infty \Theta(x) \delta(x) dx = \frac{1}{2}.   
\end{align}
It is straightforward to show that Eq.~\eqref{eq:Sepinsky_da_integrate} is equivalent to Eq.~\eqref{eq:da_MT}.

We then expand Eq.~\eqref{eq:da_MT} to compare with the forward iteration scheme, and we find
\begin{align}
    \Delta a_k = \frac{a_k \Delta e_k}{1 - e_{k}} \left[1 + \frac{\Delta e_k}{1-e_k} + ... \right],
\end{align}
where the error terms diverge at the parabolic limit. Hence, we conclude that Eqs.~\eqref{eq:de_MT} and \eqref{eq:da_MT} provide a more consistent iteration scheme for Eqs.~\eqref{eq:Sepinsky_de}, \eqref{eq:Sepinsky_da} and \eqref{eq:Sepinsky_drp} than an explicit forward iteration.

\section{\label{app:angular_momentum_loss} Angular momentum loss via wave breaking}

The mass transfer effect on the orbital elements depends heavily on the choice of parameters $\sigma_1$, $\sigma_2$, and $\sigma_3$. In particular, the sign of $\Delta e_\text{MT}$ (and also $\Delta a_\text{MT}$) is sensitive to a small fractional change in $\sigma_3$. Here, we explain the choice of $\sigma_3$ based on previous analytical work on the binary mass transfer processes where part of the angular momentum is lost.

In the idealized situation where the mass ejected by the WD can leave the binary isotropically, the loss in angular momentum follows Eq.~\eqref{eq:sigma_3} (the intermediate mode in \cite{Huang_1963}). To demonstrate this explicitly, consider that the WD angular momentum, $J_1$, is related to the total orbital angular momentum by 
\begin{align}
    J_1 = M_2 \frac{J_\text{orb}}{M}.
\end{align}
If a mass of $\Delta M$, on the WD surface, is ejected and escapes the binary, the angular momentum it carries away is given by
\begin{align}
    \Delta J_\text{orb} = \Delta M\frac{J_1}{M_1}. \label{eq:spec_ang_mom_loss}
\end{align}
This gives the relation
\begin{align}
    \frac{\Delta J_\text{orb}}{\Delta M} = \frac{M_2}{M_1}\frac{J_\text{orb}}{M},
\end{align}
i.e., $\sigma_3 = M_2/M_1$. This corresponds to the mass loss scenario for the wave-breaking process.

Notice that if the mass ejection from the binary happens at the accretor surface, via, e.g. isotropic re-emission in the ultracompact X-ray binaries \cite{vanHaaften_2012, Sberna_2021}, then $\sigma_3 = M_1/M_2$ (with $M_1$ being the donor mass and $M_2$ being the accretor mass). This can be easily shown by replacing the angular momentum loss per mass in Eq.~\eqref{eq:spec_ang_mom_loss} by that at the accretor surface.

\section{\label{app:numerical_check} Verifying the chaotic growth with numerical integration}
\begin{figure}[!tp]
    \centering
    \includegraphics[width=1.0\linewidth]{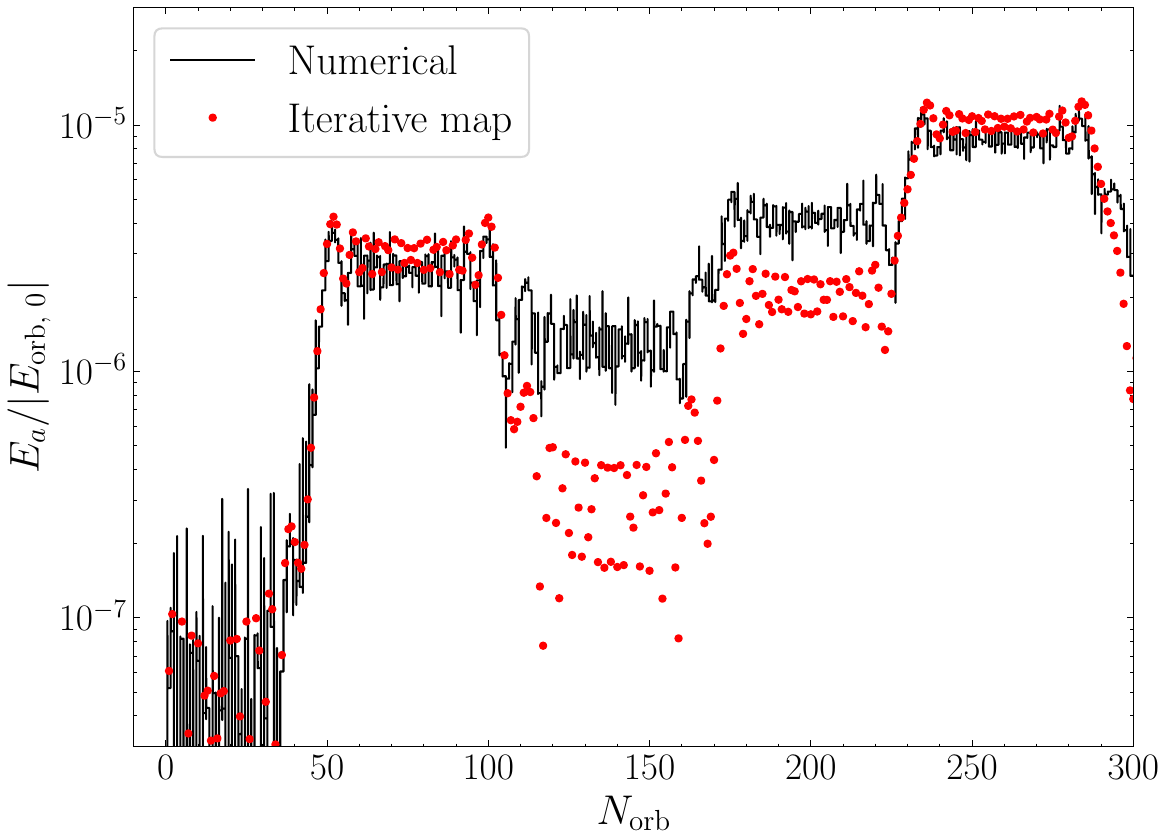}
    \caption{The mode energy evolution from numerical integration compared to that calculated with the iterative map. The initial $r_p$ is set as $3.5 R_t$. For visual clarity, the numerical solution contains only the dynamical component of the tide (see Eq.~\eqref{eq:q_eq}). The model parameters follow those in Table~\ref{tab:sources}.}
    \label{fig:num}
\end{figure}
In this appendix, we demonstrate an example of the chaotic growth of the $f$-mode driven by GW backreaction by solving the initial value problem with numerical integration.

The orbital equation of motion including the tidal and GW backreactions is written as
\begin{align}
    \ddot{r} =& -\frac{GM}{r^2} + r\dot{\Phi}^2 + f_{r, \text{tide}} + f_{r, \text{GW}}, \label{eq:num_r}\\
    r\ddot{\Phi} =& -2\dot{r}\dot{\Phi} + f_{\Phi, \text{tide}} + f_{\Phi, \text{GW}}, \label{eq:num_phi}
\end{align}
where the tidal backreaction terms are defined by \cite{Weinberg_2012}
\begin{align}
    f_{r, \text{tide}} =& -\frac{2GM}{r^2}\sum_{\substack{a \\ {\omega_a} > 0}} (\ell +1) W_{\ell m} Q_a \left(\frac{R_1}{r}\right)^{\ell} \text{Re}(q_a e^{im \Phi}),\\
    f_{\Phi, \text{tide}} =& \frac{2GM}{r^2}\sum_{\substack{a \\ {\omega_a} > 0}} m W_{\ell m} Q_a \left(\frac{R_1}{r}\right)^{\ell} \text{Re}(i q_a e^{im \Phi}), 
\end{align}
and the GW backreaction terms follow \cite{Kidder_1995}
\begin{align}
    f_{r, \text{GW}} =& -\frac{GM}{r^2}\left(A_{2.5}+ B_{2.5} \frac{\dot{r}}{c}\right) ,\\
    f_{\Phi, \text{GW}} =& -\frac{GM}{r} B_{2.5} \frac{\dot{\Phi}}{c}.
\end{align}
The explicit forms of the 2.5 PN terms, $A_{2.5}$ and $B_{2.5}$, are obtained from Eq.~(2.2) of \cite{Kidder_1995}:
\begin{align}
    A_{2.5} =& -\frac{8}{5} \eta \frac{GM\dot{r}}{rc^3} \bigg[18 \left(\frac{v}{c}\right)^2 +\frac{2}{3} \left(\frac{GM}{rc^2}\right) - 25\left(\frac{\dot{r}}{c}\right)^2\bigg],\\
    B_{2.5} =& \frac{8}{5} \eta \frac{GM}{rc^2}\left[6\left(\frac{v}{c}\right)^2 - 2\left(\frac{GM}{rc^2}\right) - 15\left(\frac{\dot{r}}{c}\right)^2\right],
\end{align}
where $v^2 = \dot{r}^2 + r^2 \dot{\Phi}^2$.

We numerically integrate Eqs.~\eqref{eq:num_r}, \eqref{eq:num_phi}, and \eqref{eq:q} for 400 orbits to compare with the iterative map results. For clarity, we subtract the equilibrium component from $q_a$, which is defined by
\begin{align}
    q_{a,\text{eq}}(t) =& U_a(t) - \frac{1}{i\omega_a}\dot{U}_a(t) + \mathcal{O}\left[\left(\frac{1}{\omega_a P}\right)^2\right], \label{eq:q_eq}
\end{align}
This is the instantaneous component of the tide, expressed as a convergent series for $ \omega_a P \gtrsim 1$. It can easily be proven through applying integration by parts on the integral solution of Eq.~\eqref{eq:q}.

Due to the steep dependence of $U_a$ on $1/r$, the equilibrium component peaks near pericenter and falls off quickly outside its vicinity. To compare with the iterative results, where the mode amplitude is evaluated at apocenter, we remove the first two leading terms in $q_{a,\text{eq}}$ in the numerical result. The comparison is shown in Fig.~\ref{fig:num}, where we choose $r_p/R_t = 3.5$, so that the chaotic evolution is driven by GW backreaction. We see that the iterative map result agrees with the numerical integration very well until after the second resonance. This deviation is caused by the strong feedback onto the orbit from the first resonance, so the second resonance cannot be treated as the mode evolution in a fixed background orbit, giving rise to nonlinearity in the evolution. The agreement is expected to get worse as $N_\text{orb}$ increases due to the chaotic nature, which causes the evolution to be very sensitive to numerical errors. Nevertheless, a longer evolution of the numerical code demonstrates a similar ``random walk" behavior and a linear scaling in $N_\text{orb}$ for the mode energy as seen in the iterative map.


\bibliography{main}

\begin{thebibliography}{93}%
\makeatletter
\providecommand \@ifxundefined [1]{%
 \@ifx{#1\undefined}
}%
\providecommand \@ifnum [1]{%
 \ifnum #1\expandafter \@firstoftwo
 \else \expandafter \@secondoftwo
 \fi
}%
\providecommand \@ifx [1]{%
 \ifx #1\expandafter \@firstoftwo
 \else \expandafter \@secondoftwo
 \fi
}%
\providecommand \natexlab [1]{#1}%
\providecommand \enquote  [1]{``#1''}%
\providecommand \bibnamefont  [1]{#1}%
\providecommand \bibfnamefont [1]{#1}%
\providecommand \citenamefont [1]{#1}%
\providecommand \href@noop [0]{\@secondoftwo}%
\providecommand \href [0]{\begingroup \@sanitize@url \@href}%
\providecommand \@href[1]{\@@startlink{#1}\@@href}%
\providecommand \@@href[1]{\endgroup#1\@@endlink}%
\providecommand \@sanitize@url [0]{\catcode `\\12\catcode `\$12\catcode `\&12\catcode `\#12\catcode `\^12\catcode `\_12\catcode `\%12\relax}%
\providecommand \@@startlink[1]{}%
\providecommand \@@endlink[0]{}%
\providecommand \url  [0]{\begingroup\@sanitize@url \@url }%
\providecommand \@url [1]{\endgroup\@href {#1}{\urlprefix }}%
\providecommand \urlprefix  [0]{URL }%
\providecommand \Eprint [0]{\href }%
\providecommand \doibase [0]{https://doi.org/}%
\providecommand \selectlanguage [0]{\@gobble}%
\providecommand \bibinfo  [0]{\@secondoftwo}%
\providecommand \bibfield  [0]{\@secondoftwo}%
\providecommand \translation [1]{[#1]}%
\providecommand \BibitemOpen [0]{}%
\providecommand \bibitemStop [0]{}%
\providecommand \bibitemNoStop [0]{.\EOS\space}%
\providecommand \EOS [0]{\spacefactor3000\relax}%
\providecommand \BibitemShut  [1]{\csname bibitem#1\endcsname}%
\let\auto@bib@innerbib\@empty
\bibitem [{\citenamefont {Amaro-Seoane}\ \emph {et~al.}(2017)\citenamefont {Amaro-Seoane}, \citenamefont {Audley}, \citenamefont {Babak}, \citenamefont {Baker}, \citenamefont {Barausse}, \citenamefont {Bender}, \citenamefont {Berti}, \citenamefont {Binetruy}, \citenamefont {Born}, \citenamefont {Bortoluzzi}, \citenamefont {Camp}, \citenamefont {Caprini}, \citenamefont {Cardoso}, \citenamefont {Colpi}, \citenamefont {Conklin}, \citenamefont {Cornish}, \citenamefont {Cutler}, \citenamefont {Danzmann}, \citenamefont {Dolesi}, \citenamefont {Ferraioli}, \citenamefont {Ferroni}, \citenamefont {Fitzsimons}, \citenamefont {Gair}, \citenamefont {Bote}, \citenamefont {Giardini}, \citenamefont {Gibert}, \citenamefont {Grimani}, \citenamefont {Halloin}, \citenamefont {Heinzel}, \citenamefont {Hertog}, \citenamefont {Hewitson}, \citenamefont {Holley-Bockelmann}, \citenamefont {Hollington}, \citenamefont {Hueller}, \citenamefont {Inchauspe}, \citenamefont {Jetzer}, \citenamefont {Karnesis}, \citenamefont {Killow},
  \citenamefont {Klein}, \citenamefont {Klipstein}, \citenamefont {Korsakova}, \citenamefont {Larson}, \citenamefont {Livas}, \citenamefont {Lloro}, \citenamefont {Man}, \citenamefont {Mance}, \citenamefont {Martino}, \citenamefont {Mateos}, \citenamefont {McKenzie}, \citenamefont {McWilliams}, \citenamefont {Miller}, \citenamefont {Mueller}, \citenamefont {Nardini}, \citenamefont {Nelemans}, \citenamefont {Nofrarias}, \citenamefont {Petiteau}, \citenamefont {Pivato}, \citenamefont {Plagnol}, \citenamefont {Porter}, \citenamefont {Reiche}, \citenamefont {Robertson}, \citenamefont {Robertson}, \citenamefont {Rossi}, \citenamefont {Russano}, \citenamefont {Schutz}, \citenamefont {Sesana}, \citenamefont {Shoemaker}, \citenamefont {Slutsky}, \citenamefont {Sopuerta}, \citenamefont {Sumner}, \citenamefont {Tamanini}, \citenamefont {Thorpe}, \citenamefont {Troebs}, \citenamefont {Vallisneri}, \citenamefont {Vecchio}, \citenamefont {Vetrugno}, \citenamefont {Vitale}, \citenamefont {Volonteri}, \citenamefont
  {Wanner}, \citenamefont {Ward}, \citenamefont {Wass}, \citenamefont {Weber}, \citenamefont {Ziemer},\ and\ \citenamefont {Zweifel}}]{AmaroSeoane_2017}%
  \BibitemOpen
  \bibfield  {author} {\bibinfo {author} {\bibfnamefont {P.}~\bibnamefont {Amaro-Seoane}}, \bibinfo {author} {\bibfnamefont {H.}~\bibnamefont {Audley}}, \bibinfo {author} {\bibfnamefont {S.}~\bibnamefont {Babak}}, \bibinfo {author} {\bibfnamefont {J.}~\bibnamefont {Baker}}, \bibinfo {author} {\bibfnamefont {E.}~\bibnamefont {Barausse}}, \bibinfo {author} {\bibfnamefont {P.}~\bibnamefont {Bender}}, \bibinfo {author} {\bibfnamefont {E.}~\bibnamefont {Berti}}, \bibinfo {author} {\bibfnamefont {P.}~\bibnamefont {Binetruy}}, \bibinfo {author} {\bibfnamefont {M.}~\bibnamefont {Born}}, \bibinfo {author} {\bibfnamefont {D.}~\bibnamefont {Bortoluzzi}}, \bibinfo {author} {\bibfnamefont {J.}~\bibnamefont {Camp}}, \bibinfo {author} {\bibfnamefont {C.}~\bibnamefont {Caprini}}, \bibinfo {author} {\bibfnamefont {V.}~\bibnamefont {Cardoso}}, \bibinfo {author} {\bibfnamefont {M.}~\bibnamefont {Colpi}}, \bibinfo {author} {\bibfnamefont {J.}~\bibnamefont {Conklin}}, \bibinfo {author} {\bibfnamefont {N.}~\bibnamefont {Cornish}},
  \bibinfo {author} {\bibfnamefont {C.}~\bibnamefont {Cutler}}, \bibinfo {author} {\bibfnamefont {K.}~\bibnamefont {Danzmann}}, \bibinfo {author} {\bibfnamefont {R.}~\bibnamefont {Dolesi}}, \bibinfo {author} {\bibfnamefont {L.}~\bibnamefont {Ferraioli}}, \bibinfo {author} {\bibfnamefont {V.}~\bibnamefont {Ferroni}}, \bibinfo {author} {\bibfnamefont {E.}~\bibnamefont {Fitzsimons}}, \bibinfo {author} {\bibfnamefont {J.}~\bibnamefont {Gair}}, \bibinfo {author} {\bibfnamefont {L.~G.}\ \bibnamefont {Bote}}, \bibinfo {author} {\bibfnamefont {D.}~\bibnamefont {Giardini}}, \bibinfo {author} {\bibfnamefont {F.}~\bibnamefont {Gibert}}, \bibinfo {author} {\bibfnamefont {C.}~\bibnamefont {Grimani}}, \bibinfo {author} {\bibfnamefont {H.}~\bibnamefont {Halloin}}, \bibinfo {author} {\bibfnamefont {G.}~\bibnamefont {Heinzel}}, \bibinfo {author} {\bibfnamefont {T.}~\bibnamefont {Hertog}}, \bibinfo {author} {\bibfnamefont {M.}~\bibnamefont {Hewitson}}, \bibinfo {author} {\bibfnamefont {K.}~\bibnamefont {Holley-Bockelmann}},
  \bibinfo {author} {\bibfnamefont {D.}~\bibnamefont {Hollington}}, \bibinfo {author} {\bibfnamefont {M.}~\bibnamefont {Hueller}}, \bibinfo {author} {\bibfnamefont {H.}~\bibnamefont {Inchauspe}}, \bibinfo {author} {\bibfnamefont {P.}~\bibnamefont {Jetzer}}, \bibinfo {author} {\bibfnamefont {N.}~\bibnamefont {Karnesis}}, \bibinfo {author} {\bibfnamefont {C.}~\bibnamefont {Killow}}, \bibinfo {author} {\bibfnamefont {A.}~\bibnamefont {Klein}}, \bibinfo {author} {\bibfnamefont {B.}~\bibnamefont {Klipstein}}, \bibinfo {author} {\bibfnamefont {N.}~\bibnamefont {Korsakova}}, \bibinfo {author} {\bibfnamefont {S.~L.}\ \bibnamefont {Larson}}, \bibinfo {author} {\bibfnamefont {J.}~\bibnamefont {Livas}}, \bibinfo {author} {\bibfnamefont {I.}~\bibnamefont {Lloro}}, \bibinfo {author} {\bibfnamefont {N.}~\bibnamefont {Man}}, \bibinfo {author} {\bibfnamefont {D.}~\bibnamefont {Mance}}, \bibinfo {author} {\bibfnamefont {J.}~\bibnamefont {Martino}}, \bibinfo {author} {\bibfnamefont {I.}~\bibnamefont {Mateos}}, \bibinfo
  {author} {\bibfnamefont {K.}~\bibnamefont {McKenzie}}, \bibinfo {author} {\bibfnamefont {S.~T.}\ \bibnamefont {McWilliams}}, \bibinfo {author} {\bibfnamefont {C.}~\bibnamefont {Miller}}, \bibinfo {author} {\bibfnamefont {G.}~\bibnamefont {Mueller}}, \bibinfo {author} {\bibfnamefont {G.}~\bibnamefont {Nardini}}, \bibinfo {author} {\bibfnamefont {G.}~\bibnamefont {Nelemans}}, \bibinfo {author} {\bibfnamefont {M.}~\bibnamefont {Nofrarias}}, \bibinfo {author} {\bibfnamefont {A.}~\bibnamefont {Petiteau}}, \bibinfo {author} {\bibfnamefont {P.}~\bibnamefont {Pivato}}, \bibinfo {author} {\bibfnamefont {E.}~\bibnamefont {Plagnol}}, \bibinfo {author} {\bibfnamefont {E.}~\bibnamefont {Porter}}, \bibinfo {author} {\bibfnamefont {J.}~\bibnamefont {Reiche}}, \bibinfo {author} {\bibfnamefont {D.}~\bibnamefont {Robertson}}, \bibinfo {author} {\bibfnamefont {N.}~\bibnamefont {Robertson}}, \bibinfo {author} {\bibfnamefont {E.}~\bibnamefont {Rossi}}, \bibinfo {author} {\bibfnamefont {G.}~\bibnamefont {Russano}}, \bibinfo
  {author} {\bibfnamefont {B.}~\bibnamefont {Schutz}}, \bibinfo {author} {\bibfnamefont {A.}~\bibnamefont {Sesana}}, \bibinfo {author} {\bibfnamefont {D.}~\bibnamefont {Shoemaker}}, \bibinfo {author} {\bibfnamefont {J.}~\bibnamefont {Slutsky}}, \bibinfo {author} {\bibfnamefont {C.~F.}\ \bibnamefont {Sopuerta}}, \bibinfo {author} {\bibfnamefont {T.}~\bibnamefont {Sumner}}, \bibinfo {author} {\bibfnamefont {N.}~\bibnamefont {Tamanini}}, \bibinfo {author} {\bibfnamefont {I.}~\bibnamefont {Thorpe}}, \bibinfo {author} {\bibfnamefont {M.}~\bibnamefont {Troebs}}, \bibinfo {author} {\bibfnamefont {M.}~\bibnamefont {Vallisneri}}, \bibinfo {author} {\bibfnamefont {A.}~\bibnamefont {Vecchio}}, \bibinfo {author} {\bibfnamefont {D.}~\bibnamefont {Vetrugno}}, \bibinfo {author} {\bibfnamefont {S.}~\bibnamefont {Vitale}}, \bibinfo {author} {\bibfnamefont {M.}~\bibnamefont {Volonteri}}, \bibinfo {author} {\bibfnamefont {G.}~\bibnamefont {Wanner}}, \bibinfo {author} {\bibfnamefont {H.}~\bibnamefont {Ward}}, \bibinfo {author}
  {\bibfnamefont {P.}~\bibnamefont {Wass}}, \bibinfo {author} {\bibfnamefont {W.}~\bibnamefont {Weber}}, \bibinfo {author} {\bibfnamefont {J.}~\bibnamefont {Ziemer}},\ and\ \bibinfo {author} {\bibfnamefont {P.}~\bibnamefont {Zweifel}},\ }\href {https://doi.org/10.48550/ARXIV.1702.00786} {\bibinfo {title} {Laser interferometer space antenna}} (\bibinfo {year} {2017})\BibitemShut {NoStop}%
\bibitem [{\citenamefont {Amaro-Seoane}(2018)}]{AmaroSeoane_2018}%
  \BibitemOpen
  \bibfield  {author} {\bibinfo {author} {\bibfnamefont {P.}~\bibnamefont {Amaro-Seoane}},\ }\bibfield  {title} {\bibinfo {title} {Relativistic dynamics and extreme mass ratio inspirals},\ }\bibfield  {journal} {\bibinfo  {journal} {Living~Rev.~Relativ.}\ }\textbf {\bibinfo {volume} {21}},\ \href {https://doi.org/10.1007/s41114-018-0013-8} {10.1007/s41114-018-0013-8} (\bibinfo {year} {2018})\BibitemShut {NoStop}%
\bibitem [{\citenamefont {Hills}(1988)}]{Hills_1988}%
  \BibitemOpen
  \bibfield  {author} {\bibinfo {author} {\bibfnamefont {J.~G.}\ \bibnamefont {Hills}},\ }\bibfield  {title} {\bibinfo {title} {Hyper-velocity and tidal stars from binaries disrupted by a massive galactic black hole},\ }\href {https://doi.org/10.1038/331687a0} {\bibfield  {journal} {\bibinfo  {journal} {Nature}\ }\textbf {\bibinfo {volume} {331}},\ \bibinfo {pages} {687–689} (\bibinfo {year} {1988})}\BibitemShut {NoStop}%
\bibitem [{\citenamefont {Yu}\ and\ \citenamefont {Tremaine}(2003)}]{Yu_2003}%
  \BibitemOpen
  \bibfield  {author} {\bibinfo {author} {\bibfnamefont {Q.}~\bibnamefont {Yu}}\ and\ \bibinfo {author} {\bibfnamefont {S.}~\bibnamefont {Tremaine}},\ }\bibfield  {title} {\bibinfo {title} {Ejection of hypervelocity stars by the (binary) black hole in the galactic center},\ }\href {https://doi.org/10.1086/379546} {\bibfield  {journal} {\bibinfo  {journal} {Astrophys.~J.}\ }\textbf {\bibinfo {volume} {599}},\ \bibinfo {pages} {1129–1138} (\bibinfo {year} {2003})}\BibitemShut {NoStop}%
\bibitem [{\citenamefont {Miller}\ \emph {et~al.}(2005)\citenamefont {Miller}, \citenamefont {Freitag}, \citenamefont {Hamilton},\ and\ \citenamefont {Lauburg}}]{Miller_2005}%
  \BibitemOpen
  \bibfield  {author} {\bibinfo {author} {\bibfnamefont {M.~C.}\ \bibnamefont {Miller}}, \bibinfo {author} {\bibfnamefont {M.}~\bibnamefont {Freitag}}, \bibinfo {author} {\bibfnamefont {D.~P.}\ \bibnamefont {Hamilton}},\ and\ \bibinfo {author} {\bibfnamefont {V.~M.}\ \bibnamefont {Lauburg}},\ }\bibfield  {title} {\bibinfo {title} {Binary encounters with supermassive black holes: Zero-eccentricity lisa events},\ }\href {https://doi.org/10.1086/497335} {\bibfield  {journal} {\bibinfo  {journal} {Astrophys.~J.~Lett.}\ }\textbf {\bibinfo {volume} {631}},\ \bibinfo {pages} {L117–L120} (\bibinfo {year} {2005})}\BibitemShut {NoStop}%
\bibitem [{\citenamefont {Bromley}\ \emph {et~al.}(2012)\citenamefont {Bromley}, \citenamefont {Kenyon}, \citenamefont {Geller},\ and\ \citenamefont {Brown}}]{Bromley_2012}%
  \BibitemOpen
  \bibfield  {author} {\bibinfo {author} {\bibfnamefont {B.~C.}\ \bibnamefont {Bromley}}, \bibinfo {author} {\bibfnamefont {S.~J.}\ \bibnamefont {Kenyon}}, \bibinfo {author} {\bibfnamefont {M.~J.}\ \bibnamefont {Geller}},\ and\ \bibinfo {author} {\bibfnamefont {W.~R.}\ \bibnamefont {Brown}},\ }\bibfield  {title} {\bibinfo {title} {Binary disruption by massive black holes: Hypervelocity stars, s stars, and tidal disruption events},\ }\href {https://doi.org/10.1088/2041-8205/749/2/l42} {\bibfield  {journal} {\bibinfo  {journal} {Astrophys.~J.~Lett.}\ }\textbf {\bibinfo {volume} {749}},\ \bibinfo {pages} {L42} (\bibinfo {year} {2012})}\BibitemShut {NoStop}%
\bibitem [{\citenamefont {Hopman}\ and\ \citenamefont {Alexander}(2005)}]{Hopman_2005}%
  \BibitemOpen
  \bibfield  {author} {\bibinfo {author} {\bibfnamefont {C.}~\bibnamefont {Hopman}}\ and\ \bibinfo {author} {\bibfnamefont {T.}~\bibnamefont {Alexander}},\ }\bibfield  {title} {\bibinfo {title} {The orbital statistics of stellar inspiral and relaxation near a massive black hole: Characterizing gravitational wave sources},\ }\href {https://doi.org/10.1086/431475} {\bibfield  {journal} {\bibinfo  {journal} {Astrophys.~J.}\ }\textbf {\bibinfo {volume} {629}},\ \bibinfo {pages} {362–372} (\bibinfo {year} {2005})}\BibitemShut {NoStop}%
\bibitem [{\citenamefont {Amaro-Seoane}\ \emph {et~al.}(2007)\citenamefont {Amaro-Seoane}, \citenamefont {Gair}, \citenamefont {Freitag}, \citenamefont {Miller}, \citenamefont {Mandel}, \citenamefont {Cutler},\ and\ \citenamefont {Babak}}]{AmaroSeoane_2007}%
  \BibitemOpen
  \bibfield  {author} {\bibinfo {author} {\bibfnamefont {P.}~\bibnamefont {Amaro-Seoane}}, \bibinfo {author} {\bibfnamefont {J.~R.}\ \bibnamefont {Gair}}, \bibinfo {author} {\bibfnamefont {M.}~\bibnamefont {Freitag}}, \bibinfo {author} {\bibfnamefont {M.~C.}\ \bibnamefont {Miller}}, \bibinfo {author} {\bibfnamefont {I.}~\bibnamefont {Mandel}}, \bibinfo {author} {\bibfnamefont {C.~J.}\ \bibnamefont {Cutler}},\ and\ \bibinfo {author} {\bibfnamefont {S.}~\bibnamefont {Babak}},\ }\bibfield  {title} {\bibinfo {title} {Intermediate and extreme mass-ratio inspirals—astrophysics, science applications and detection using lisa},\ }\href {https://doi.org/10.1088/0264-9381/24/17/r01} {\bibfield  {journal} {\bibinfo  {journal} {Classical~Quantum~Gravity}\ }\textbf {\bibinfo {volume} {24}},\ \bibinfo {pages} {R113–R169} (\bibinfo {year} {2007})}\BibitemShut {NoStop}%
\bibitem [{\citenamefont {Gair}\ \emph {et~al.}(2017)\citenamefont {Gair}, \citenamefont {Babak}, \citenamefont {Sesana}, \citenamefont {Amaro-Seoane}, \citenamefont {Barausse}, \citenamefont {Berry}, \citenamefont {Berti},\ and\ \citenamefont {Sopuerta}}]{Gair_2017}%
  \BibitemOpen
  \bibfield  {author} {\bibinfo {author} {\bibfnamefont {J.~R.}\ \bibnamefont {Gair}}, \bibinfo {author} {\bibfnamefont {S.}~\bibnamefont {Babak}}, \bibinfo {author} {\bibfnamefont {A.}~\bibnamefont {Sesana}}, \bibinfo {author} {\bibfnamefont {P.}~\bibnamefont {Amaro-Seoane}}, \bibinfo {author} {\bibfnamefont {E.}~\bibnamefont {Barausse}}, \bibinfo {author} {\bibfnamefont {C.~P.~L.}\ \bibnamefont {Berry}}, \bibinfo {author} {\bibfnamefont {E.}~\bibnamefont {Berti}},\ and\ \bibinfo {author} {\bibfnamefont {C.}~\bibnamefont {Sopuerta}},\ }\bibfield  {title} {\bibinfo {title} {Prospects for observing extreme-mass-ratio inspirals with lisa},\ }\href {https://doi.org/10.1088/1742-6596/840/1/012021} {\bibfield  {journal} {\bibinfo  {journal} {J.~Phys.~Conf.~Ser.}\ }\textbf {\bibinfo {volume} {840}},\ \bibinfo {pages} {012021} (\bibinfo {year} {2017})}\BibitemShut {NoStop}%
\bibitem [{\citenamefont {Babak}\ \emph {et~al.}(2017)\citenamefont {Babak}, \citenamefont {Gair}, \citenamefont {Sesana}, \citenamefont {Barausse}, \citenamefont {Sopuerta}, \citenamefont {Berry}, \citenamefont {Berti}, \citenamefont {Amaro-Seoane}, \citenamefont {Petiteau},\ and\ \citenamefont {Klein}}]{Babak_2017}%
  \BibitemOpen
  \bibfield  {author} {\bibinfo {author} {\bibfnamefont {S.}~\bibnamefont {Babak}}, \bibinfo {author} {\bibfnamefont {J.}~\bibnamefont {Gair}}, \bibinfo {author} {\bibfnamefont {A.}~\bibnamefont {Sesana}}, \bibinfo {author} {\bibfnamefont {E.}~\bibnamefont {Barausse}}, \bibinfo {author} {\bibfnamefont {C.~F.}\ \bibnamefont {Sopuerta}}, \bibinfo {author} {\bibfnamefont {C.~P.~L.}\ \bibnamefont {Berry}}, \bibinfo {author} {\bibfnamefont {E.}~\bibnamefont {Berti}}, \bibinfo {author} {\bibfnamefont {P.}~\bibnamefont {Amaro-Seoane}}, \bibinfo {author} {\bibfnamefont {A.}~\bibnamefont {Petiteau}},\ and\ \bibinfo {author} {\bibfnamefont {A.}~\bibnamefont {Klein}},\ }\bibfield  {title} {\bibinfo {title} {Science with the space-based interferometer lisa. v. extreme mass-ratio inspirals},\ }\href {https://doi.org/10.1103/PhysRevD.95.103012} {\bibfield  {journal} {\bibinfo  {journal} {Phys. Rev. D}\ }\textbf {\bibinfo {volume} {95}},\ \bibinfo {pages} {103012} (\bibinfo {year} {2017})}\BibitemShut {NoStop}%
\bibitem [{\citenamefont {Pan}\ and\ \citenamefont {Yang}(2021)}]{Pan_2021}%
  \BibitemOpen
  \bibfield  {author} {\bibinfo {author} {\bibfnamefont {Z.}~\bibnamefont {Pan}}\ and\ \bibinfo {author} {\bibfnamefont {H.}~\bibnamefont {Yang}},\ }\bibfield  {title} {\bibinfo {title} {Formation rate of extreme mass ratio inspirals in active galactic nuclei},\ }\href {https://doi.org/10.1103/PhysRevD.103.103018} {\bibfield  {journal} {\bibinfo  {journal} {Phys. Rev. D}\ }\textbf {\bibinfo {volume} {103}},\ \bibinfo {pages} {103018} (\bibinfo {year} {2021})}\BibitemShut {NoStop}%
\bibitem [{\citenamefont {Pan}\ \emph {et~al.}(2021)\citenamefont {Pan}, \citenamefont {Lyu},\ and\ \citenamefont {Yang}}]{Pan_2021_a}%
  \BibitemOpen
  \bibfield  {author} {\bibinfo {author} {\bibfnamefont {Z.}~\bibnamefont {Pan}}, \bibinfo {author} {\bibfnamefont {Z.}~\bibnamefont {Lyu}},\ and\ \bibinfo {author} {\bibfnamefont {H.}~\bibnamefont {Yang}},\ }\bibfield  {title} {\bibinfo {title} {Wet extreme mass ratio inspirals may be more common for spaceborne gravitational wave detection},\ }\href {https://doi.org/10.1103/PhysRevD.104.063007} {\bibfield  {journal} {\bibinfo  {journal} {Phys. Rev. D}\ }\textbf {\bibinfo {volume} {104}},\ \bibinfo {pages} {063007} (\bibinfo {year} {2021})}\BibitemShut {NoStop}%
\bibitem [{\citenamefont {Li}\ \emph {et~al.}(2025)\citenamefont {Li}, \citenamefont {Sun}, \citenamefont {Zou},\ and\ \citenamefont {Yang}}]{Li_2025}%
  \BibitemOpen
  \bibfield  {author} {\bibinfo {author} {\bibfnamefont {X.}~\bibnamefont {Li}}, \bibinfo {author} {\bibfnamefont {H.}~\bibnamefont {Sun}}, \bibinfo {author} {\bibfnamefont {Y.-C.}\ \bibnamefont {Zou}},\ and\ \bibinfo {author} {\bibfnamefont {H.}~\bibnamefont {Yang}},\ }\href {https://doi.org/10.48550/ARXIV.2501.13702} {\bibinfo {title} {Micro-tidal disruption events at galactic centers}} (\bibinfo {year} {2025})\BibitemShut {NoStop}%
\bibitem [{\citenamefont {Miniutti}\ \emph {et~al.}(2019)\citenamefont {Miniutti}, \citenamefont {Saxton}, \citenamefont {Giustini}, \citenamefont {Alexander}, \citenamefont {Fender}, \citenamefont {Heywood}, \citenamefont {Monageng}, \citenamefont {Coriat}, \citenamefont {Tzioumis}, \citenamefont {Read}, \citenamefont {Knigge}, \citenamefont {Gandhi}, \citenamefont {Pretorius},\ and\ \citenamefont {Agís-González}}]{Miniutti_2019}%
  \BibitemOpen
  \bibfield  {author} {\bibinfo {author} {\bibfnamefont {G.}~\bibnamefont {Miniutti}}, \bibinfo {author} {\bibfnamefont {R.~D.}\ \bibnamefont {Saxton}}, \bibinfo {author} {\bibfnamefont {M.}~\bibnamefont {Giustini}}, \bibinfo {author} {\bibfnamefont {K.~D.}\ \bibnamefont {Alexander}}, \bibinfo {author} {\bibfnamefont {R.~P.}\ \bibnamefont {Fender}}, \bibinfo {author} {\bibfnamefont {I.}~\bibnamefont {Heywood}}, \bibinfo {author} {\bibfnamefont {I.}~\bibnamefont {Monageng}}, \bibinfo {author} {\bibfnamefont {M.}~\bibnamefont {Coriat}}, \bibinfo {author} {\bibfnamefont {A.~K.}\ \bibnamefont {Tzioumis}}, \bibinfo {author} {\bibfnamefont {A.~M.}\ \bibnamefont {Read}}, \bibinfo {author} {\bibfnamefont {C.}~\bibnamefont {Knigge}}, \bibinfo {author} {\bibfnamefont {P.}~\bibnamefont {Gandhi}}, \bibinfo {author} {\bibfnamefont {M.~L.}\ \bibnamefont {Pretorius}},\ and\ \bibinfo {author} {\bibfnamefont {B.}~\bibnamefont {Agís-González}},\ }\bibfield  {title} {\bibinfo {title} {Nine-hour x-ray quasi-periodic eruptions
  from a low-mass black hole galactic nucleus},\ }\href {https://doi.org/10.1038/s41586-019-1556-x} {\bibfield  {journal} {\bibinfo  {journal} {Nature}\ }\textbf {\bibinfo {volume} {573}},\ \bibinfo {pages} {381–384} (\bibinfo {year} {2019})}\BibitemShut {NoStop}%
\bibitem [{\citenamefont {Giustini}\ \emph {et~al.}(2020)\citenamefont {Giustini}, \citenamefont {Miniutti},\ and\ \citenamefont {Saxton}}]{Giustini_2020}%
  \BibitemOpen
  \bibfield  {author} {\bibinfo {author} {\bibfnamefont {M.}~\bibnamefont {Giustini}}, \bibinfo {author} {\bibfnamefont {G.}~\bibnamefont {Miniutti}},\ and\ \bibinfo {author} {\bibfnamefont {R.~D.}\ \bibnamefont {Saxton}},\ }\bibfield  {title} {\bibinfo {title} {X-ray quasi-periodic eruptions from the galactic nucleus of rx j1301.9+2747},\ }\href {https://doi.org/10.1051/0004-6361/202037610} {\bibfield  {journal} {\bibinfo  {journal} {Astron.~Astrophys.}\ }\textbf {\bibinfo {volume} {636}},\ \bibinfo {pages} {L2} (\bibinfo {year} {2020})}\BibitemShut {NoStop}%
\bibitem [{\citenamefont {Arcodia}\ \emph {et~al.}(2021)\citenamefont {Arcodia}, \citenamefont {Merloni}, \citenamefont {Nandra}, \citenamefont {Buchner}, \citenamefont {Salvato}, \citenamefont {Pasham}, \citenamefont {Remillard}, \citenamefont {Comparat}, \citenamefont {Lamer}, \citenamefont {Ponti}, \citenamefont {Malyali}, \citenamefont {Wolf}, \citenamefont {Arzoumanian}, \citenamefont {Bogensberger}, \citenamefont {Buckley}, \citenamefont {Gendreau}, \citenamefont {Gromadzki}, \citenamefont {Kara}, \citenamefont {Krumpe}, \citenamefont {Markwardt}, \citenamefont {Ramos-Ceja}, \citenamefont {Rau}, \citenamefont {Schramm},\ and\ \citenamefont {Schwope}}]{Arcodia_2021}%
  \BibitemOpen
  \bibfield  {author} {\bibinfo {author} {\bibfnamefont {R.}~\bibnamefont {Arcodia}}, \bibinfo {author} {\bibfnamefont {A.}~\bibnamefont {Merloni}}, \bibinfo {author} {\bibfnamefont {K.}~\bibnamefont {Nandra}}, \bibinfo {author} {\bibfnamefont {J.}~\bibnamefont {Buchner}}, \bibinfo {author} {\bibfnamefont {M.}~\bibnamefont {Salvato}}, \bibinfo {author} {\bibfnamefont {D.}~\bibnamefont {Pasham}}, \bibinfo {author} {\bibfnamefont {R.}~\bibnamefont {Remillard}}, \bibinfo {author} {\bibfnamefont {J.}~\bibnamefont {Comparat}}, \bibinfo {author} {\bibfnamefont {G.}~\bibnamefont {Lamer}}, \bibinfo {author} {\bibfnamefont {G.}~\bibnamefont {Ponti}}, \bibinfo {author} {\bibfnamefont {A.}~\bibnamefont {Malyali}}, \bibinfo {author} {\bibfnamefont {J.}~\bibnamefont {Wolf}}, \bibinfo {author} {\bibfnamefont {Z.}~\bibnamefont {Arzoumanian}}, \bibinfo {author} {\bibfnamefont {D.}~\bibnamefont {Bogensberger}}, \bibinfo {author} {\bibfnamefont {D.~A.~H.}\ \bibnamefont {Buckley}}, \bibinfo {author} {\bibfnamefont
  {K.}~\bibnamefont {Gendreau}}, \bibinfo {author} {\bibfnamefont {M.}~\bibnamefont {Gromadzki}}, \bibinfo {author} {\bibfnamefont {E.}~\bibnamefont {Kara}}, \bibinfo {author} {\bibfnamefont {M.}~\bibnamefont {Krumpe}}, \bibinfo {author} {\bibfnamefont {C.}~\bibnamefont {Markwardt}}, \bibinfo {author} {\bibfnamefont {M.~E.}\ \bibnamefont {Ramos-Ceja}}, \bibinfo {author} {\bibfnamefont {A.}~\bibnamefont {Rau}}, \bibinfo {author} {\bibfnamefont {M.}~\bibnamefont {Schramm}},\ and\ \bibinfo {author} {\bibfnamefont {A.}~\bibnamefont {Schwope}},\ }\bibfield  {title} {\bibinfo {title} {X-ray quasi-periodic eruptions from two previously quiescent galaxies},\ }\href {https://doi.org/10.1038/s41586-021-03394-6} {\bibfield  {journal} {\bibinfo  {journal} {Nature}\ }\textbf {\bibinfo {volume} {592}},\ \bibinfo {pages} {704–707} (\bibinfo {year} {2021})}\BibitemShut {NoStop}%
\bibitem [{\citenamefont {Arcodia}\ \emph {et~al.}(2022)\citenamefont {Arcodia}, \citenamefont {Miniutti}, \citenamefont {Ponti}, \citenamefont {Buchner}, \citenamefont {Giustini}, \citenamefont {Merloni}, \citenamefont {Nandra}, \citenamefont {Vincentelli}, \citenamefont {Kara}, \citenamefont {Salvato},\ and\ \citenamefont {Pasham}}]{Arcodia_2022}%
  \BibitemOpen
  \bibfield  {author} {\bibinfo {author} {\bibfnamefont {R.}~\bibnamefont {Arcodia}}, \bibinfo {author} {\bibfnamefont {G.}~\bibnamefont {Miniutti}}, \bibinfo {author} {\bibfnamefont {G.}~\bibnamefont {Ponti}}, \bibinfo {author} {\bibfnamefont {J.}~\bibnamefont {Buchner}}, \bibinfo {author} {\bibfnamefont {M.}~\bibnamefont {Giustini}}, \bibinfo {author} {\bibfnamefont {A.}~\bibnamefont {Merloni}}, \bibinfo {author} {\bibfnamefont {K.}~\bibnamefont {Nandra}}, \bibinfo {author} {\bibfnamefont {F.}~\bibnamefont {Vincentelli}}, \bibinfo {author} {\bibfnamefont {E.}~\bibnamefont {Kara}}, \bibinfo {author} {\bibfnamefont {M.}~\bibnamefont {Salvato}},\ and\ \bibinfo {author} {\bibfnamefont {D.}~\bibnamefont {Pasham}},\ }\bibfield  {title} {\bibinfo {title} {The complex time and energy evolution of quasi-periodic eruptions in ero-qpe1},\ }\href {https://doi.org/10.1051/0004-6361/202243259} {\bibfield  {journal} {\bibinfo  {journal} {Astron.~Astrophys.}\ }\textbf {\bibinfo {volume} {662}},\ \bibinfo {pages} {A49}
  (\bibinfo {year} {2022})}\BibitemShut {NoStop}%
\bibitem [{\citenamefont {Arcodia}\ \emph {et~al.}(2024)\citenamefont {Arcodia}, \citenamefont {Liu}, \citenamefont {Merloni}, \citenamefont {Malyali}, \citenamefont {Rau}, \citenamefont {Chakraborty}, \citenamefont {Goodwin}, \citenamefont {Buckley}, \citenamefont {Brink}, \citenamefont {Gromadzki}, \citenamefont {Arzoumanian}, \citenamefont {Buchner}, \citenamefont {Kara}, \citenamefont {Nandra}, \citenamefont {Ponti}, \citenamefont {Salvato}, \citenamefont {Anderson}, \citenamefont {Baldini}, \citenamefont {Grotova}, \citenamefont {Krumpe}, \citenamefont {Maitra}, \citenamefont {Miller-Jones},\ and\ \citenamefont {Ramos-Ceja}}]{Arcodia_2024}%
  \BibitemOpen
  \bibfield  {author} {\bibinfo {author} {\bibfnamefont {R.}~\bibnamefont {Arcodia}}, \bibinfo {author} {\bibfnamefont {Z.}~\bibnamefont {Liu}}, \bibinfo {author} {\bibfnamefont {A.}~\bibnamefont {Merloni}}, \bibinfo {author} {\bibfnamefont {A.}~\bibnamefont {Malyali}}, \bibinfo {author} {\bibfnamefont {A.}~\bibnamefont {Rau}}, \bibinfo {author} {\bibfnamefont {J.}~\bibnamefont {Chakraborty}}, \bibinfo {author} {\bibfnamefont {A.}~\bibnamefont {Goodwin}}, \bibinfo {author} {\bibfnamefont {D.}~\bibnamefont {Buckley}}, \bibinfo {author} {\bibfnamefont {J.}~\bibnamefont {Brink}}, \bibinfo {author} {\bibfnamefont {M.}~\bibnamefont {Gromadzki}}, \bibinfo {author} {\bibfnamefont {Z.}~\bibnamefont {Arzoumanian}}, \bibinfo {author} {\bibfnamefont {J.}~\bibnamefont {Buchner}}, \bibinfo {author} {\bibfnamefont {E.}~\bibnamefont {Kara}}, \bibinfo {author} {\bibfnamefont {K.}~\bibnamefont {Nandra}}, \bibinfo {author} {\bibfnamefont {G.}~\bibnamefont {Ponti}}, \bibinfo {author} {\bibfnamefont {M.}~\bibnamefont {Salvato}},
  \bibinfo {author} {\bibfnamefont {G.}~\bibnamefont {Anderson}}, \bibinfo {author} {\bibfnamefont {P.}~\bibnamefont {Baldini}}, \bibinfo {author} {\bibfnamefont {I.}~\bibnamefont {Grotova}}, \bibinfo {author} {\bibfnamefont {M.}~\bibnamefont {Krumpe}}, \bibinfo {author} {\bibfnamefont {C.}~\bibnamefont {Maitra}}, \bibinfo {author} {\bibfnamefont {J.~C.~A.}\ \bibnamefont {Miller-Jones}},\ and\ \bibinfo {author} {\bibfnamefont {M.~E.}\ \bibnamefont {Ramos-Ceja}},\ }\bibfield  {title} {\bibinfo {title} {The more the merrier: Srg/erosita discovers two further galaxies showing x-ray quasi-periodic eruptions},\ }\href {https://doi.org/10.1051/0004-6361/202348881} {\bibfield  {journal} {\bibinfo  {journal} {Astron.~Astrophys.}\ }\textbf {\bibinfo {volume} {684}},\ \bibinfo {pages} {A64} (\bibinfo {year} {2024})}\BibitemShut {NoStop}%
\bibitem [{\citenamefont {Evans}\ \emph {et~al.}(2023)\citenamefont {Evans}, \citenamefont {Nixon}, \citenamefont {Campana}, \citenamefont {Charalampopoulos}, \citenamefont {Perley}, \citenamefont {Breeveld}, \citenamefont {Page}, \citenamefont {Oates}, \citenamefont {Eyles-Ferris}, \citenamefont {Malesani}, \citenamefont {Izzo}, \citenamefont {Goad}, \citenamefont {O’Brien}, \citenamefont {Osborne},\ and\ \citenamefont {Sbarufatti}}]{Evans_2023}%
  \BibitemOpen
  \bibfield  {author} {\bibinfo {author} {\bibfnamefont {P.~A.}\ \bibnamefont {Evans}}, \bibinfo {author} {\bibfnamefont {C.~J.}\ \bibnamefont {Nixon}}, \bibinfo {author} {\bibfnamefont {S.}~\bibnamefont {Campana}}, \bibinfo {author} {\bibfnamefont {P.}~\bibnamefont {Charalampopoulos}}, \bibinfo {author} {\bibfnamefont {D.~A.}\ \bibnamefont {Perley}}, \bibinfo {author} {\bibfnamefont {A.~A.}\ \bibnamefont {Breeveld}}, \bibinfo {author} {\bibfnamefont {K.~L.}\ \bibnamefont {Page}}, \bibinfo {author} {\bibfnamefont {S.~R.}\ \bibnamefont {Oates}}, \bibinfo {author} {\bibfnamefont {R.~A.~J.}\ \bibnamefont {Eyles-Ferris}}, \bibinfo {author} {\bibfnamefont {D.~B.}\ \bibnamefont {Malesani}}, \bibinfo {author} {\bibfnamefont {L.}~\bibnamefont {Izzo}}, \bibinfo {author} {\bibfnamefont {M.~R.}\ \bibnamefont {Goad}}, \bibinfo {author} {\bibfnamefont {P.~T.}\ \bibnamefont {O’Brien}}, \bibinfo {author} {\bibfnamefont {J.~P.}\ \bibnamefont {Osborne}},\ and\ \bibinfo {author} {\bibfnamefont {B.}~\bibnamefont
  {Sbarufatti}},\ }\bibfield  {title} {\bibinfo {title} {Monthly quasi-periodic eruptions from repeated stellar disruption by a massive black hole},\ }\href {https://doi.org/10.1038/s41550-023-02073-y} {\bibfield  {journal} {\bibinfo  {journal} {Nat.~Astron.}\ }\textbf {\bibinfo {volume} {7}},\ \bibinfo {pages} {1368–1375} (\bibinfo {year} {2023})}\BibitemShut {NoStop}%
\bibitem [{\citenamefont {Guolo}\ \emph {et~al.}(2024)\citenamefont {Guolo}, \citenamefont {Pasham}, \citenamefont {Zajaček}, \citenamefont {Coughlin}, \citenamefont {Gezari}, \citenamefont {Suková}, \citenamefont {Wevers}, \citenamefont {Witzany}, \citenamefont {Tombesi}, \citenamefont {van Velzen}, \citenamefont {Alexander}, \citenamefont {Yao}, \citenamefont {Arcodia}, \citenamefont {Karas}, \citenamefont {Miller-Jones}, \citenamefont {Remillard}, \citenamefont {Gendreau},\ and\ \citenamefont {Ferrara}}]{Guolo_2024}%
  \BibitemOpen
  \bibfield  {author} {\bibinfo {author} {\bibfnamefont {M.}~\bibnamefont {Guolo}}, \bibinfo {author} {\bibfnamefont {D.~R.}\ \bibnamefont {Pasham}}, \bibinfo {author} {\bibfnamefont {M.}~\bibnamefont {Zajaček}}, \bibinfo {author} {\bibfnamefont {E.~R.}\ \bibnamefont {Coughlin}}, \bibinfo {author} {\bibfnamefont {S.}~\bibnamefont {Gezari}}, \bibinfo {author} {\bibfnamefont {P.}~\bibnamefont {Suková}}, \bibinfo {author} {\bibfnamefont {T.}~\bibnamefont {Wevers}}, \bibinfo {author} {\bibfnamefont {V.}~\bibnamefont {Witzany}}, \bibinfo {author} {\bibfnamefont {F.}~\bibnamefont {Tombesi}}, \bibinfo {author} {\bibfnamefont {S.}~\bibnamefont {van Velzen}}, \bibinfo {author} {\bibfnamefont {K.~D.}\ \bibnamefont {Alexander}}, \bibinfo {author} {\bibfnamefont {Y.}~\bibnamefont {Yao}}, \bibinfo {author} {\bibfnamefont {R.}~\bibnamefont {Arcodia}}, \bibinfo {author} {\bibfnamefont {V.}~\bibnamefont {Karas}}, \bibinfo {author} {\bibfnamefont {J.~C.~A.}\ \bibnamefont {Miller-Jones}}, \bibinfo {author} {\bibfnamefont
  {R.}~\bibnamefont {Remillard}}, \bibinfo {author} {\bibfnamefont {K.}~\bibnamefont {Gendreau}},\ and\ \bibinfo {author} {\bibfnamefont {E.~C.}\ \bibnamefont {Ferrara}},\ }\bibfield  {title} {\bibinfo {title} {X-ray eruptions every 22 days from the nucleus of a nearby galaxy},\ }\href {https://doi.org/10.1038/s41550-023-02178-4} {\bibfield  {journal} {\bibinfo  {journal} {Nat.~Astron.}\ }\textbf {\bibinfo {volume} {8}},\ \bibinfo {pages} {347–358} (\bibinfo {year} {2024})}\BibitemShut {NoStop}%
\bibitem [{\citenamefont {Nicholl}\ \emph {et~al.}(2024)\citenamefont {Nicholl}, \citenamefont {Pasham}, \citenamefont {Mummery}, \citenamefont {Guolo}, \citenamefont {Gendreau}, \citenamefont {Dewangan}, \citenamefont {Ferrara}, \citenamefont {Remillard}, \citenamefont {Bonnerot}, \citenamefont {Chakraborty}, \citenamefont {Hajela}, \citenamefont {Dhillon}, \citenamefont {Gillan}, \citenamefont {Greenwood}, \citenamefont {Huber}, \citenamefont {Janiuk}, \citenamefont {Salvesen}, \citenamefont {van Velzen}, \citenamefont {Aamer}, \citenamefont {Alexander}, \citenamefont {Angus}, \citenamefont {Arzoumanian}, \citenamefont {Auchettl}, \citenamefont {Berger}, \citenamefont {de~Boer}, \citenamefont {Cendes}, \citenamefont {Chambers}, \citenamefont {Chen}, \citenamefont {Chornock}, \citenamefont {Fulton}, \citenamefont {Gao}, \citenamefont {Gillanders}, \citenamefont {Gomez}, \citenamefont {Gompertz}, \citenamefont {Fabian}, \citenamefont {Herman}, \citenamefont {Ingram}, \citenamefont {Kara}, \citenamefont
  {Laskar}, \citenamefont {Lawrence}, \citenamefont {Lin}, \citenamefont {Lowe}, \citenamefont {Magnier}, \citenamefont {Margutti}, \citenamefont {McGee}, \citenamefont {Minguez}, \citenamefont {Moore}, \citenamefont {Nathan}, \citenamefont {Oates}, \citenamefont {Patra}, \citenamefont {Ramsden}, \citenamefont {Ravi}, \citenamefont {Ridley}, \citenamefont {Sheng}, \citenamefont {Smartt}, \citenamefont {Smith}, \citenamefont {Srivastav}, \citenamefont {Stein}, \citenamefont {Stevance}, \citenamefont {Turner}, \citenamefont {Wainscoat}, \citenamefont {Weston}, \citenamefont {Wevers},\ and\ \citenamefont {Young}}]{Nicholl_2024}%
  \BibitemOpen
  \bibfield  {author} {\bibinfo {author} {\bibfnamefont {M.}~\bibnamefont {Nicholl}}, \bibinfo {author} {\bibfnamefont {D.~R.}\ \bibnamefont {Pasham}}, \bibinfo {author} {\bibfnamefont {A.}~\bibnamefont {Mummery}}, \bibinfo {author} {\bibfnamefont {M.}~\bibnamefont {Guolo}}, \bibinfo {author} {\bibfnamefont {K.}~\bibnamefont {Gendreau}}, \bibinfo {author} {\bibfnamefont {G.~C.}\ \bibnamefont {Dewangan}}, \bibinfo {author} {\bibfnamefont {E.~C.}\ \bibnamefont {Ferrara}}, \bibinfo {author} {\bibfnamefont {R.}~\bibnamefont {Remillard}}, \bibinfo {author} {\bibfnamefont {C.}~\bibnamefont {Bonnerot}}, \bibinfo {author} {\bibfnamefont {J.}~\bibnamefont {Chakraborty}}, \bibinfo {author} {\bibfnamefont {A.}~\bibnamefont {Hajela}}, \bibinfo {author} {\bibfnamefont {V.~S.}\ \bibnamefont {Dhillon}}, \bibinfo {author} {\bibfnamefont {A.~F.}\ \bibnamefont {Gillan}}, \bibinfo {author} {\bibfnamefont {J.}~\bibnamefont {Greenwood}}, \bibinfo {author} {\bibfnamefont {M.~E.}\ \bibnamefont {Huber}}, \bibinfo {author}
  {\bibfnamefont {A.}~\bibnamefont {Janiuk}}, \bibinfo {author} {\bibfnamefont {G.}~\bibnamefont {Salvesen}}, \bibinfo {author} {\bibfnamefont {S.}~\bibnamefont {van Velzen}}, \bibinfo {author} {\bibfnamefont {A.}~\bibnamefont {Aamer}}, \bibinfo {author} {\bibfnamefont {K.~D.}\ \bibnamefont {Alexander}}, \bibinfo {author} {\bibfnamefont {C.~R.}\ \bibnamefont {Angus}}, \bibinfo {author} {\bibfnamefont {Z.}~\bibnamefont {Arzoumanian}}, \bibinfo {author} {\bibfnamefont {K.}~\bibnamefont {Auchettl}}, \bibinfo {author} {\bibfnamefont {E.}~\bibnamefont {Berger}}, \bibinfo {author} {\bibfnamefont {T.}~\bibnamefont {de~Boer}}, \bibinfo {author} {\bibfnamefont {Y.}~\bibnamefont {Cendes}}, \bibinfo {author} {\bibfnamefont {K.~C.}\ \bibnamefont {Chambers}}, \bibinfo {author} {\bibfnamefont {T.-W.}\ \bibnamefont {Chen}}, \bibinfo {author} {\bibfnamefont {R.}~\bibnamefont {Chornock}}, \bibinfo {author} {\bibfnamefont {M.~D.}\ \bibnamefont {Fulton}}, \bibinfo {author} {\bibfnamefont {H.}~\bibnamefont {Gao}}, \bibinfo
  {author} {\bibfnamefont {J.~H.}\ \bibnamefont {Gillanders}}, \bibinfo {author} {\bibfnamefont {S.}~\bibnamefont {Gomez}}, \bibinfo {author} {\bibfnamefont {B.~P.}\ \bibnamefont {Gompertz}}, \bibinfo {author} {\bibfnamefont {A.~C.}\ \bibnamefont {Fabian}}, \bibinfo {author} {\bibfnamefont {J.}~\bibnamefont {Herman}}, \bibinfo {author} {\bibfnamefont {A.}~\bibnamefont {Ingram}}, \bibinfo {author} {\bibfnamefont {E.}~\bibnamefont {Kara}}, \bibinfo {author} {\bibfnamefont {T.}~\bibnamefont {Laskar}}, \bibinfo {author} {\bibfnamefont {A.}~\bibnamefont {Lawrence}}, \bibinfo {author} {\bibfnamefont {C.-C.}\ \bibnamefont {Lin}}, \bibinfo {author} {\bibfnamefont {T.~B.}\ \bibnamefont {Lowe}}, \bibinfo {author} {\bibfnamefont {E.~A.}\ \bibnamefont {Magnier}}, \bibinfo {author} {\bibfnamefont {R.}~\bibnamefont {Margutti}}, \bibinfo {author} {\bibfnamefont {S.~L.}\ \bibnamefont {McGee}}, \bibinfo {author} {\bibfnamefont {P.}~\bibnamefont {Minguez}}, \bibinfo {author} {\bibfnamefont {T.}~\bibnamefont {Moore}}, \bibinfo
  {author} {\bibfnamefont {E.}~\bibnamefont {Nathan}}, \bibinfo {author} {\bibfnamefont {S.~R.}\ \bibnamefont {Oates}}, \bibinfo {author} {\bibfnamefont {K.~C.}\ \bibnamefont {Patra}}, \bibinfo {author} {\bibfnamefont {P.}~\bibnamefont {Ramsden}}, \bibinfo {author} {\bibfnamefont {V.}~\bibnamefont {Ravi}}, \bibinfo {author} {\bibfnamefont {E.~J.}\ \bibnamefont {Ridley}}, \bibinfo {author} {\bibfnamefont {X.}~\bibnamefont {Sheng}}, \bibinfo {author} {\bibfnamefont {S.~J.}\ \bibnamefont {Smartt}}, \bibinfo {author} {\bibfnamefont {K.~W.}\ \bibnamefont {Smith}}, \bibinfo {author} {\bibfnamefont {S.}~\bibnamefont {Srivastav}}, \bibinfo {author} {\bibfnamefont {R.}~\bibnamefont {Stein}}, \bibinfo {author} {\bibfnamefont {H.~F.}\ \bibnamefont {Stevance}}, \bibinfo {author} {\bibfnamefont {S.~G.~D.}\ \bibnamefont {Turner}}, \bibinfo {author} {\bibfnamefont {R.~J.}\ \bibnamefont {Wainscoat}}, \bibinfo {author} {\bibfnamefont {J.}~\bibnamefont {Weston}}, \bibinfo {author} {\bibfnamefont {T.}~\bibnamefont {Wevers}},\
  and\ \bibinfo {author} {\bibfnamefont {D.~R.}\ \bibnamefont {Young}},\ }\bibfield  {title} {\bibinfo {title} {Quasi-periodic x-ray eruptions years after a nearby tidal disruption event},\ }\href {https://doi.org/10.1038/s41586-024-08023-6} {\bibfield  {journal} {\bibinfo  {journal} {Nature}\ }\textbf {\bibinfo {volume} {634}},\ \bibinfo {pages} {804–808} (\bibinfo {year} {2024})}\BibitemShut {NoStop}%
\bibitem [{\citenamefont {Chakraborty}\ \emph {et~al.}(2025)\citenamefont {Chakraborty}, \citenamefont {Kara}, \citenamefont {Arcodia}, \citenamefont {Buchner}, \citenamefont {Giustini}, \citenamefont {Hernández-García}, \citenamefont {Linial}, \citenamefont {Masterson}, \citenamefont {Miniutti}, \citenamefont {Mummery}, \citenamefont {Panagiotou}, \citenamefont {Quintin},\ and\ \citenamefont {Sánchez-Sáez}}]{Chakraborty_2025}%
  \BibitemOpen
  \bibfield  {author} {\bibinfo {author} {\bibfnamefont {J.}~\bibnamefont {Chakraborty}}, \bibinfo {author} {\bibfnamefont {E.}~\bibnamefont {Kara}}, \bibinfo {author} {\bibfnamefont {R.}~\bibnamefont {Arcodia}}, \bibinfo {author} {\bibfnamefont {J.}~\bibnamefont {Buchner}}, \bibinfo {author} {\bibfnamefont {M.}~\bibnamefont {Giustini}}, \bibinfo {author} {\bibfnamefont {L.}~\bibnamefont {Hernández-García}}, \bibinfo {author} {\bibfnamefont {I.}~\bibnamefont {Linial}}, \bibinfo {author} {\bibfnamefont {M.}~\bibnamefont {Masterson}}, \bibinfo {author} {\bibfnamefont {G.}~\bibnamefont {Miniutti}}, \bibinfo {author} {\bibfnamefont {A.}~\bibnamefont {Mummery}}, \bibinfo {author} {\bibfnamefont {C.}~\bibnamefont {Panagiotou}}, \bibinfo {author} {\bibfnamefont {E.}~\bibnamefont {Quintin}},\ and\ \bibinfo {author} {\bibfnamefont {P.}~\bibnamefont {Sánchez-Sáez}},\ }\href {https://doi.org/10.48550/ARXIV.2503.19013} {\bibinfo {title} {Discovery of quasi-periodic eruptions in the tidal disruption event and extreme
  coronal line emitter at2022upj: implications for the qpe/tde fraction and a connection to ecles}} (\bibinfo {year} {2025})\BibitemShut {NoStop}%
\bibitem [{\citenamefont {Kara}\ and\ \citenamefont {García}(2025)}]{Kara_2025}%
  \BibitemOpen
  \bibfield  {author} {\bibinfo {author} {\bibfnamefont {E.}~\bibnamefont {Kara}}\ and\ \bibinfo {author} {\bibfnamefont {J.}~\bibnamefont {García}},\ }\href {https://doi.org/10.48550/ARXIV.2503.22791} {\bibinfo {title} {Supermassive black holes in x-rays: From standard accretion to extreme transients}} (\bibinfo {year} {2025})\BibitemShut {NoStop}%
\bibitem [{\citenamefont {Chakraborty}\ \emph {et~al.}(2021)\citenamefont {Chakraborty}, \citenamefont {Kara}, \citenamefont {Masterson}, \citenamefont {Giustini}, \citenamefont {Miniutti},\ and\ \citenamefont {Saxton}}]{Chakraborty_2021}%
  \BibitemOpen
  \bibfield  {author} {\bibinfo {author} {\bibfnamefont {J.}~\bibnamefont {Chakraborty}}, \bibinfo {author} {\bibfnamefont {E.}~\bibnamefont {Kara}}, \bibinfo {author} {\bibfnamefont {M.}~\bibnamefont {Masterson}}, \bibinfo {author} {\bibfnamefont {M.}~\bibnamefont {Giustini}}, \bibinfo {author} {\bibfnamefont {G.}~\bibnamefont {Miniutti}},\ and\ \bibinfo {author} {\bibfnamefont {R.}~\bibnamefont {Saxton}},\ }\bibfield  {title} {\bibinfo {title} {Possible x-ray quasi-periodic eruptions in a tidal disruption event candidate},\ }\href {https://doi.org/10.3847/2041-8213/ac313b} {\bibfield  {journal} {\bibinfo  {journal} {Astrophys.~J.~Lett.}\ }\textbf {\bibinfo {volume} {921}},\ \bibinfo {pages} {L40} (\bibinfo {year} {2021})}\BibitemShut {NoStop}%
\bibitem [{\citenamefont {Quintin}\ \emph {et~al.}(2023)\citenamefont {Quintin}, \citenamefont {Webb}, \citenamefont {Guillot}, \citenamefont {Miniutti}, \citenamefont {Kammoun}, \citenamefont {Giustini}, \citenamefont {Arcodia}, \citenamefont {Soucail}, \citenamefont {Clerc}, \citenamefont {Amato},\ and\ \citenamefont {Markwardt}}]{Quintin_2023}%
  \BibitemOpen
  \bibfield  {author} {\bibinfo {author} {\bibfnamefont {E.}~\bibnamefont {Quintin}}, \bibinfo {author} {\bibfnamefont {N.~A.}\ \bibnamefont {Webb}}, \bibinfo {author} {\bibfnamefont {S.}~\bibnamefont {Guillot}}, \bibinfo {author} {\bibfnamefont {G.}~\bibnamefont {Miniutti}}, \bibinfo {author} {\bibfnamefont {E.~S.}\ \bibnamefont {Kammoun}}, \bibinfo {author} {\bibfnamefont {M.}~\bibnamefont {Giustini}}, \bibinfo {author} {\bibfnamefont {R.}~\bibnamefont {Arcodia}}, \bibinfo {author} {\bibfnamefont {G.}~\bibnamefont {Soucail}}, \bibinfo {author} {\bibfnamefont {N.}~\bibnamefont {Clerc}}, \bibinfo {author} {\bibfnamefont {R.}~\bibnamefont {Amato}},\ and\ \bibinfo {author} {\bibfnamefont {C.~B.}\ \bibnamefont {Markwardt}},\ }\bibfield  {title} {\bibinfo {title} {Tormund’s return: Hints of quasi-periodic eruption features from a recent optical tidal disruption event},\ }\href {https://doi.org/10.1051/0004-6361/202346440} {\bibfield  {journal} {\bibinfo  {journal} {Astron.~Astrophys.}\ }\textbf {\bibinfo
  {volume} {675}},\ \bibinfo {pages} {A152} (\bibinfo {year} {2023})}\BibitemShut {NoStop}%
\bibitem [{\citenamefont {Bykov}\ \emph {et~al.}(2024)\citenamefont {Bykov}, \citenamefont {Gilfanov}, \citenamefont {Sunyaev},\ and\ \citenamefont {Medvedev}}]{Bykov_2024}%
  \BibitemOpen
  \bibfield  {author} {\bibinfo {author} {\bibfnamefont {S.}~\bibnamefont {Bykov}}, \bibinfo {author} {\bibfnamefont {M.}~\bibnamefont {Gilfanov}}, \bibinfo {author} {\bibfnamefont {R.}~\bibnamefont {Sunyaev}},\ and\ \bibinfo {author} {\bibfnamefont {P.}~\bibnamefont {Medvedev}},\ }\href {https://doi.org/10.48550/ARXIV.2409.16908} {\bibinfo {title} {Further evidence of quasiperiodic eruptions in a tidal disruption event at2019vcb by srg/erosita}} (\bibinfo {year} {2024})\BibitemShut {NoStop}%
\bibitem [{\citenamefont {Suková}\ \emph {et~al.}(2021)\citenamefont {Suková}, \citenamefont {Zajaček}, \citenamefont {Witzany},\ and\ \citenamefont {Karas}}]{Suková_2021}%
  \BibitemOpen
  \bibfield  {author} {\bibinfo {author} {\bibfnamefont {P.}~\bibnamefont {Suková}}, \bibinfo {author} {\bibfnamefont {M.}~\bibnamefont {Zajaček}}, \bibinfo {author} {\bibfnamefont {V.}~\bibnamefont {Witzany}},\ and\ \bibinfo {author} {\bibfnamefont {V.}~\bibnamefont {Karas}},\ }\bibfield  {title} {\bibinfo {title} {Stellar transits across a magnetized accretion torus as a mechanism for plasmoid ejection},\ }\href {https://doi.org/10.3847/1538-4357/ac05c6} {\bibfield  {journal} {\bibinfo  {journal} {Astrophys.~J.}\ }\textbf {\bibinfo {volume} {917}},\ \bibinfo {pages} {43} (\bibinfo {year} {2021})}\BibitemShut {NoStop}%
\bibitem [{\citenamefont {Xian}\ \emph {et~al.}(2021)\citenamefont {Xian}, \citenamefont {Zhang}, \citenamefont {Dou}, \citenamefont {He},\ and\ \citenamefont {Shu}}]{Xian_2021}%
  \BibitemOpen
  \bibfield  {author} {\bibinfo {author} {\bibfnamefont {J.}~\bibnamefont {Xian}}, \bibinfo {author} {\bibfnamefont {F.}~\bibnamefont {Zhang}}, \bibinfo {author} {\bibfnamefont {L.}~\bibnamefont {Dou}}, \bibinfo {author} {\bibfnamefont {J.}~\bibnamefont {He}},\ and\ \bibinfo {author} {\bibfnamefont {X.}~\bibnamefont {Shu}},\ }\bibfield  {title} {\bibinfo {title} {X-ray quasi-periodic eruptions driven by star–disk collisions: Application to gsn069 and probing the spin of massive black holes},\ }\href {https://doi.org/10.3847/2041-8213/ac31aa} {\bibfield  {journal} {\bibinfo  {journal} {Astrophys.~J.~Lett.}\ }\textbf {\bibinfo {volume} {921}},\ \bibinfo {pages} {L32} (\bibinfo {year} {2021})}\BibitemShut {NoStop}%
\bibitem [{\citenamefont {Franchini}\ \emph {et~al.}(2023)\citenamefont {Franchini}, \citenamefont {Bonetti}, \citenamefont {Lupi}, \citenamefont {Miniutti}, \citenamefont {Bortolas}, \citenamefont {Giustini}, \citenamefont {Dotti}, \citenamefont {Sesana}, \citenamefont {Arcodia},\ and\ \citenamefont {Ryu}}]{Franchini_2023}%
  \BibitemOpen
  \bibfield  {author} {\bibinfo {author} {\bibfnamefont {A.}~\bibnamefont {Franchini}}, \bibinfo {author} {\bibfnamefont {M.}~\bibnamefont {Bonetti}}, \bibinfo {author} {\bibfnamefont {A.}~\bibnamefont {Lupi}}, \bibinfo {author} {\bibfnamefont {G.}~\bibnamefont {Miniutti}}, \bibinfo {author} {\bibfnamefont {E.}~\bibnamefont {Bortolas}}, \bibinfo {author} {\bibfnamefont {M.}~\bibnamefont {Giustini}}, \bibinfo {author} {\bibfnamefont {M.}~\bibnamefont {Dotti}}, \bibinfo {author} {\bibfnamefont {A.}~\bibnamefont {Sesana}}, \bibinfo {author} {\bibfnamefont {R.}~\bibnamefont {Arcodia}},\ and\ \bibinfo {author} {\bibfnamefont {T.}~\bibnamefont {Ryu}},\ }\bibfield  {title} {\bibinfo {title} {Quasi-periodic eruptions from impacts between the secondary and a rigidly precessing accretion disc in an extreme mass-ratio inspiral system},\ }\href {https://doi.org/10.1051/0004-6361/202346565} {\bibfield  {journal} {\bibinfo  {journal} {Astron.~Astrophys.}\ }\textbf {\bibinfo {volume} {675}},\ \bibinfo {pages} {A100}
  (\bibinfo {year} {2023})}\BibitemShut {NoStop}%
\bibitem [{\citenamefont {Linial}\ and\ \citenamefont {Metzger}(2023)}]{Linial_2023}%
  \BibitemOpen
  \bibfield  {author} {\bibinfo {author} {\bibfnamefont {I.}~\bibnamefont {Linial}}\ and\ \bibinfo {author} {\bibfnamefont {B.~D.}\ \bibnamefont {Metzger}},\ }\bibfield  {title} {\bibinfo {title} {Emri + tde = qpe: Periodic x-ray flares from star–disk collisions in galactic nuclei},\ }\href {https://doi.org/10.3847/1538-4357/acf65b} {\bibfield  {journal} {\bibinfo  {journal} {Astrophys.~J.}\ }\textbf {\bibinfo {volume} {957}},\ \bibinfo {pages} {34} (\bibinfo {year} {2023})}\BibitemShut {NoStop}%
\bibitem [{\citenamefont {Zhou}\ \emph {et~al.}(2024{\natexlab{a}})\citenamefont {Zhou}, \citenamefont {Huang}, \citenamefont {Guo}, \citenamefont {Li},\ and\ \citenamefont {Pan}}]{Zhou_2024}%
  \BibitemOpen
  \bibfield  {author} {\bibinfo {author} {\bibfnamefont {C.}~\bibnamefont {Zhou}}, \bibinfo {author} {\bibfnamefont {L.}~\bibnamefont {Huang}}, \bibinfo {author} {\bibfnamefont {K.}~\bibnamefont {Guo}}, \bibinfo {author} {\bibfnamefont {Y.-P.}\ \bibnamefont {Li}},\ and\ \bibinfo {author} {\bibfnamefont {Z.}~\bibnamefont {Pan}},\ }\bibfield  {title} {\bibinfo {title} {Probing orbits of stellar mass objects deep in galactic nuclei with quasiperiodic eruptions},\ }\href {https://doi.org/10.1103/PhysRevD.109.103031} {\bibfield  {journal} {\bibinfo  {journal} {Phys. Rev. D}\ }\textbf {\bibinfo {volume} {109}},\ \bibinfo {pages} {103031} (\bibinfo {year} {2024}{\natexlab{a}})}\BibitemShut {NoStop}%
\bibitem [{\citenamefont {Zhou}\ \emph {et~al.}(2024{\natexlab{b}})\citenamefont {Zhou}, \citenamefont {Zhong}, \citenamefont {Zeng}, \citenamefont {Huang},\ and\ \citenamefont {Pan}}]{Zhou_2024_2}%
  \BibitemOpen
  \bibfield  {author} {\bibinfo {author} {\bibfnamefont {C.}~\bibnamefont {Zhou}}, \bibinfo {author} {\bibfnamefont {B.}~\bibnamefont {Zhong}}, \bibinfo {author} {\bibfnamefont {Y.}~\bibnamefont {Zeng}}, \bibinfo {author} {\bibfnamefont {L.}~\bibnamefont {Huang}},\ and\ \bibinfo {author} {\bibfnamefont {Z.}~\bibnamefont {Pan}},\ }\bibfield  {title} {\bibinfo {title} {Probing orbits of stellar mass objects deep in galactic nuclei with quasiperiodic eruptions. ii. population analysis},\ }\href {https://doi.org/10.1103/PhysRevD.110.083019} {\bibfield  {journal} {\bibinfo  {journal} {Phys. Rev. D}\ }\textbf {\bibinfo {volume} {110}},\ \bibinfo {pages} {083019} (\bibinfo {year} {2024}{\natexlab{b}})}\BibitemShut {NoStop}%
\bibitem [{\citenamefont {Miniutti}\ \emph {et~al.}(2025)\citenamefont {Miniutti}, \citenamefont {Franchini}, \citenamefont {Bonetti}, \citenamefont {Giustini}, \citenamefont {Chakraborty}, \citenamefont {Arcodia}, \citenamefont {Saxton}, \citenamefont {Quintin}, \citenamefont {Kosec}, \citenamefont {Linial},\ and\ \citenamefont {Sesana}}]{Miniutti_2025}%
  \BibitemOpen
  \bibfield  {author} {\bibinfo {author} {\bibfnamefont {G.}~\bibnamefont {Miniutti}}, \bibinfo {author} {\bibfnamefont {A.}~\bibnamefont {Franchini}}, \bibinfo {author} {\bibfnamefont {M.}~\bibnamefont {Bonetti}}, \bibinfo {author} {\bibfnamefont {M.}~\bibnamefont {Giustini}}, \bibinfo {author} {\bibfnamefont {J.}~\bibnamefont {Chakraborty}}, \bibinfo {author} {\bibfnamefont {R.}~\bibnamefont {Arcodia}}, \bibinfo {author} {\bibfnamefont {R.}~\bibnamefont {Saxton}}, \bibinfo {author} {\bibfnamefont {E.}~\bibnamefont {Quintin}}, \bibinfo {author} {\bibfnamefont {P.}~\bibnamefont {Kosec}}, \bibinfo {author} {\bibfnamefont {I.}~\bibnamefont {Linial}},\ and\ \bibinfo {author} {\bibfnamefont {A.}~\bibnamefont {Sesana}},\ }\bibfield  {title} {\bibinfo {title} {Eppur si muove: Evidence of disc precession or a sub-milliparsec smbh binary in the qpe-emitting galaxy gsn 069},\ }\href {https://doi.org/10.1051/0004-6361/202452400} {\bibfield  {journal} {\bibinfo  {journal} {Astron.~Astrophys.}\ }\textbf {\bibinfo
  {volume} {693}},\ \bibinfo {pages} {A179} (\bibinfo {year} {2025})}\BibitemShut {NoStop}%
\bibitem [{\citenamefont {King}(2020)}]{King_2020}%
  \BibitemOpen
  \bibfield  {author} {\bibinfo {author} {\bibfnamefont {A.}~\bibnamefont {King}},\ }\bibfield  {title} {\bibinfo {title} {Gsn 069 – a tidal disruption near miss},\ }\href {https://doi.org/10.1093/mnrasl/slaa020} {\bibfield  {journal} {\bibinfo  {journal} {Mon.~Not.~R.~Astron.~Soc.:~Lett.}\ }\textbf {\bibinfo {volume} {493}},\ \bibinfo {pages} {L120–L123} (\bibinfo {year} {2020})}\BibitemShut {NoStop}%
\bibitem [{\citenamefont {Krolik}\ and\ \citenamefont {Linial}(2022)}]{Krolik_2022}%
  \BibitemOpen
  \bibfield  {author} {\bibinfo {author} {\bibfnamefont {J.~H.}\ \bibnamefont {Krolik}}\ and\ \bibinfo {author} {\bibfnamefont {I.}~\bibnamefont {Linial}},\ }\bibfield  {title} {\bibinfo {title} {Quasiperiodic erupters: A stellar mass-transfer model for the radiation},\ }\href {https://doi.org/10.3847/1538-4357/ac9eb6} {\bibfield  {journal} {\bibinfo  {journal} {Astrophys.~J.}\ }\textbf {\bibinfo {volume} {941}},\ \bibinfo {pages} {24} (\bibinfo {year} {2022})}\BibitemShut {NoStop}%
\bibitem [{\citenamefont {Wang}\ \emph {et~al.}(2022)\citenamefont {Wang}, \citenamefont {Yin}, \citenamefont {Ma},\ and\ \citenamefont {Wu}}]{Wang_2022}%
  \BibitemOpen
  \bibfield  {author} {\bibinfo {author} {\bibfnamefont {M.}~\bibnamefont {Wang}}, \bibinfo {author} {\bibfnamefont {J.}~\bibnamefont {Yin}}, \bibinfo {author} {\bibfnamefont {Y.}~\bibnamefont {Ma}},\ and\ \bibinfo {author} {\bibfnamefont {Q.}~\bibnamefont {Wu}},\ }\bibfield  {title} {\bibinfo {title} {A model for the possible connection between a tidal disruption event and quasi-periodic eruption in gsn 069},\ }\href {https://doi.org/10.3847/1538-4357/ac75e6} {\bibfield  {journal} {\bibinfo  {journal} {Astrophys.~J.}\ }\textbf {\bibinfo {volume} {933}},\ \bibinfo {pages} {225} (\bibinfo {year} {2022})}\BibitemShut {NoStop}%
\bibitem [{\citenamefont {Metzger}\ \emph {et~al.}(2022)\citenamefont {Metzger}, \citenamefont {Stone},\ and\ \citenamefont {Gilbaum}}]{Metzger_2022}%
  \BibitemOpen
  \bibfield  {author} {\bibinfo {author} {\bibfnamefont {B.~D.}\ \bibnamefont {Metzger}}, \bibinfo {author} {\bibfnamefont {N.~C.}\ \bibnamefont {Stone}},\ and\ \bibinfo {author} {\bibfnamefont {S.}~\bibnamefont {Gilbaum}},\ }\bibfield  {title} {\bibinfo {title} {Interacting stellar emris as sources of quasi-periodic eruptions in galactic nuclei},\ }\href {https://doi.org/10.3847/1538-4357/ac3ee1} {\bibfield  {journal} {\bibinfo  {journal} {Astrophys.~J.}\ }\textbf {\bibinfo {volume} {926}},\ \bibinfo {pages} {101} (\bibinfo {year} {2022})}\BibitemShut {NoStop}%
\bibitem [{\citenamefont {Pan}\ \emph {et~al.}(2022)\citenamefont {Pan}, \citenamefont {Li}, \citenamefont {Cao}, \citenamefont {Miniutti},\ and\ \citenamefont {Gu}}]{Pan_2022}%
  \BibitemOpen
  \bibfield  {author} {\bibinfo {author} {\bibfnamefont {X.}~\bibnamefont {Pan}}, \bibinfo {author} {\bibfnamefont {S.-L.}\ \bibnamefont {Li}}, \bibinfo {author} {\bibfnamefont {X.}~\bibnamefont {Cao}}, \bibinfo {author} {\bibfnamefont {G.}~\bibnamefont {Miniutti}},\ and\ \bibinfo {author} {\bibfnamefont {M.}~\bibnamefont {Gu}},\ }\bibfield  {title} {\bibinfo {title} {A disk instability model for the quasi-periodic eruptions of gsn 069},\ }\href {https://doi.org/10.3847/2041-8213/ac5faf} {\bibfield  {journal} {\bibinfo  {journal} {Astrophys.~J.~Lett.}\ }\textbf {\bibinfo {volume} {928}},\ \bibinfo {pages} {L18} (\bibinfo {year} {2022})}\BibitemShut {NoStop}%
\bibitem [{\citenamefont {Pan}\ \emph {et~al.}(2023)\citenamefont {Pan}, \citenamefont {Li},\ and\ \citenamefont {Cao}}]{Pan_2023}%
  \BibitemOpen
  \bibfield  {author} {\bibinfo {author} {\bibfnamefont {X.}~\bibnamefont {Pan}}, \bibinfo {author} {\bibfnamefont {S.-L.}\ \bibnamefont {Li}},\ and\ \bibinfo {author} {\bibfnamefont {X.}~\bibnamefont {Cao}},\ }\bibfield  {title} {\bibinfo {title} {Application of the disk instability model to all quasiperiodic eruptions},\ }\href {https://doi.org/10.3847/1538-4357/acd180} {\bibfield  {journal} {\bibinfo  {journal} {Astrophys.~J.}\ }\textbf {\bibinfo {volume} {952}},\ \bibinfo {pages} {32} (\bibinfo {year} {2023})}\BibitemShut {NoStop}%
\bibitem [{\citenamefont {Ingram}\ \emph {et~al.}(2021)\citenamefont {Ingram}, \citenamefont {Motta}, \citenamefont {Aigrain},\ and\ \citenamefont {Karastergiou}}]{Ingram_2021}%
  \BibitemOpen
  \bibfield  {author} {\bibinfo {author} {\bibfnamefont {A.}~\bibnamefont {Ingram}}, \bibinfo {author} {\bibfnamefont {S.~E.}\ \bibnamefont {Motta}}, \bibinfo {author} {\bibfnamefont {S.}~\bibnamefont {Aigrain}},\ and\ \bibinfo {author} {\bibfnamefont {A.}~\bibnamefont {Karastergiou}},\ }\bibfield  {title} {\bibinfo {title} {A self-lensing binary massive black hole interpretation of quasi-periodic eruptions},\ }\href {https://doi.org/10.1093/mnras/stab609} {\bibfield  {journal} {\bibinfo  {journal} {Mon.~Not.~R.~Astron.~Soc.}\ }\textbf {\bibinfo {volume} {503}},\ \bibinfo {pages} {1703–1716} (\bibinfo {year} {2021})}\BibitemShut {NoStop}%
\bibitem [{\citenamefont {Miniutti}\ \emph {et~al.}(2023{\natexlab{a}})\citenamefont {Miniutti}, \citenamefont {Giustini}, \citenamefont {Arcodia}, \citenamefont {Saxton}, \citenamefont {Read}, \citenamefont {Bianchi},\ and\ \citenamefont {Alexander}}]{Miniutti_2023}%
  \BibitemOpen
  \bibfield  {author} {\bibinfo {author} {\bibfnamefont {G.}~\bibnamefont {Miniutti}}, \bibinfo {author} {\bibfnamefont {M.}~\bibnamefont {Giustini}}, \bibinfo {author} {\bibfnamefont {R.}~\bibnamefont {Arcodia}}, \bibinfo {author} {\bibfnamefont {R.~D.}\ \bibnamefont {Saxton}}, \bibinfo {author} {\bibfnamefont {A.~M.}\ \bibnamefont {Read}}, \bibinfo {author} {\bibfnamefont {S.}~\bibnamefont {Bianchi}},\ and\ \bibinfo {author} {\bibfnamefont {K.~D.}\ \bibnamefont {Alexander}},\ }\bibfield  {title} {\bibinfo {title} {Repeating tidal disruptions in gsn 069: Long-term evolution and constraints on quasi-periodic eruptions’ models},\ }\href {https://doi.org/10.1051/0004-6361/202244512} {\bibfield  {journal} {\bibinfo  {journal} {Astron.~Astrophys.}\ }\textbf {\bibinfo {volume} {670}},\ \bibinfo {pages} {A93} (\bibinfo {year} {2023}{\natexlab{a}})}\BibitemShut {NoStop}%
\bibitem [{\citenamefont {Miniutti}\ \emph {et~al.}(2023{\natexlab{b}})\citenamefont {Miniutti}, \citenamefont {Giustini}, \citenamefont {Arcodia}, \citenamefont {Saxton}, \citenamefont {Chakraborty}, \citenamefont {Read},\ and\ \citenamefont {Kara}}]{Miniutti_2023_a}%
  \BibitemOpen
  \bibfield  {author} {\bibinfo {author} {\bibfnamefont {G.}~\bibnamefont {Miniutti}}, \bibinfo {author} {\bibfnamefont {M.}~\bibnamefont {Giustini}}, \bibinfo {author} {\bibfnamefont {R.}~\bibnamefont {Arcodia}}, \bibinfo {author} {\bibfnamefont {R.~D.}\ \bibnamefont {Saxton}}, \bibinfo {author} {\bibfnamefont {J.}~\bibnamefont {Chakraborty}}, \bibinfo {author} {\bibfnamefont {A.~M.}\ \bibnamefont {Read}},\ and\ \bibinfo {author} {\bibfnamefont {E.}~\bibnamefont {Kara}},\ }\bibfield  {title} {\bibinfo {title} {Alive and kicking: A new qpe phase in gsn 069 revealing a quiescent luminosity threshold for qpes},\ }\href {https://doi.org/10.1051/0004-6361/202346653} {\bibfield  {journal} {\bibinfo  {journal} {Astron.~Astrophys.}\ }\textbf {\bibinfo {volume} {674}},\ \bibinfo {pages} {L1} (\bibinfo {year} {2023}{\natexlab{b}})}\BibitemShut {NoStop}%
\bibitem [{\citenamefont {Pasham}\ \emph {et~al.}(2024)\citenamefont {Pasham}, \citenamefont {Coughlin}, \citenamefont {Zajaček}, \citenamefont {Linial}, \citenamefont {Suková}, \citenamefont {Nixon}, \citenamefont {Janiuk}, \citenamefont {Sniegowska}, \citenamefont {Witzany}, \citenamefont {Karas}, \citenamefont {Krumpe}, \citenamefont {Altamirano}, \citenamefont {Wevers},\ and\ \citenamefont {Arcodia}}]{Pasham_2024}%
  \BibitemOpen
  \bibfield  {author} {\bibinfo {author} {\bibfnamefont {D.~R.}\ \bibnamefont {Pasham}}, \bibinfo {author} {\bibfnamefont {E.~R.}\ \bibnamefont {Coughlin}}, \bibinfo {author} {\bibfnamefont {M.}~\bibnamefont {Zajaček}}, \bibinfo {author} {\bibfnamefont {I.}~\bibnamefont {Linial}}, \bibinfo {author} {\bibfnamefont {P.}~\bibnamefont {Suková}}, \bibinfo {author} {\bibfnamefont {C.~J.}\ \bibnamefont {Nixon}}, \bibinfo {author} {\bibfnamefont {A.}~\bibnamefont {Janiuk}}, \bibinfo {author} {\bibfnamefont {M.}~\bibnamefont {Sniegowska}}, \bibinfo {author} {\bibfnamefont {V.}~\bibnamefont {Witzany}}, \bibinfo {author} {\bibfnamefont {V.}~\bibnamefont {Karas}}, \bibinfo {author} {\bibfnamefont {M.}~\bibnamefont {Krumpe}}, \bibinfo {author} {\bibfnamefont {D.}~\bibnamefont {Altamirano}}, \bibinfo {author} {\bibfnamefont {T.}~\bibnamefont {Wevers}},\ and\ \bibinfo {author} {\bibfnamefont {R.}~\bibnamefont {Arcodia}},\ }\bibfield  {title} {\bibinfo {title} {Alive but barely kicking: News from 3+ yr of swift and
  xmm-newton x-ray monitoring of quasiperiodic eruptions from ero-qpe1},\ }\href {https://doi.org/10.3847/2041-8213/ad2a5c} {\bibfield  {journal} {\bibinfo  {journal} {Astrophys.~J.~Lett.}\ }\textbf {\bibinfo {volume} {963}},\ \bibinfo {pages} {L47} (\bibinfo {year} {2024})}\BibitemShut {NoStop}%
\bibitem [{\citenamefont {Suková}\ \emph {et~al.}(2024)\citenamefont {Suková}, \citenamefont {Tombesi}, \citenamefont {Pasham}, \citenamefont {Zajaček}, \citenamefont {Wevers}, \citenamefont {Ryu}, \citenamefont {Linial},\ and\ \citenamefont {Franchini}}]{Suková_2024}%
  \BibitemOpen
  \bibfield  {author} {\bibinfo {author} {\bibfnamefont {P.}~\bibnamefont {Suková}}, \bibinfo {author} {\bibfnamefont {F.}~\bibnamefont {Tombesi}}, \bibinfo {author} {\bibfnamefont {D.~R.}\ \bibnamefont {Pasham}}, \bibinfo {author} {\bibfnamefont {M.}~\bibnamefont {Zajaček}}, \bibinfo {author} {\bibfnamefont {T.}~\bibnamefont {Wevers}}, \bibinfo {author} {\bibfnamefont {T.}~\bibnamefont {Ryu}}, \bibinfo {author} {\bibfnamefont {I.}~\bibnamefont {Linial}},\ and\ \bibinfo {author} {\bibfnamefont {A.}~\bibnamefont {Franchini}},\ }\href {https://doi.org/10.48550/ARXIV.2411.04592} {\bibinfo {title} {Repeating transients in galactic nuclei: confronting observations with theory}} (\bibinfo {year} {2024})\BibitemShut {NoStop}%
\bibitem [{\citenamefont {Coughlin}\ and\ \citenamefont {Nixon}(2019)}]{Coughlin_2019}%
  \BibitemOpen
  \bibfield  {author} {\bibinfo {author} {\bibfnamefont {E.~R.}\ \bibnamefont {Coughlin}}\ and\ \bibinfo {author} {\bibfnamefont {C.~J.}\ \bibnamefont {Nixon}},\ }\bibfield  {title} {\bibinfo {title} {Partial stellar disruption by a supermassive black hole: Is the light curve really proportional to $t^{-9/4}$?},\ }\href {https://doi.org/10.3847/2041-8213/ab412d} {\bibfield  {journal} {\bibinfo  {journal} {Astrophys.~J.~Lett.}\ }\textbf {\bibinfo {volume} {883}},\ \bibinfo {pages} {L17} (\bibinfo {year} {2019})}\BibitemShut {NoStop}%
\bibitem [{\citenamefont {Gezari}(2021)}]{Gezari_2021}%
  \BibitemOpen
  \bibfield  {author} {\bibinfo {author} {\bibfnamefont {S.}~\bibnamefont {Gezari}},\ }\bibfield  {title} {\bibinfo {title} {Tidal disruption events},\ }\href {https://doi.org/10.1146/annurev-astro-111720-030029} {\bibfield  {journal} {\bibinfo  {journal} {Annu.~Rev.~Astron.~Astrophys.}\ }\textbf {\bibinfo {volume} {59}},\ \bibinfo {pages} {21–58} (\bibinfo {year} {2021})}\BibitemShut {NoStop}%
\bibitem [{\citenamefont {Rees}(1988)}]{Rees_1988}%
  \BibitemOpen
  \bibfield  {author} {\bibinfo {author} {\bibfnamefont {M.~J.}\ \bibnamefont {Rees}},\ }\bibfield  {title} {\bibinfo {title} {Tidal disruption of stars by black holes of 106–108 solar masses in nearby galaxies},\ }\href {https://doi.org/10.1038/333523a0} {\bibfield  {journal} {\bibinfo  {journal} {Nature}\ }\textbf {\bibinfo {volume} {333}},\ \bibinfo {pages} {523–528} (\bibinfo {year} {1988})}\BibitemShut {NoStop}%
\bibitem [{\citenamefont {{Phinney}}(1989)}]{Phinney_1989}%
  \BibitemOpen
  \bibfield  {author} {\bibinfo {author} {\bibfnamefont {E.~S.}\ \bibnamefont {{Phinney}}},\ }\bibfield  {title} {\bibinfo {title} {{Manifestations of a Massive Black Hole in the Galactic Center}},\ }in\ \href@noop {} {\emph {\bibinfo {booktitle} {The Center of the Galaxy}}},\ \bibinfo {series} {IAU Symposium}, Vol.\ \bibinfo {volume} {136},\ \bibinfo {editor} {edited by\ \bibinfo {editor} {\bibfnamefont {M.}~\bibnamefont {{Morris}}}}\ (\bibinfo {year} {1989})\ p.\ \bibinfo {pages} {543}\BibitemShut {NoStop}%
\bibitem [{\citenamefont {Nicholl}\ \emph {et~al.}(2020)\citenamefont {Nicholl}, \citenamefont {Wevers}, \citenamefont {Oates}, \citenamefont {Alexander}, \citenamefont {Leloudas}, \citenamefont {Onori}, \citenamefont {Jerkstrand}, \citenamefont {Gomez}, \citenamefont {Campana}, \citenamefont {Arcavi}, \citenamefont {Charalampopoulos}, \citenamefont {Gromadzki}, \citenamefont {Ihanec}, \citenamefont {Jonker}, \citenamefont {Lawrence}, \citenamefont {Mandel}, \citenamefont {Schulze}, \citenamefont {Short}, \citenamefont {Burke}, \citenamefont {McCully}, \citenamefont {Hiramatsu}, \citenamefont {Howell}, \citenamefont {Pellegrino}, \citenamefont {Abbot}, \citenamefont {Anderson}, \citenamefont {Berger}, \citenamefont {Blanchard}, \citenamefont {Cannizzaro}, \citenamefont {Chen}, \citenamefont {Dennefeld}, \citenamefont {Galbany}, \citenamefont {Gonz\`{a}lez-Gait\`{a}n}, \citenamefont {Hosseinzadeh}, \citenamefont {Inserra}, \citenamefont {Irani}, \citenamefont {Kuin}, \citenamefont {M\"{u}ller-Bravo},
  \citenamefont {Pineda}, \citenamefont {Ross}, \citenamefont {Roy}, \citenamefont {Smartt}, \citenamefont {Smith}, \citenamefont {Tucker}, \citenamefont {Wyrzykowski},\ and\ \citenamefont {Young}}]{Nicholl_2020}%
  \BibitemOpen
  \bibfield  {author} {\bibinfo {author} {\bibfnamefont {M.}~\bibnamefont {Nicholl}}, \bibinfo {author} {\bibfnamefont {T.}~\bibnamefont {Wevers}}, \bibinfo {author} {\bibfnamefont {S.~R.}\ \bibnamefont {Oates}}, \bibinfo {author} {\bibfnamefont {K.~D.}\ \bibnamefont {Alexander}}, \bibinfo {author} {\bibfnamefont {G.}~\bibnamefont {Leloudas}}, \bibinfo {author} {\bibfnamefont {F.}~\bibnamefont {Onori}}, \bibinfo {author} {\bibfnamefont {A.}~\bibnamefont {Jerkstrand}}, \bibinfo {author} {\bibfnamefont {S.}~\bibnamefont {Gomez}}, \bibinfo {author} {\bibfnamefont {S.}~\bibnamefont {Campana}}, \bibinfo {author} {\bibfnamefont {I.}~\bibnamefont {Arcavi}}, \bibinfo {author} {\bibfnamefont {P.}~\bibnamefont {Charalampopoulos}}, \bibinfo {author} {\bibfnamefont {M.}~\bibnamefont {Gromadzki}}, \bibinfo {author} {\bibfnamefont {N.}~\bibnamefont {Ihanec}}, \bibinfo {author} {\bibfnamefont {P.~G.}\ \bibnamefont {Jonker}}, \bibinfo {author} {\bibfnamefont {A.}~\bibnamefont {Lawrence}}, \bibinfo {author} {\bibfnamefont
  {I.}~\bibnamefont {Mandel}}, \bibinfo {author} {\bibfnamefont {S.}~\bibnamefont {Schulze}}, \bibinfo {author} {\bibfnamefont {P.}~\bibnamefont {Short}}, \bibinfo {author} {\bibfnamefont {J.}~\bibnamefont {Burke}}, \bibinfo {author} {\bibfnamefont {C.}~\bibnamefont {McCully}}, \bibinfo {author} {\bibfnamefont {D.}~\bibnamefont {Hiramatsu}}, \bibinfo {author} {\bibfnamefont {D.~A.}\ \bibnamefont {Howell}}, \bibinfo {author} {\bibfnamefont {C.}~\bibnamefont {Pellegrino}}, \bibinfo {author} {\bibfnamefont {H.}~\bibnamefont {Abbot}}, \bibinfo {author} {\bibfnamefont {J.~P.}\ \bibnamefont {Anderson}}, \bibinfo {author} {\bibfnamefont {E.}~\bibnamefont {Berger}}, \bibinfo {author} {\bibfnamefont {P.~K.}\ \bibnamefont {Blanchard}}, \bibinfo {author} {\bibfnamefont {G.}~\bibnamefont {Cannizzaro}}, \bibinfo {author} {\bibfnamefont {T.-W.}\ \bibnamefont {Chen}}, \bibinfo {author} {\bibfnamefont {M.}~\bibnamefont {Dennefeld}}, \bibinfo {author} {\bibfnamefont {L.}~\bibnamefont {Galbany}}, \bibinfo {author}
  {\bibfnamefont {S.}~\bibnamefont {Gonz\`{a}lez-Gait\`{a}n}}, \bibinfo {author} {\bibfnamefont {G.}~\bibnamefont {Hosseinzadeh}}, \bibinfo {author} {\bibfnamefont {C.}~\bibnamefont {Inserra}}, \bibinfo {author} {\bibfnamefont {I.}~\bibnamefont {Irani}}, \bibinfo {author} {\bibfnamefont {P.}~\bibnamefont {Kuin}}, \bibinfo {author} {\bibfnamefont {T.}~\bibnamefont {M\"{u}ller-Bravo}}, \bibinfo {author} {\bibfnamefont {J.}~\bibnamefont {Pineda}}, \bibinfo {author} {\bibfnamefont {N.~P.}\ \bibnamefont {Ross}}, \bibinfo {author} {\bibfnamefont {R.}~\bibnamefont {Roy}}, \bibinfo {author} {\bibfnamefont {S.~J.}\ \bibnamefont {Smartt}}, \bibinfo {author} {\bibfnamefont {K.~W.}\ \bibnamefont {Smith}}, \bibinfo {author} {\bibfnamefont {B.}~\bibnamefont {Tucker}}, \bibinfo {author} {\bibfnamefont {{\L}.}~\bibnamefont {Wyrzykowski}},\ and\ \bibinfo {author} {\bibfnamefont {D.~R.}\ \bibnamefont {Young}},\ }\bibfield  {title} {\bibinfo {title} {An outflow powers the optical rise of the nearby, fast-evolving tidal
  disruption event at2019qiz},\ }\href {https://doi.org/10.1093/mnras/staa2824} {\bibfield  {journal} {\bibinfo  {journal} {Mon.~Not.~R.~Astron.~Soc.}\ }\textbf {\bibinfo {volume} {499}},\ \bibinfo {pages} {482–504} (\bibinfo {year} {2020})}\BibitemShut {NoStop}%
\bibitem [{\citenamefont {Jiang}\ and\ \citenamefont {Pan}(2025)}]{Jiang_2025}%
  \BibitemOpen
  \bibfield  {author} {\bibinfo {author} {\bibfnamefont {N.}~\bibnamefont {Jiang}}\ and\ \bibinfo {author} {\bibfnamefont {Z.}~\bibnamefont {Pan}},\ }\href {https://doi.org/10.48550/ARXIV.2503.17609} {\bibinfo {title} {Embers of active galactic nuclei: Tidal disruption events and quasi-periodic eruptions}} (\bibinfo {year} {2025})\BibitemShut {NoStop}%
\bibitem [{\citenamefont {Zalamea}\ \emph {et~al.}(2010)\citenamefont {Zalamea}, \citenamefont {Menou},\ and\ \citenamefont {Beloborodov}}]{Zalamea_2010}%
  \BibitemOpen
  \bibfield  {author} {\bibinfo {author} {\bibfnamefont {I.}~\bibnamefont {Zalamea}}, \bibinfo {author} {\bibfnamefont {K.}~\bibnamefont {Menou}},\ and\ \bibinfo {author} {\bibfnamefont {A.~M.}\ \bibnamefont {Beloborodov}},\ }\bibfield  {title} {\bibinfo {title} {White dwarfs stripped by massive black holes: sources of coincident gravitational and electromagnetic radiation},\ }\href {https://doi.org/10.1111/j.1745-3933.2010.00930.x} {\bibfield  {journal} {\bibinfo  {journal} {Mon.~Not.~R.~Astron.~Soc.:~Lett.}\ }\textbf {\bibinfo {volume} {409}},\ \bibinfo {pages} {L25–L29} (\bibinfo {year} {2010})}\BibitemShut {NoStop}%
\bibitem [{\citenamefont {Press}\ and\ \citenamefont {Teukolsky}(1977)}]{Press_1977}%
  \BibitemOpen
  \bibfield  {author} {\bibinfo {author} {\bibfnamefont {W.~H.}\ \bibnamefont {Press}}\ and\ \bibinfo {author} {\bibfnamefont {S.~A.}\ \bibnamefont {Teukolsky}},\ }\bibfield  {title} {\bibinfo {title} {On formation of close binaries by two-body tidal capture},\ }\href {https://doi.org/10.1086/155143} {\bibfield  {journal} {\bibinfo  {journal} {Astrophys.~J.}\ }\textbf {\bibinfo {volume} {213}},\ \bibinfo {pages} {183} (\bibinfo {year} {1977})}\BibitemShut {NoStop}%
\bibitem [{\citenamefont {Fuller}\ and\ \citenamefont {Lai}(2012)}]{Fuller_2012}%
  \BibitemOpen
  \bibfield  {author} {\bibinfo {author} {\bibfnamefont {J.}~\bibnamefont {Fuller}}\ and\ \bibinfo {author} {\bibfnamefont {D.}~\bibnamefont {Lai}},\ }\bibfield  {title} {\bibinfo {title} {Tidal novae in compact binary white dwarfs},\ }\href {https://doi.org/10.1088/2041-8205/756/1/l17} {\bibfield  {journal} {\bibinfo  {journal} {Astrophys.~J.}\ }\textbf {\bibinfo {volume} {756}},\ \bibinfo {pages} {L17} (\bibinfo {year} {2012})}\BibitemShut {NoStop}%
\bibitem [{\citenamefont {Lamberts}\ \emph {et~al.}(2019)\citenamefont {Lamberts}, \citenamefont {Blunt}, \citenamefont {Littenberg}, \citenamefont {Garrison-Kimmel}, \citenamefont {Kupfer},\ and\ \citenamefont {Sanderson}}]{Lamberts_2019}%
  \BibitemOpen
  \bibfield  {author} {\bibinfo {author} {\bibfnamefont {A.}~\bibnamefont {Lamberts}}, \bibinfo {author} {\bibfnamefont {S.}~\bibnamefont {Blunt}}, \bibinfo {author} {\bibfnamefont {T.~B.}\ \bibnamefont {Littenberg}}, \bibinfo {author} {\bibfnamefont {S.}~\bibnamefont {Garrison-Kimmel}}, \bibinfo {author} {\bibfnamefont {T.}~\bibnamefont {Kupfer}},\ and\ \bibinfo {author} {\bibfnamefont {R.~E.}\ \bibnamefont {Sanderson}},\ }\bibfield  {title} {\bibinfo {title} {Predicting the lisa white dwarf binary population in the milky way with cosmological simulations},\ }\href {https://doi.org/10.1093/mnras/stz2834} {\bibfield  {journal} {\bibinfo  {journal} {Mon.~Not.~R.~Astron.~Soc.}\ }\textbf {\bibinfo {volume} {490}},\ \bibinfo {pages} {5888–5903} (\bibinfo {year} {2019})}\BibitemShut {NoStop}%
\bibitem [{\citenamefont {Eggleton}(1983)}]{Eggleton_1983}%
  \BibitemOpen
  \bibfield  {author} {\bibinfo {author} {\bibfnamefont {P.~P.}\ \bibnamefont {Eggleton}},\ }\bibfield  {title} {\bibinfo {title} {Approximations to the radii of roche lobes},\ }\href {https://doi.org/10.1086/160960} {\bibfield  {journal} {\bibinfo  {journal} {Astrophys.~J.}\ }\textbf {\bibinfo {volume} {268}},\ \bibinfo {pages} {368} (\bibinfo {year} {1983})}\BibitemShut {NoStop}%
\bibitem [{\citenamefont {Sepinsky}\ \emph {et~al.}(2007)\citenamefont {Sepinsky}, \citenamefont {Willems},\ and\ \citenamefont {Kalogera}}]{Sepinsky_2007}%
  \BibitemOpen
  \bibfield  {author} {\bibinfo {author} {\bibfnamefont {J.~F.}\ \bibnamefont {Sepinsky}}, \bibinfo {author} {\bibfnamefont {B.}~\bibnamefont {Willems}},\ and\ \bibinfo {author} {\bibfnamefont {V.}~\bibnamefont {Kalogera}},\ }\bibfield  {title} {\bibinfo {title} {Equipotential surfaces and lagrangian points in nonsynchronous, eccentric binary and planetary systems},\ }\href {https://doi.org/10.1086/513736} {\bibfield  {journal} {\bibinfo  {journal} {Astrophys.~J.}\ }\textbf {\bibinfo {volume} {660}},\ \bibinfo {pages} {1624–1635} (\bibinfo {year} {2007})}\BibitemShut {NoStop}%
\bibitem [{\citenamefont {Guillochon}\ \emph {et~al.}(2011)\citenamefont {Guillochon}, \citenamefont {Ramirez-Ruiz},\ and\ \citenamefont {Lin}}]{Guillochon_2011}%
  \BibitemOpen
  \bibfield  {author} {\bibinfo {author} {\bibfnamefont {J.}~\bibnamefont {Guillochon}}, \bibinfo {author} {\bibfnamefont {E.}~\bibnamefont {Ramirez-Ruiz}},\ and\ \bibinfo {author} {\bibfnamefont {D.}~\bibnamefont {Lin}},\ }\bibfield  {title} {\bibinfo {title} {Consequences of the ejection and disruption of giant planets},\ }\href {https://doi.org/10.1088/0004-637x/732/2/74} {\bibfield  {journal} {\bibinfo  {journal} {Astrophys.~J.}\ }\textbf {\bibinfo {volume} {732}},\ \bibinfo {pages} {74} (\bibinfo {year} {2011})}\BibitemShut {NoStop}%
\bibitem [{\citenamefont {Liu}\ \emph {et~al.}(2012)\citenamefont {Liu}, \citenamefont {Guillochon}, \citenamefont {Lin},\ and\ \citenamefont {Ramirez-Ruiz}}]{Liu_2012}%
  \BibitemOpen
  \bibfield  {author} {\bibinfo {author} {\bibfnamefont {S.-F.}\ \bibnamefont {Liu}}, \bibinfo {author} {\bibfnamefont {J.}~\bibnamefont {Guillochon}}, \bibinfo {author} {\bibfnamefont {D.~N.~C.}\ \bibnamefont {Lin}},\ and\ \bibinfo {author} {\bibfnamefont {E.}~\bibnamefont {Ramirez-Ruiz}},\ }\bibfield  {title} {\bibinfo {title} {On the survivability and metamorphism of tidally disrupted giant planets: The role of dense cores},\ }\href {https://doi.org/10.1088/0004-637x/762/1/37} {\bibfield  {journal} {\bibinfo  {journal} {Astrophys.~J.}\ }\textbf {\bibinfo {volume} {762}},\ \bibinfo {pages} {37} (\bibinfo {year} {2012})}\BibitemShut {NoStop}%
\bibitem [{\citenamefont {Mardling}(1995)}]{Mardling_1995}%
  \BibitemOpen
  \bibfield  {author} {\bibinfo {author} {\bibfnamefont {R.~A.}\ \bibnamefont {Mardling}},\ }\bibfield  {title} {\bibinfo {title} {The role of chaos in the circularization of tidal capture binaries. i. the chaos boundary},\ }\href {https://doi.org/10.1086/176178} {\bibfield  {journal} {\bibinfo  {journal} {Astrophys.~J.}\ }\textbf {\bibinfo {volume} {450}},\ \bibinfo {pages} {722} (\bibinfo {year} {1995})}\BibitemShut {NoStop}%
\bibitem [{\citenamefont {Ivanov}\ and\ \citenamefont {Papaloizou}(2004)}]{Ivanov_2004}%
  \BibitemOpen
  \bibfield  {author} {\bibinfo {author} {\bibfnamefont {P.~B.}\ \bibnamefont {Ivanov}}\ and\ \bibinfo {author} {\bibfnamefont {J.~C.~B.}\ \bibnamefont {Papaloizou}},\ }\bibfield  {title} {\bibinfo {title} {On the tidal interaction of massive extrasolar planets on highly eccentric orbits},\ }\href {https://doi.org/10.1111/j.1365-2966.2004.07238.x} {\bibfield  {journal} {\bibinfo  {journal} {Mon.~Not.~R.~Astron.~Soc.}\ }\textbf {\bibinfo {volume} {347}},\ \bibinfo {pages} {437–453} (\bibinfo {year} {2004})}\BibitemShut {NoStop}%
\bibitem [{\citenamefont {Ivanov}\ and\ \citenamefont {Papaloizou}(2007)}]{Ivanov_2007}%
  \BibitemOpen
  \bibfield  {author} {\bibinfo {author} {\bibfnamefont {P.~B.}\ \bibnamefont {Ivanov}}\ and\ \bibinfo {author} {\bibfnamefont {J.~C.~B.}\ \bibnamefont {Papaloizou}},\ }\bibfield  {title} {\bibinfo {title} {Orbital circularisation of white dwarfs and the formation of gravitational radiation sources in star clusters containing an intermediate mass black hole},\ }\href {https://doi.org/10.1051/0004-6361:20077105} {\bibfield  {journal} {\bibinfo  {journal} {Astron.~Astrophys.}\ }\textbf {\bibinfo {volume} {476}},\ \bibinfo {pages} {121–135} (\bibinfo {year} {2007})}\BibitemShut {NoStop}%
\bibitem [{\citenamefont {Vick}\ and\ \citenamefont {Lai}(2018)}]{Vick_2018}%
  \BibitemOpen
  \bibfield  {author} {\bibinfo {author} {\bibfnamefont {M.}~\bibnamefont {Vick}}\ and\ \bibinfo {author} {\bibfnamefont {D.}~\bibnamefont {Lai}},\ }\bibfield  {title} {\bibinfo {title} {Dynamical tides in highly eccentric binaries: chaos, dissipation, and quasi-steady state},\ }\href {https://doi.org/10.1093/mnras/sty225} {\bibfield  {journal} {\bibinfo  {journal} {Mon.~Not.~R.~Astron.~Soc.}\ }\textbf {\bibinfo {volume} {476}},\ \bibinfo {pages} {482–495} (\bibinfo {year} {2018})}\BibitemShut {NoStop}%
\bibitem [{\citenamefont {Wu}(2018)}]{Wu_2018}%
  \BibitemOpen
  \bibfield  {author} {\bibinfo {author} {\bibfnamefont {Y.}~\bibnamefont {Wu}},\ }\bibfield  {title} {\bibinfo {title} {Diffusive tidal evolution for migrating hot jupiters},\ }\href {https://doi.org/10.3847/1538-3881/aaa970} {\bibfield  {journal} {\bibinfo  {journal} {Astrophys.~J.}\ }\textbf {\bibinfo {volume} {155}},\ \bibinfo {pages} {118} (\bibinfo {year} {2018})}\BibitemShut {NoStop}%
\bibitem [{\citenamefont {Yu}\ \emph {et~al.}(2021)\citenamefont {Yu}, \citenamefont {Weinberg},\ and\ \citenamefont {Arras}}]{Yu_2021}%
  \BibitemOpen
  \bibfield  {author} {\bibinfo {author} {\bibfnamefont {H.}~\bibnamefont {Yu}}, \bibinfo {author} {\bibfnamefont {N.~N.}\ \bibnamefont {Weinberg}},\ and\ \bibinfo {author} {\bibfnamefont {P.}~\bibnamefont {Arras}},\ }\bibfield  {title} {\bibinfo {title} {Tides in the high-eccentricity migration of hot jupiters: Triggering diffusive growth by nonlinear mode interactions},\ }\href {https://doi.org/10.3847/1538-4357/ac0a79} {\bibfield  {journal} {\bibinfo  {journal} {Astrophys.~J.}\ }\textbf {\bibinfo {volume} {917}},\ \bibinfo {pages} {31} (\bibinfo {year} {2021})}\BibitemShut {NoStop}%
\bibitem [{\citenamefont {Yu}\ \emph {et~al.}(2022)\citenamefont {Yu}, \citenamefont {Weinberg},\ and\ \citenamefont {Arras}}]{Yu_2022}%
  \BibitemOpen
  \bibfield  {author} {\bibinfo {author} {\bibfnamefont {H.}~\bibnamefont {Yu}}, \bibinfo {author} {\bibfnamefont {N.~N.}\ \bibnamefont {Weinberg}},\ and\ \bibinfo {author} {\bibfnamefont {P.}~\bibnamefont {Arras}},\ }\bibfield  {title} {\bibinfo {title} {Tidal evolution and diffusive growth during high-eccentricity planet migration: Revisiting the eccentricity distribution of hot jupiters},\ }\href {https://doi.org/10.3847/1538-4357/ac5627} {\bibfield  {journal} {\bibinfo  {journal} {Astrophys.~J.}\ }\textbf {\bibinfo {volume} {928}},\ \bibinfo {pages} {140} (\bibinfo {year} {2022})}\BibitemShut {NoStop}%
\bibitem [{\citenamefont {Vick}\ \emph {et~al.}(2019)\citenamefont {Vick}, \citenamefont {Lai},\ and\ \citenamefont {Anderson}}]{Vick_2019}%
  \BibitemOpen
  \bibfield  {author} {\bibinfo {author} {\bibfnamefont {M.}~\bibnamefont {Vick}}, \bibinfo {author} {\bibfnamefont {D.}~\bibnamefont {Lai}},\ and\ \bibinfo {author} {\bibfnamefont {K.~R.}\ \bibnamefont {Anderson}},\ }\bibfield  {title} {\bibinfo {title} {Chaotic tides in migrating gas giants: Forming hot and transient warm jupiters via lidov-kozai migration},\ }\bibfield  {journal} {\bibinfo  {journal} {Mon.~Not.~R.~Astron.~Soc.}\ }\href {https://doi.org/10.1093/mnras/stz354} {10.1093/mnras/stz354} (\bibinfo {year} {2019})\BibitemShut {NoStop}%
\bibitem [{\citenamefont {MacLeod}\ \emph {et~al.}(2022)\citenamefont {MacLeod}, \citenamefont {Vick},\ and\ \citenamefont {Loeb}}]{MacLeod_2022}%
  \BibitemOpen
  \bibfield  {author} {\bibinfo {author} {\bibfnamefont {M.}~\bibnamefont {MacLeod}}, \bibinfo {author} {\bibfnamefont {M.}~\bibnamefont {Vick}},\ and\ \bibinfo {author} {\bibfnamefont {A.}~\bibnamefont {Loeb}},\ }\bibfield  {title} {\bibinfo {title} {Tidal wave breaking in the eccentric lead-in to mass transfer and common envelope phases},\ }\href {https://doi.org/10.3847/1538-4357/ac8aff} {\bibfield  {journal} {\bibinfo  {journal} {Astrophys.~J.}\ }\textbf {\bibinfo {volume} {937}},\ \bibinfo {pages} {37} (\bibinfo {year} {2022})}\BibitemShut {NoStop}%
\bibitem [{\citenamefont {Chakraborty}\ \emph {et~al.}(2024)\citenamefont {Chakraborty}, \citenamefont {Arcodia}, \citenamefont {Kara}, \citenamefont {Miniutti}, \citenamefont {Giustini}, \citenamefont {Tetarenko}, \citenamefont {Rhodes}, \citenamefont {Franchini}, \citenamefont {Bonetti}, \citenamefont {Burdge}, \citenamefont {Goodwin}, \citenamefont {Maccarone}, \citenamefont {Merloni}, \citenamefont {Ponti}, \citenamefont {Remillard},\ and\ \citenamefont {Saxton}}]{Chakraborty_2024}%
  \BibitemOpen
  \bibfield  {author} {\bibinfo {author} {\bibfnamefont {J.}~\bibnamefont {Chakraborty}}, \bibinfo {author} {\bibfnamefont {R.}~\bibnamefont {Arcodia}}, \bibinfo {author} {\bibfnamefont {E.}~\bibnamefont {Kara}}, \bibinfo {author} {\bibfnamefont {G.}~\bibnamefont {Miniutti}}, \bibinfo {author} {\bibfnamefont {M.}~\bibnamefont {Giustini}}, \bibinfo {author} {\bibfnamefont {A.~J.}\ \bibnamefont {Tetarenko}}, \bibinfo {author} {\bibfnamefont {L.}~\bibnamefont {Rhodes}}, \bibinfo {author} {\bibfnamefont {A.}~\bibnamefont {Franchini}}, \bibinfo {author} {\bibfnamefont {M.}~\bibnamefont {Bonetti}}, \bibinfo {author} {\bibfnamefont {K.~B.}\ \bibnamefont {Burdge}}, \bibinfo {author} {\bibfnamefont {A.~J.}\ \bibnamefont {Goodwin}}, \bibinfo {author} {\bibfnamefont {T.~J.}\ \bibnamefont {Maccarone}}, \bibinfo {author} {\bibfnamefont {A.}~\bibnamefont {Merloni}}, \bibinfo {author} {\bibfnamefont {G.}~\bibnamefont {Ponti}}, \bibinfo {author} {\bibfnamefont {R.~A.}\ \bibnamefont {Remillard}},\ and\ \bibinfo {author}
  {\bibfnamefont {R.~D.}\ \bibnamefont {Saxton}},\ }\bibfield  {title} {\bibinfo {title} {Testing emri models for quasi-periodic eruptions with 3.5 yr of monitoring ero-qpe1},\ }\href {https://doi.org/10.3847/1538-4357/ad2941} {\bibfield  {journal} {\bibinfo  {journal} {Astrophys.~J.}\ }\textbf {\bibinfo {volume} {965}},\ \bibinfo {pages} {12} (\bibinfo {year} {2024})}\BibitemShut {NoStop}%
\bibitem [{\citenamefont {Rubbo}\ \emph {et~al.}(2006)\citenamefont {Rubbo}, \citenamefont {Holley-Bockelmann},\ and\ \citenamefont {Finn}}]{Rubbo_2006}%
  \BibitemOpen
  \bibfield  {author} {\bibinfo {author} {\bibfnamefont {L.~J.}\ \bibnamefont {Rubbo}}, \bibinfo {author} {\bibfnamefont {K.}~\bibnamefont {Holley-Bockelmann}},\ and\ \bibinfo {author} {\bibfnamefont {L.~S.}\ \bibnamefont {Finn}},\ }\bibfield  {title} {\bibinfo {title} {Event rate for extreme mass ratio burst signals in the lisa band},\ }in\ \href {https://doi.org/10.1063/1.2405057} {\emph {\bibinfo {booktitle} {AIP Conference Proceedings}}},\ Vol.\ \bibinfo {volume} {873}\ (\bibinfo  {publisher} {AIP},\ \bibinfo {year} {2006})\ p.\ \bibinfo {pages} {284–288}\BibitemShut {NoStop}%
\bibitem [{\citenamefont {Hopman}\ \emph {et~al.}(2007)\citenamefont {Hopman}, \citenamefont {Freitag},\ and\ \citenamefont {Larson}}]{Hopman_2007}%
  \BibitemOpen
  \bibfield  {author} {\bibinfo {author} {\bibfnamefont {C.}~\bibnamefont {Hopman}}, \bibinfo {author} {\bibfnamefont {M.}~\bibnamefont {Freitag}},\ and\ \bibinfo {author} {\bibfnamefont {S.~L.}\ \bibnamefont {Larson}},\ }\bibfield  {title} {\bibinfo {title} {Gravitational wave bursts from the galactic massive black hole},\ }\href {https://doi.org/10.1111/j.1365-2966.2007.11758.x} {\bibfield  {journal} {\bibinfo  {journal} {Mon.~Not.~R.~Astron.~Soc.}\ }\textbf {\bibinfo {volume} {378}},\ \bibinfo {pages} {129–136} (\bibinfo {year} {2007})}\BibitemShut {NoStop}%
\bibitem [{\citenamefont {Berry}\ and\ \citenamefont {Gair}(2012)}]{Berry_2012}%
  \BibitemOpen
  \bibfield  {author} {\bibinfo {author} {\bibfnamefont {C.~P.~L.}\ \bibnamefont {Berry}}\ and\ \bibinfo {author} {\bibfnamefont {J.~R.}\ \bibnamefont {Gair}},\ }\bibfield  {title} {\bibinfo {title} {Observing the galaxy’s massive black hole with gravitational wave bursts},\ }\href {https://doi.org/10.1093/mnras/sts360} {\bibfield  {journal} {\bibinfo  {journal} {Mon.~Not.~R.~Astron.~Soc.}\ }\textbf {\bibinfo {volume} {429}},\ \bibinfo {pages} {589–612} (\bibinfo {year} {2012})}\BibitemShut {NoStop}%
\bibitem [{\citenamefont {Maguire}\ \emph {et~al.}(2020)\citenamefont {Maguire}, \citenamefont {Eracleous}, \citenamefont {Jonker}, \citenamefont {MacLeod},\ and\ \citenamefont {Rosswog}}]{Maguire_2020}%
  \BibitemOpen
  \bibfield  {author} {\bibinfo {author} {\bibfnamefont {K.}~\bibnamefont {Maguire}}, \bibinfo {author} {\bibfnamefont {M.}~\bibnamefont {Eracleous}}, \bibinfo {author} {\bibfnamefont {P.~G.}\ \bibnamefont {Jonker}}, \bibinfo {author} {\bibfnamefont {M.}~\bibnamefont {MacLeod}},\ and\ \bibinfo {author} {\bibfnamefont {S.}~\bibnamefont {Rosswog}},\ }\bibfield  {title} {\bibinfo {title} {Tidal disruptions of white dwarfs: Theoretical models and observational prospects},\ }\bibfield  {journal} {\bibinfo  {journal} {Space~Sci.~Rev.}\ }\textbf {\bibinfo {volume} {216}},\ \href {https://doi.org/10.1007/s11214-020-00661-2} {10.1007/s11214-020-00661-2} (\bibinfo {year} {2020})\BibitemShut {NoStop}%
\bibitem [{\citenamefont {Coughlin}\ and\ \citenamefont {Nixon}(2022)}]{Coughlin_2022}%
  \BibitemOpen
  \bibfield  {author} {\bibinfo {author} {\bibfnamefont {E.~R.}\ \bibnamefont {Coughlin}}\ and\ \bibinfo {author} {\bibfnamefont {C.~J.}\ \bibnamefont {Nixon}},\ }\bibfield  {title} {\bibinfo {title} {A simple and accurate prescription for the tidal disruption radius of a star and the peak accretion rate in tidal disruption events},\ }\href {https://doi.org/10.1093/mnrasl/slac106} {\bibfield  {journal} {\bibinfo  {journal} {Mon.~Not.~R.~Astron.~Soc.:~Lett.}\ }\textbf {\bibinfo {volume} {517}},\ \bibinfo {pages} {L26–L30} (\bibinfo {year} {2022})}\BibitemShut {NoStop}%
\bibitem [{\citenamefont {Manukian}\ \emph {et~al.}(2013)\citenamefont {Manukian}, \citenamefont {Guillochon}, \citenamefont {Ramirez-Ruiz},\ and\ \citenamefont {O’Leary}}]{Manukian_2013}%
  \BibitemOpen
  \bibfield  {author} {\bibinfo {author} {\bibfnamefont {H.}~\bibnamefont {Manukian}}, \bibinfo {author} {\bibfnamefont {J.}~\bibnamefont {Guillochon}}, \bibinfo {author} {\bibfnamefont {E.}~\bibnamefont {Ramirez-Ruiz}},\ and\ \bibinfo {author} {\bibfnamefont {R.~M.}\ \bibnamefont {O’Leary}},\ }\bibfield  {title} {\bibinfo {title} {Turbovelocity stars: Kicks resulting from the tidal disruption of solitary stars},\ }\href {https://doi.org/10.1088/2041-8205/771/2/l28} {\bibfield  {journal} {\bibinfo  {journal} {Astrophys.~J.~Lett.}\ }\textbf {\bibinfo {volume} {771}},\ \bibinfo {pages} {L28} (\bibinfo {year} {2013})}\BibitemShut {NoStop}%
\bibitem [{\citenamefont {Coughlin}\ and\ \citenamefont {Nixon}(2025)}]{Coughlin_2025}%
  \BibitemOpen
  \bibfield  {author} {\bibinfo {author} {\bibfnamefont {E.~R.}\ \bibnamefont {Coughlin}}\ and\ \bibinfo {author} {\bibfnamefont {C.~J.}\ \bibnamefont {Nixon}},\ }\href {https://doi.org/10.48550/ARXIV.2503.19018} {\bibinfo {title} {What's kickin' in partial tidal disruption events?}} (\bibinfo {year} {2025})\BibitemShut {NoStop}%
\bibitem [{\citenamefont {Yu}\ and\ \citenamefont {Lau}()}]{Yu_prep}%
  \BibitemOpen
  \bibfield  {author} {\bibinfo {author} {\bibfnamefont {H.}~\bibnamefont {Yu}}\ and\ \bibinfo {author} {\bibfnamefont {S.~Y.}\ \bibnamefont {Lau}},\ }\bibinfo {title} {in preparation}\BibitemShut {NoStop}%
\bibitem [{\citenamefont {Huang}(1963)}]{Huang_1963}%
  \BibitemOpen
\bibfield  {title} {  }\bibfield  {author} {\bibinfo {author} {\bibfnamefont {S.-S.}\ \bibnamefont {Huang}},\ }\bibfield  {title} {\bibinfo {title} {Modes of mass ejection by binary stars and the effect on their orbital periods.},\ }\href {https://doi.org/10.1086/147659} {\bibfield  {journal} {\bibinfo  {journal} {Astrophys.~J.}\ }\textbf {\bibinfo {volume} {138}},\ \bibinfo {pages} {471} (\bibinfo {year} {1963})}\BibitemShut {NoStop}%
\bibitem [{\citenamefont {MacLeod}\ \emph {et~al.}(2018)\citenamefont {MacLeod}, \citenamefont {Ostriker},\ and\ \citenamefont {Stone}}]{MacLeod_2018}%
  \BibitemOpen
  \bibfield  {author} {\bibinfo {author} {\bibfnamefont {M.}~\bibnamefont {MacLeod}}, \bibinfo {author} {\bibfnamefont {E.~C.}\ \bibnamefont {Ostriker}},\ and\ \bibinfo {author} {\bibfnamefont {J.~M.}\ \bibnamefont {Stone}},\ }\bibfield  {title} {\bibinfo {title} {Runaway coalescence at the onset of common envelope episodes},\ }\href {https://doi.org/10.3847/1538-4357/aacf08} {\bibfield  {journal} {\bibinfo  {journal} {Astrophys.~J.}\ }\textbf {\bibinfo {volume} {863}},\ \bibinfo {pages} {5} (\bibinfo {year} {2018})}\BibitemShut {NoStop}%
\bibitem [{\citenamefont {Schenk}\ \emph {et~al.}(2001)\citenamefont {Schenk}, \citenamefont {Arras}, \citenamefont {Flanagan}, \citenamefont {Teukolsky},\ and\ \citenamefont {Wasserman}}]{Schenk_2001}%
  \BibitemOpen
  \bibfield  {author} {\bibinfo {author} {\bibfnamefont {A.~K.}\ \bibnamefont {Schenk}}, \bibinfo {author} {\bibfnamefont {P.}~\bibnamefont {Arras}}, \bibinfo {author} {\bibfnamefont {E.~E.}\ \bibnamefont {Flanagan}}, \bibinfo {author} {\bibfnamefont {S.~A.}\ \bibnamefont {Teukolsky}},\ and\ \bibinfo {author} {\bibfnamefont {I.}~\bibnamefont {Wasserman}},\ }\bibfield  {title} {\bibinfo {title} {Nonlinear mode coupling in rotating stars and the r-mode instability in neutron stars},\ }\href {https://doi.org/10.1103/PhysRevD.65.024001} {\bibfield  {journal} {\bibinfo  {journal} {Phys. Rev. D}\ }\textbf {\bibinfo {volume} {65}},\ \bibinfo {pages} {024001} (\bibinfo {year} {2001})}\BibitemShut {NoStop}%
\bibitem [{\citenamefont {Weinberg}\ \emph {et~al.}(2012)\citenamefont {Weinberg}, \citenamefont {Arras}, \citenamefont {Quataert},\ and\ \citenamefont {Burkart}}]{Weinberg_2012}%
  \BibitemOpen
  \bibfield  {author} {\bibinfo {author} {\bibfnamefont {N.~N.}\ \bibnamefont {Weinberg}}, \bibinfo {author} {\bibfnamefont {P.}~\bibnamefont {Arras}}, \bibinfo {author} {\bibfnamefont {E.}~\bibnamefont {Quataert}},\ and\ \bibinfo {author} {\bibfnamefont {J.}~\bibnamefont {Burkart}},\ }\bibfield  {title} {\bibinfo {title} {Nonlinear tides in close binary systems},\ }\href {https://doi.org/10.1088/0004-637x/751/2/136} {\bibfield  {journal} {\bibinfo  {journal} {Astrophys.~J.}\ }\textbf {\bibinfo {volume} {751}},\ \bibinfo {pages} {136} (\bibinfo {year} {2012})}\BibitemShut {NoStop}%
\bibitem [{\citenamefont {Lai}(1997)}]{Lai_1997}%
  \BibitemOpen
  \bibfield  {author} {\bibinfo {author} {\bibfnamefont {D.}~\bibnamefont {Lai}},\ }\bibfield  {title} {\bibinfo {title} {Dynamical tides in rotating binary stars},\ }\href {https://doi.org/10.1086/304899} {\bibfield  {journal} {\bibinfo  {journal} {Astrophys.~J.}\ }\textbf {\bibinfo {volume} {490}},\ \bibinfo {pages} {847–862} (\bibinfo {year} {1997})}\BibitemShut {NoStop}%
\bibitem [{\citenamefont {Peters}\ and\ \citenamefont {Mathews}(1963)}]{Peters_1963}%
  \BibitemOpen
  \bibfield  {author} {\bibinfo {author} {\bibfnamefont {P.~C.}\ \bibnamefont {Peters}}\ and\ \bibinfo {author} {\bibfnamefont {J.}~\bibnamefont {Mathews}},\ }\bibfield  {title} {\bibinfo {title} {Gravitational radiation from point masses in a keplerian orbit},\ }\href {https://doi.org/10.1103/PhysRev.131.435} {\bibfield  {journal} {\bibinfo  {journal} {Phys. Rev.}\ }\textbf {\bibinfo {volume} {131}},\ \bibinfo {pages} {435} (\bibinfo {year} {1963})}\BibitemShut {NoStop}%
\bibitem [{\citenamefont {Lee}\ and\ \citenamefont {Ostriker}(1986)}]{Lee_1986}%
  \BibitemOpen
  \bibfield  {author} {\bibinfo {author} {\bibfnamefont {H.~M.}\ \bibnamefont {Lee}}\ and\ \bibinfo {author} {\bibfnamefont {J.~P.}\ \bibnamefont {Ostriker}},\ }\bibfield  {title} {\bibinfo {title} {Cross sections for tidal capture binary formation and stellar merger},\ }\href {https://doi.org/10.1086/164674} {\bibfield  {journal} {\bibinfo  {journal} {Astrophys.~J.}\ }\textbf {\bibinfo {volume} {310}},\ \bibinfo {pages} {176} (\bibinfo {year} {1986})}\BibitemShut {NoStop}%
\bibitem [{\citenamefont {García-Berro}\ \emph {et~al.}(2006)\citenamefont {García-Berro}, \citenamefont {Lorén-Aguilar}, \citenamefont {Córsico}, \citenamefont {Althaus}, \citenamefont {Lobo},\ and\ \citenamefont {Isern}}]{GarcaBerro_2006}%
  \BibitemOpen
  \bibfield  {author} {\bibinfo {author} {\bibfnamefont {E.}~\bibnamefont {García-Berro}}, \bibinfo {author} {\bibfnamefont {P.}~\bibnamefont {Lorén-Aguilar}}, \bibinfo {author} {\bibfnamefont {A.~H.}\ \bibnamefont {Córsico}}, \bibinfo {author} {\bibfnamefont {L.~G.}\ \bibnamefont {Althaus}}, \bibinfo {author} {\bibfnamefont {J.~A.}\ \bibnamefont {Lobo}},\ and\ \bibinfo {author} {\bibfnamefont {J.}~\bibnamefont {Isern}},\ }\bibfield  {title} {\bibinfo {title} {The gravitational wave radiation of pulsating white dwarfs revisited: the case of bpm 37093 and pg 1159-035},\ }\href {https://doi.org/10.1051/0004-6361:20053781} {\bibfield  {journal} {\bibinfo  {journal} {Astron.~Astrophys.}\ }\textbf {\bibinfo {volume} {446}},\ \bibinfo {pages} {259–266} (\bibinfo {year} {2006})}\BibitemShut {NoStop}%
\bibitem [{\citenamefont {Sepinsky}\ \emph {et~al.}(2009)\citenamefont {Sepinsky}, \citenamefont {Willems}, \citenamefont {Kalogera},\ and\ \citenamefont {Rasio}}]{Sepinsky_2009}%
  \BibitemOpen
  \bibfield  {author} {\bibinfo {author} {\bibfnamefont {J.~F.}\ \bibnamefont {Sepinsky}}, \bibinfo {author} {\bibfnamefont {B.}~\bibnamefont {Willems}}, \bibinfo {author} {\bibfnamefont {V.}~\bibnamefont {Kalogera}},\ and\ \bibinfo {author} {\bibfnamefont {F.~A.}\ \bibnamefont {Rasio}},\ }\bibfield  {title} {\bibinfo {title} {Interacting binaries with eccentric orbits. ii. secular orbital evolution due to non-conservative mass transfer},\ }\href {https://doi.org/10.1088/0004-637x/702/2/1387} {\bibfield  {journal} {\bibinfo  {journal} {Astrophys.~J.}\ }\textbf {\bibinfo {volume} {702}},\ \bibinfo {pages} {1387–1392} (\bibinfo {year} {2009})}\BibitemShut {NoStop}%
\bibitem [{\citenamefont {Yu}\ and\ \citenamefont {Dai}(2024)}]{Yu_2024}%
  \BibitemOpen
  \bibfield  {author} {\bibinfo {author} {\bibfnamefont {H.}~\bibnamefont {Yu}}\ and\ \bibinfo {author} {\bibfnamefont {F.}~\bibnamefont {Dai}},\ }\bibfield  {title} {\bibinfo {title} {Are wasp-107-like systems consistent with high-eccentricity migration?},\ }\href {https://doi.org/10.3847/1538-4357/ad5ffb} {\bibfield  {journal} {\bibinfo  {journal} {Astrophys.~J.}\ }\textbf {\bibinfo {volume} {972}},\ \bibinfo {pages} {159} (\bibinfo {year} {2024})}\BibitemShut {NoStop}%
\bibitem [{\citenamefont {Ulmer}(1999)}]{Ulmer_1999}%
  \BibitemOpen
  \bibfield  {author} {\bibinfo {author} {\bibfnamefont {A.}~\bibnamefont {Ulmer}},\ }\bibfield  {title} {\bibinfo {title} {Flares from the tidal disruption of stars by massive black holes},\ }\href {https://doi.org/10.1086/306909} {\bibfield  {journal} {\bibinfo  {journal} {Astrophys.~J.}\ }\textbf {\bibinfo {volume} {514}},\ \bibinfo {pages} {180–187} (\bibinfo {year} {1999})}\BibitemShut {NoStop}%
\bibitem [{\citenamefont {Chandrasekhar}(1963)}]{Chandrasekhar_1963}%
  \BibitemOpen
  \bibfield  {author} {\bibinfo {author} {\bibfnamefont {S.}~\bibnamefont {Chandrasekhar}},\ }\bibfield  {title} {\bibinfo {title} {The equilibrium and the stability of the roche ellipsoids.},\ }\href {https://doi.org/10.1086/147716} {\bibfield  {journal} {\bibinfo  {journal} {Astrophys.~J.}\ }\textbf {\bibinfo {volume} {138}},\ \bibinfo {pages} {1182} (\bibinfo {year} {1963})}\BibitemShut {NoStop}%
\bibitem [{\citenamefont {{Diener}}\ \emph {et~al.}(1995)\citenamefont {{Diener}}, \citenamefont {{Kosovichev}}, \citenamefont {{Kotok}}, \citenamefont {{Novikov}},\ and\ \citenamefont {{Pethick}}}]{Diener_1995}%
  \BibitemOpen
  \bibfield  {author} {\bibinfo {author} {\bibfnamefont {P.}~\bibnamefont {{Diener}}}, \bibinfo {author} {\bibfnamefont {A.~G.}\ \bibnamefont {{Kosovichev}}}, \bibinfo {author} {\bibfnamefont {E.~V.}\ \bibnamefont {{Kotok}}}, \bibinfo {author} {\bibfnamefont {I.~D.}\ \bibnamefont {{Novikov}}},\ and\ \bibinfo {author} {\bibfnamefont {C.~J.}\ \bibnamefont {{Pethick}}},\ }\bibfield  {title} {\bibinfo {title} {{Non-linear effects at tidal capture of stars by a massive black hole - II. Compressible affine models and tidal interaction after capture}},\ }\href {https://doi.org/10.1093/mnras/275.2.498} {\bibfield  {journal} {\bibinfo  {journal} {Mon.~Not.~R.~Astron.~Soc.}\ }\textbf {\bibinfo {volume} {275}},\ \bibinfo {pages} {498} (\bibinfo {year} {1995})}\BibitemShut {NoStop}%
\bibitem [{\citenamefont {{Pribulla}}(1998)}]{Pribulla_1998}%
  \BibitemOpen
  \bibfield  {author} {\bibinfo {author} {\bibfnamefont {T.}~\bibnamefont {{Pribulla}}},\ }\bibfield  {title} {\bibinfo {title} {{Efficiency of mass transfer and outflow in close binaries}},\ }\href {https://ui.adsabs.harvard.edu/abs/1998CoSka..28..101P} {\bibfield  {journal} {\bibinfo  {journal} {Contrib.~Astron.~Obs.~Skalnate~Pleso}\ }\textbf {\bibinfo {volume} {28}},\ \bibinfo {pages} {101} (\bibinfo {year} {1998})}\BibitemShut {NoStop}%
\bibitem [{\citenamefont {van Haaften}\ \emph {et~al.}(2012)\citenamefont {van Haaften}, \citenamefont {Nelemans}, \citenamefont {Voss}, \citenamefont {Wood},\ and\ \citenamefont {Kuijpers}}]{vanHaaften_2012}%
  \BibitemOpen
  \bibfield  {author} {\bibinfo {author} {\bibfnamefont {L.~M.}\ \bibnamefont {van Haaften}}, \bibinfo {author} {\bibfnamefont {G.}~\bibnamefont {Nelemans}}, \bibinfo {author} {\bibfnamefont {R.}~\bibnamefont {Voss}}, \bibinfo {author} {\bibfnamefont {M.~A.}\ \bibnamefont {Wood}},\ and\ \bibinfo {author} {\bibfnamefont {J.}~\bibnamefont {Kuijpers}},\ }\bibfield  {title} {\bibinfo {title} {The evolution of ultracompact x-ray binaries},\ }\href {https://doi.org/10.1051/0004-6361/201117880} {\bibfield  {journal} {\bibinfo  {journal} {Astron.~Astrophys.}\ }\textbf {\bibinfo {volume} {537}},\ \bibinfo {pages} {A104} (\bibinfo {year} {2012})}\BibitemShut {NoStop}%
\bibitem [{\citenamefont {Sberna}\ \emph {et~al.}(2021)\citenamefont {Sberna}, \citenamefont {Toubiana},\ and\ \citenamefont {Miller}}]{Sberna_2021}%
  \BibitemOpen
  \bibfield  {author} {\bibinfo {author} {\bibfnamefont {L.}~\bibnamefont {Sberna}}, \bibinfo {author} {\bibfnamefont {A.}~\bibnamefont {Toubiana}},\ and\ \bibinfo {author} {\bibfnamefont {M.~C.}\ \bibnamefont {Miller}},\ }\bibfield  {title} {\bibinfo {title} {Golden galactic binaries for lisa: Mass-transferring white dwarf black hole binaries},\ }\href {https://doi.org/10.3847/1538-4357/abccc7} {\bibfield  {journal} {\bibinfo  {journal} {Astrophys.~J.}\ }\textbf {\bibinfo {volume} {908}},\ \bibinfo {pages} {1} (\bibinfo {year} {2021})}\BibitemShut {NoStop}%
\bibitem [{\citenamefont {Kidder}(1995)}]{Kidder_1995}%
  \BibitemOpen
  \bibfield  {author} {\bibinfo {author} {\bibfnamefont {L.~E.}\ \bibnamefont {Kidder}},\ }\bibfield  {title} {\bibinfo {title} {Coalescing binary systems of compact objects to (post${)}^{5/2}$-newtonian order. v. spin effects},\ }\href {https://doi.org/10.1103/PhysRevD.52.821} {\bibfield  {journal} {\bibinfo  {journal} {Phys. Rev. D}\ }\textbf {\bibinfo {volume} {52}},\ \bibinfo {pages} {821} (\bibinfo {year} {1995})}\BibitemShut {NoStop}%
\end{thebibliography}%

\end{document}